%% file: draft.tex
\pdfoutput=1
\documentclass[letter,11pt]{article}
\usepackage{latexsym}
\usepackage{amsmath,amssymb,bm,ascmac,color,calc}
\usepackage{slashed}
\usepackage[mathscr]{eucal}
\usepackage{graphicx}
\usepackage{cite}
\usepackage{lineno}
\usepackage{jheppub}
\usepackage{multirow}
\usepackage{subfigure}
\usepackage[normalem]{ulem}
\usepackage[utf8x]{inputenc}
\usepackage[T1]{fontenc}

\usepackage{color}

\usepackage{makecell}

\allowdisplaybreaks
\newcommand{\ignore}[1]{}
\renewcommand{\bf}{\textbf}

\def\beq{\begin{equation}}
\def\eeq{\end{equation}}
\newcommand{\ba}{\begin{array}}
\newcommand{\ea}{\end{array}}
\newcommand{\bea}{\begin{eqnarray}}
\newcommand{\eea}{\end{eqnarray} }
\newcommand{\bal}{\begin{align}}
\newcommand{\eal}{\end{align}}
\def\bi{\begin{itemize}}
\def\ei{\end{itemize}}
\def\ben{\begin{enumerate}}
\def\een{\end{enumerate}}
\def\beq{\begin{equation}}
\def\eeq{\end{equation}}
\def\bc{\begin{center}}
\def\ec{\end{center}}
\def\bt{\begin{table}}
\def\et{\end{table}}
\def\btb{\begin{tabular}}
\def\etb{\end{tabular}}

\usepackage{xcolor}
\usepackage{pifont}
\usepackage{subfigure}

\definecolor{red1}{cmyk}{0,1,1,0.1}

\newcommand{\mpt}{{\;/\!\!\!\! {P}_T}} 
\newcommand{\mptvec}{{\;/\!\!\!\! \vec{P}_T}}

\newcommand{\amc}{{\sc MadGraph5\textunderscore}a{\sc MC@NLO}}

\title{Portraying Double Higgs at the Large Hadron Collider II } 

\author[a,b]{Li Huang,}
\author[c]{Su-beom Kang,}
\author[d,e,f]{Jeong Han Kim,}
\author[g]{Kyoungchul Kong,}
\author[d]{Jun Seung Pi}

\affiliation[a]{International Centre for Theoretical Physics Asia-Pacific,
University of Chinese Academy of Sciences, 100190 Beijing, China}
\affiliation[b]{Taiji Laboratory for Gravitational Wave Universe, University of Chinese Academy of Sciences, 100049 Beijing, China}
\affiliation[c]{Department of Physics, Sungkyunkwan University, Suwon, Gyeonggi-do 16419, Korea}
\affiliation[d]{Department of Physics, Chungbuk National University, Cheongju, Chungbuk 28644, Korea}
\affiliation[e]{Center for Theoretical Physics of the Universe, Institute for Basic Science, Daejeon 34126, Korea}
\affiliation[f]{School of Physics, KIAS, Seoul 02455, Korea}
\affiliation[g]{Department of Physics and Astronomy, University of Kansas, Lawrence, KS 66045, USA}

\emailAdd{huangli@ucas.ac.cn}
\emailAdd{subeom527@gmail.com}
\emailAdd{jeonghan.kim@cbu.ac.kr}
\emailAdd{kckong@ku.edu}
\emailAdd{junseung.pi@cbu.ac.kr}

\abstract{
The Higgs potential is vital to understand the electroweak symmetry breaking mechanism, and probing the Higgs self-interaction is arguably one of the most important physics targets at current and upcoming collider experiments. In particular, the triple Higgs coupling may be accessible at the HL-LHC by combining results in multiple channels, which motivates to study all possible decay modes for the double Higgs production. In this paper, we revisit the double Higgs production at the HL-LHC in the final state with two $b$-tagged jets, two leptons and missing transverse momentum. 
We focus on the performance of various neural network architectures with different input features: low-level (four momenta), high-level (kinematic variables) and image-based. 
We find it possible to bring a modest increase in the signal sensitivity over existing results via careful optimization of machine learning algorithms making a full use of novel kinematic variables.}

\begin{document} 
\maketitle

\section{Introduction}
\label{sec:intro}
\input{1_intro.tex}

\section{Theoretical setup and simulation}
\label{sec:eventgen}
\input{2_EventGen.tex}

\section{Pre-processing input data}
\label{sec:input}
\input{3_InputData.tex}

\section{Performance of machine learning algorithms}
\label{sec:analysis}
\input{4_analysis.tex}

\section{Comparison of different networks}
\label{sec:results}

\input{5_results.tex}

\section{Discussion and outlook}
\label{sec:outlook}
\input{6_conclusion.tex}

\bigskip\bigskip
\emph{Acknowledgements:} 
We thank Minho Kim, Myeonghun Park, and Konstantin Matchev for useful discussion and collaboration at the early stage of the work. This work is supported by Chungbuk National University Korea National University Development Project (2020). KK acknowledges support from the US DOE, Office of Science under contract DE-SC0021447. LH is supported by the Fundamental Research Funds for the Central Universities, and the Bureau of International Cooperation, Chinese Academy of Sciences.
\bigskip

 \appendix
 \input{7_appendix.tex}


\bibliographystyle{JHEP}
\bibliography{references}

\end{document}

%% file: 1_intro.tex
The discovery of the Higgs boson at the Large Hadron Collider (LHC) launched a comprehensive program of the precision measurements of all Higgs couplings. While the current data shows that the Higgs couplings to fermions and gauge bosons appear to be consistent with the predictions of the Standard Model (SM) \cite{Khachatryan:2016vau}, the Higgs self-couplings are yet to be probed at the LHC and at future colliders. The measurement of the Higgs self-couplings is vital to understand the electroweak symmetry breaking mechanism. 
In particular, the triple Higgs coupling is a guaranteed physics target that can be probed at the high luminosity (HL) LHC and the succeeding experimental bounds on the self-couplings will have an immediate and long-lasting impact on model-building for new physics beyond the SM. 

The Higgs ($h$) self-interaction is parameterized as 
\begin{equation}
V = \frac{m_h^2}{2} h^2 + \kappa_3 \lambda_{3}^{\rm SM} v h^3 + \frac{1}{4}\kappa_4  \lambda_{4}^{\rm SM} h^4 \label{eq:Vh} \, ,
\end{equation}
where $m_h$ is the Higgs mass, and $v$ is the vacuum expectation value of the Higgs field. 
$\lambda_{3}^{\rm SM} = \lambda_{4}^{\rm SM} = \frac{m_h^2}{2 v^2}$ are the SM Higgs triple and quartic couplings, while $\kappa_i$ ($i=3,4$) parameterize the deviation from the corresponding SM coupling. 
In order to access the triple (quartic) Higgs coupling, one has to measure the double (triple) Higgs boson production. 
In this paper we focus on probing the triple Higgs coupling at the HL-LHC, which is likely achievable when combining both ATLAS and CMS data \cite{DiMicco:2019ngk,ATLAS:2018rvj,CMS:2018ccd,Cepeda:2019klc} with a potential improvement on each decay channel\footnote{See Refs. \cite{Fuks:2017zkg,Fuks:2015hna,Papaefstathiou:2015paa, Chen:2015gva,Kilian:2017nio,Liu:2018peg,Bizon:2018syu,Borowka:2018pxx,Papaefstathiou:2019ofh,Chiesa:2020awd,Chen:2021pqi,Maltoni:2018ttu} for the quartic Higgs coupling at future colliders, and Refs. \cite{Adhikary:2020fqf,Agrawal:2019bpm,Park:2020yps} for the triple Higgs coupling at future colliders.}.

Double Higgs ($hh$) production has been extensively discussed in many different channels, such as 
$b \bar b \gamma \gamma$ \cite{CMS:2020tkr,ATLAS:2022qjq,ATLAS:2022okt,ATLAS:2021tyg,  CMS-PAS-FTR-15-002, ATL-PHYS-PUB-2014-019, Kim:2018uty, Kling:2016lay, Baur:2003gp, Baglio:2012np, Huang:2015tdv, Azatov:2015oxa,Cao:2015oaa,Cao:2016zob,Alves:2017ued, Barger:2013jfa, Chang:2018uwu,Chang:2019ncg}, 
$b \bar b \tau\tau$ \cite{CMS:2020010,ATLAS:2021hvg,ATLAS:2022okt,ATLAS:2021tyg,CMS-PAS-FTR-15-002,  Kim:2018uty, Baur:2003gpa, Goertz:2014qta, Dolan:2012rv}, 
$b \bar b b \bar b$ \cite{CMS:2022cpr,ATLAS:2021tyg,ATLAS:2018combi,Amacker:2020bmn, Aaboud:2018knk,  deLima:2014dta, Wardrope:2014kya, Behr:2015oqq}, 
$b \bar b W^+ W^- /ZZ$ \cite{Aaboud:2018zhh, ATLAS:2019mwn, CMS:2017ums, CMS:2015nat, CMS:2017cwx, Kim:2018cxf,  Papaefstathiou:2012qe, Huang:2017jws, CMS:2020gxr}, 
$W^+ W^- W^+ W^-$ \cite{CMS:2022ngs,Aaboud:2018ksn}, $W^+ W^- \tau \tau$ \cite{CMS:2022ngs}, and $\tau \tau \tau \tau$ \cite{CMS:2022ngs} (see also Refs. \cite{DiMicco:2019ngk,Cepeda:2019klc} and references therein). 
On the other hand, less attention was given to the final state with two $b$-tagged jets, two leptons and missing transverse momentum, as it suffers from large SM backgrounds, primarily due to the top quark pair production ($t\bar{t}$).  
Therefore several existing studies in this channel made use of sophisticated algorithms ($e.g.$ neural network (NN) \cite{CMS:2015nat}, boosted decision tree (BDT) \cite{Adhikary:2017jtu,CMS:2017cwx}, and deep neural network (DNN) \cite{Sirunyan:2017guj,ATLAS:2019vwv}) to increase the signal sensitivity, although they lead to somewhat pessimistic results, with a significance much smaller than $1\sigma$ at the HL-LHC with 3 ab$^{-1}$ luminosity. 
More recent studies claim that the significance can be greatly improved by utilizing novel kinematic methods \cite{Kim:2018cxf}, or by adopting more complex NNs such as convolutional neural networks (CNN) with jet images \cite{Kim:2019wns} and message passing neural networks (MPNN) with four-momentum information \cite{Abdughani:2020xfo}.

In particular, Ref. \cite{Kim:2018cxf} introduced two new kinematic variables (Topness and Higgsness) and investigated their impact on the reduction of $t\bar t$ background, along with $M_{T2}$ and $\hat{s}_{\rm min}$. They showed that the substantial improvement on the signal significance was due to the use of the kinematic variables which contain the mass information. In their follow-up study, Ref. \cite{Kim:2019wns}, authors performed more dedicated analysis using Delphes simulation, and included $tW$ background which was missing in the previous study. Ref. \cite{Kim:2019wns} utilized a simple convolutional neural network with those newly introduced kinematic variables along with existing ones (16 variables in total) to improve the signal significance. They also studied jet images (with charged and neutral hadrons only) in double Higgs production and showed that the additional improvement was possible.

The goal in this article is to extend the scope of the existing studies on the double Higgs production at the HL-LHC in the final state with $(h\to b{\bar b})(h\to W^\pm W^{*\mp}\to \ell^+\nu_\ell {\ell^\prime}^- \bar{\nu}_{\ell^\prime})$, by studying performance of various NN architectures. In particular, we would like to address the following important points, which were not answered properly in Refs. \cite{Kim:2018cxf,Kim:2019wns}. 
\begin{enumerate}
\item The performance of NNs with different types of input features: low-level (four momenta), high-level (kinematic variables), and image-based inputs. 
\item Ref. \cite{Kim:2019wns} used CNN with the jet images, which are the energy deposits of charged and neutral hadrons in the hadronic calorimeter. How robust are these results? How much error do we make due to different choice of hyper-parameters?  

\item In principle, the lepton momenta are correlated with the momenta of two $b$-quarks (and therefore their hadronic activities), so that one could consider the image of leptons simultaneously. Can the image-based NNs catch the non-trivial correlation between leptons and $b$-quarks?
\item Ref. \cite{Kim:2018cxf} introduces two novel kinematic variables (Topness and Higgsness), which provide a good signal-background separation. As a byproduct, one obtains the momentum of two missing neutrinos. What would be the best way to utilize the neutrino momentum information along with visible particles? 
Would the ``image'' of neutrinos provide an additional handle for the signal-background separation?
\item What are the signal efficiency and background rejection rate of different NN algorithms? 
\item How much improvement does the $bbWW$ channel bring in the combination of individual channels? 

\end{enumerate}

This paper is organized as follows. 
We begin in section~\ref{sec:eventgen} our discussion on the event generation for the signal and backgrounds, followed by data preparation for NN analysis in section \ref{sec:input}. 
In section \ref{sec:analysis}, we examine several NN architectures including deep neural networks (DNNs), convolutional neural networks (CNNs), residual neural networks (ResNets), graph neural networks (GNNs), capsule neural networks (CapsNets) and matrix capsule networks.
We will study their performances with the low-level (four momenta), high-level (kinematic variables), and image-based input data, which is summarized in section \ref{sec:results}.
Section \ref{sec:outlook} is reserved for a discussion and outlook. 
We provide a brief review on various kinematic variables in appendix \ref{appen:variable}.

%% file: 2_EventGen.tex
The signal ($hh$ with $\kappa_3 = 1$) and backgrounds are generated for a center-of-mass energy of $\sqrt{s} = 14$ TeV, using the \amc{} \cite{Alwall:2014hca,Hirschi:2015iia} which allows for both tree- and loop-level event generations. We use the default NNPDF2.3QED parton distribution functions \cite{Ball:2013hta} using dynamical renormalization and factorization scales.
We normalize the double Higgs production cross section to 40.7 fb at the next-to-next-to-leading order (NNLO) accuracy in QCD \cite{Grigo:2014jma}. The dominant background is $t\bar{t}$ production whose tree-level cross section is rescaled to the NNLO cross section 953.6 pb~\cite{Czakon:2013goa}. We consider only the leptonic decays of tops $t \bar{t} \rightarrow (b \ell^+ \nu) (\bar{b} \ell^- \bar{\nu})$ with $\ell$ being  $e, \mu, \tau$, that includes off-shell effects for the top and the $W$. The next dominant background is $tW+j$ production matched (five-flavor scheme) up to one additional jet in order to partially include the next-to-leading order (NLO) effects
\footnote{
The NLO $tW$ process mixes with LO $t\bar{t}$, which requires a careful treatment to avoid a double-counting. Several methods have been presented in the literature \cite{Frixione:2008yi,Belyaev:1998dn,White:2009yt,Tait:1999cf} to resolve the issue, including the diagram removal (DR) adopted in the ATLAS \cite{ATLAS:2019vwv} and CMS \cite{CMS:2011wds} analyses.
In this paper, we have followed the same DR scheme where
the $tW$ background is generated up to one additional $j$ excluding diagrams that overlap with $t \bar{t}$ process.}.
Both top and $W$ are decayed leptonically as for the $t\bar{t}$ sample. 
The sub-dominant backgrounds consist of $t\bar{t}h$ and $t\bar{t}V ~(V = W^{\pm}, Z)$ whose cross sections are normalized to 611.3~fb~\cite{Dittmaier:2011ti} and 1.71~pb~\cite{deFlorian:2016spz} at the NLO, respectively. We also include Drell-Yan backgrounds $\ell \ell b j$ and $\tau \tau b b$, where $j$ denotes partons in the five-flavor scheme. The NNLO k-factor of the Drell-Yan production \cite{deFlorian:2018wcj} is close to unity (k-factor $\approx$1).
The hard scattering events are decayed (unless mentioned otherwise), showered, and hadronized using {\sc Pythia8} \cite{Sjostrand:2014zea}. Detector effects are simulated with {\sc Delphes} 3.4.1 \cite{deFavereau:2013fsa} based on modified ATLAS configurations \cite{Kim:2019wns}.

Jets are reconstructed by {\sc Fastjet} 3.3.1 \cite{Cacciari:2011ma} implementation using the anti-$k_T$ algorithm \cite{Cacciari:2008gp} and a cone radius of $r = 0.4$. 
We take advantage of the improved $b$-tagging efficiency reported by ATLAS, associated with the central tracking system for the operation at the HL-LHC~\cite{CERN-LHCC-2017-021}. 
We use a flat $b$-tag rate of $\epsilon_{b \rightarrow b} = 0.8$, and a mistag rate that a $c$-jet (light-flavor jet) is misidentified as a $b$-jet, $\epsilon_{c \rightarrow b} = 0.2$ ($\epsilon_{j \rightarrow b} = 0.01$). 
Events with exactly two $b$-tagged jets which pass minimum cuts $p_T(b) > 30$ GeV and $|\eta(b)| < 2.5$ are considered. Two $b$-tagged jets are further required to satisfy a proximity cut $\Delta R_{bb} < 2.5$ and an invariant mass cut $70$ GeV $< m_{bb}<  140$ GeV.

A lepton is declared to be isolated if it satisfies $p_T(\ell)/(p_T(\ell)  + \sum_i p_{T_i}) > 0.7$ where $\sum_i p_{T_i}$ is the sum of the transverse momenta of nearby particles with $p_{T_i} > 0.5$ GeV and $\Delta R_{i\ell} < 0.3$.
Events with exactly two isolated leptons which pass minimum cuts $p_T(\ell) > 20$ GeV and $|\eta(\ell)| < 2.5$ are selected. Two leptons are further required to pass a proximity cut $\Delta R_{\ell \ell} < 1.5$ and an invariant mass cut $m_{\ell \ell}<  70$ GeV.
Events are required to pass the minimum missing transverse momentum (defined as in Ref. \cite{Kim:2019wns}) $\mpt = | \mptvec | > 20 $ GeV. 
After this baseline selection, the signal and background cross sections are summarized in Table \ref{tab:BGD}, including appropriate k-factors and taking into account the improved $b$-tagging efficiency and fake rates. 
The dominant background is $t\bar t$ (97\%), followed by $tW$ (2\%).
The background-to-signal cross section ratio is about 9250 after the baseline selection. 
Throughout the study in this paper, we will assume ${\cal L}$ = 3 ab$^{-1}$ for the integrated luminosity, unless otherwise mentioned.

\begin{table}[t]
\centering
  \renewcommand{\arraystretch}{1.1}
  \setlength\tabcolsep{6pt}
  \begin{tabular}{|c|c|} \hline
                   & Cross sections [fb] \\
   \hline
   \hline
      $hh$ ($\kappa_3 = 1$) & $2.81 \times 10^{-2}$\\ \hline
      $t \bar{t}$ & $2.52 \times 10^2$\\ \hline
      $tW + j$     & $5.73$\\ \hline
      $t \bar{t} h$     & $2.53 \times 10^{-1}$\\ \hline
      $t \bar{t} V$     & $3.18 \times 10^{-1}$\\ \hline
      $\ell \ell b j$     & $1.61$\\ \hline
       $\tau \tau b b$     & $1.49 \times 10^{-2}$\\ \hline
  \end{tabular}
  \caption{Signal and background cross sections after the baseline selection described in section \ref{sec:eventgen}, including appropriate k-factors as well as taking into account the improved $b$-tagging efficiency and fake rates.}
  \label{tab:BGD}
\end{table}

%% file: 3_InputData.tex
Data preparation or preprocessing is an important part of ML analysis. 
In particular, to fully understand performance of NNs with different types of inputs, we categorize input features used in this paper as follows.
The most basic information (low-level features) is four-momenta of four visible particles (two leptons and two $b$-jets)
\begin{eqnarray} 
V^{(\text{vis})}_{ p_{\mu} } &=& \{ p_\mu (\ell_1), p_\mu (\ell_2), p_\mu (b_1), p_\mu (b_2) \} \, ,
\label{eq:vis_4momenta}
\end{eqnarray}
where the dimension is $dim(V^{(\text{vis})}_{ p_{\mu} } ) = 16$.

There are various kinematic methods which provide approximate momenta of missing neutrinos. 
In this paper, we adopt Topness (in Eq.(\ref{eq:Tness1})) \cite{Graesser:2012qy,Kim:2018cxf} and Higgsness (in Eq.(\ref{eq:Hness})) \cite{Kim:2018cxf}, which are proposed for the double Higgs production in particular. As a result of minimization procedures, these momenta carry some kinematic features of the missing neutrinos approximately. 
For example, the neutrino momenta obtained using Topness are consistent with the top and $W$ mass constraints, while those obtained using Higgsness are consistent with the double Higgs production. Therefore we include them in our input variables as
\begin{eqnarray}  
V^{(\nu_{\text{T}})}_{p_{\mu}}    &=&  \{ p_\mu(\nu^{\text{T}}), p_\mu (\bar{\nu}^{\text{T}})  \}  , \label{eq:Tneu_4momenta} \\
V^{(\nu_{\text{H}})}_{ p_{\mu}}  &=&  \{ p_\mu(\nu^{\text{H}}), p_\mu (\bar{\nu}^{\text{H}})  \}  ,\label{eq:Hneu_4momenta} \\
V^{(\text{inv})}_{ p_{\mu} }  &=& V^{(\nu_{\text{T}})}_{p_{\mu}}   \,\, \cup  \,\,    V^{(\nu_{\text{H}})}_{ p_{\mu}} . \label{eq:inv}
\end{eqnarray}
Note that $dim(V^{(\nu_{\text{T}})}_{p_{\mu}} ) = dim(V^{(\nu_{\text{H}})}_{ p_{\mu}}) = 8$.

With those basic four-momenta information, we can construct 11 and 15 low-level kinematic variables as
\begin{eqnarray}  
V_{\text{11-kin}} &=& \{ p_T(\ell_1), p_T(\ell_2), \mpt, m_{\ell\ell}, m_{bb}, \Delta R_{\ell\ell}, \label{eq:11_kin} 
                                      \Delta R _{bb}, p_{Tbb}, p_{T\ell\ell}, \Delta\phi_{bb,\ell\ell}, \text{min}[\Delta R_{b\ell}]    \}  \, ,  \label{eq:11kin} \\
V_{\text{15-kin}} &=& V_{\text{11-kin}} \,\, \cup  \,\, \{  \Delta R^{\text{H}}_{\nu \nu}, m^{\text{H}}_{\nu \nu},  \Delta R^{\text{T}}_{\nu \nu}, m^{\text{T}}_{\nu \nu}   \} \, , \label{eq:15_kin}
\end{eqnarray}
where $\text{min}[\Delta R_{b\ell}]$ denotes the smallest angular distance between a $b$-jet and a lepton, $\Delta R^{\text{H}}_{\nu \nu}$ ($\Delta R^{\text{T}}_{\nu \nu}$) and $m^{\text{H}}_{\nu \nu}$ ($m^{\text{T}}_{\nu \nu}$) represent an angular distance and an invariant mass between two neutrinos reconstructed using Higgsness (H) or Topness (T) variables. 
We confirmed that distributions of the first 10 variables in $V_{\text{11-kin}}$ are similar to those in Ref. \cite{Kim:2019wns}. 
Although Ref. \cite{Kim:2019wns} did not utilize the last 5 variables in Eq.(\ref{eq:15_kin}) ($\{ \text{min}[\Delta R_{b\ell}], \Delta R^{\text{H}}_{\nu \nu}, m^{\text{H}}_{\nu \nu},  \Delta R^{\text{T}}_{\nu \nu}, m^{\text{T}}_{\nu \nu}   \} $), the $\text{min}[\Delta R_{b\ell}]$ delivers a direct angular information between a $b$-jet and a lepton, and the other 4 variables provide additional information on the neutrino sector. Therefore we include them in our analysis (see Fig. \ref{fig:5dists} for their distributions).
Note that some kinematic variables in $V_{\text{15-kin}}$ are strongly correlated with each other due to the pencil-like event shape of the $hh$ production, as compared to the isotropic dominant background ($t\bar t$) \cite{Kim:2018cxf,Kim:2019wns}. See Appendix for more details on the event shape variables for the signal and backgrounds. 
Once a few cuts are imposed, the effect of remaining cuts is significantly diminished. 
Therefore, it is important to investigate new kinematic variables, which are less correlated to those introduced in the early literatures. 

Among the low-level kinematic variables defined in $V_{\text{15-kin}}$, the invariant mass is the only mass variable. 
For more efficient background suppression utilizing mass information of top quark and Higgs, we use the following six high-level kinematic variables introduced in Refs. \cite{Kim:2018cxf,Kim:2019wns}:
\begin{eqnarray} 
V_{\text{6-kin}} &=& \{ \sqrt{\hat{s}}_{\text{min}}^{(bb\ell \ell)},  \sqrt{\hat{s}}_{\text{min}}^{(\ell \ell)}, M_{T2}^{(b)}, M_{T2}^{(\ell)}, {\rm H}, {\rm T} \}  \, , \label{eq:6_kin}
\end{eqnarray}
where $\sqrt{\hat s}^{(bb\ell\ell)}_{\text{min}}$ and $\sqrt{\hat s}^{(\ell\ell)}_{\text{min}}$ are the minimum energy required to be consistent with the produced events in the specified subsystem, $M_{T2}^{(b)}$, $M_{T2}^{(\ell)}$ are the stransverse mass variables, and T/H denote Topness and Higgsness. These six high-level variables are defined in Appendix \ref{appen:variable}. 
It has been shown that these high-level variables allow neural networks to learn key features of the processes faster and more accurately. We will use these 6 high-level kinematic variables along with $V_{\text{15-kin}}$,
\begin{eqnarray}
V_{\text{21-kin}} &=& V_{\text{15-kin}} \,\, \cup  \,\, V_{\text{6-kin}} \, . \label{eq:21}
\end{eqnarray}

\begin{figure}
\centering
\includegraphics[scale=0.292]{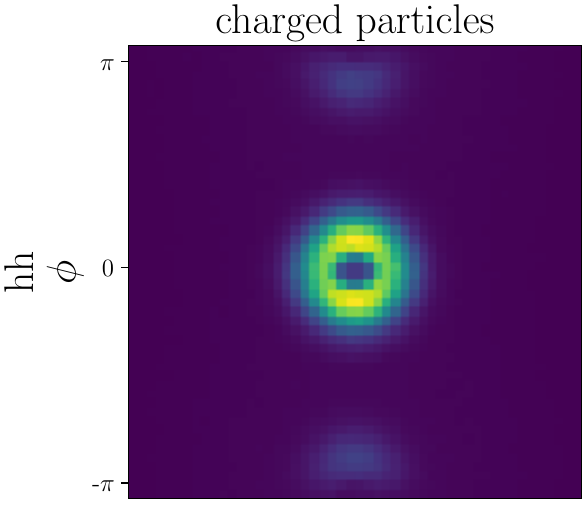}
\includegraphics[scale=0.292]{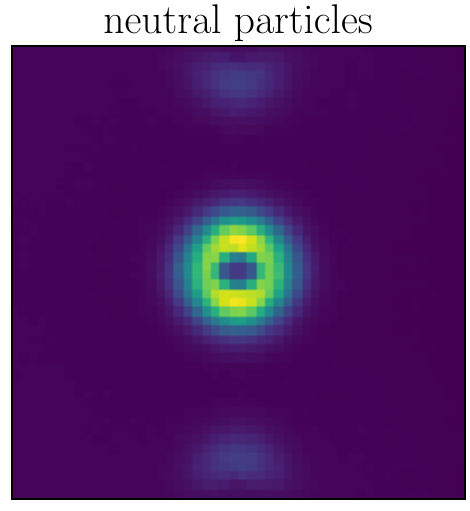}
\includegraphics[scale=0.292]{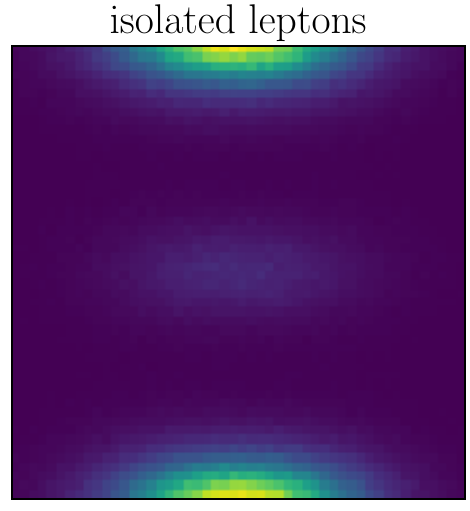}
\includegraphics[scale=0.292]{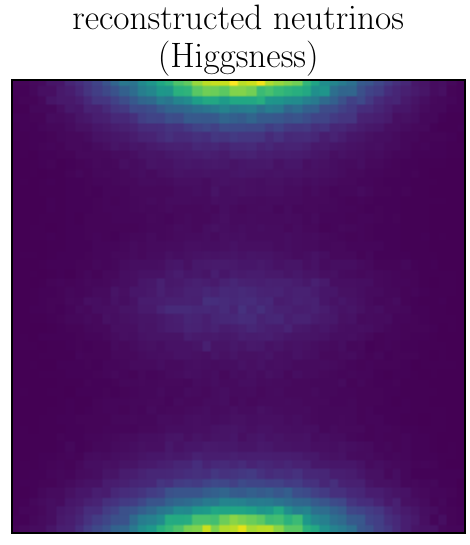}
\includegraphics[scale=0.292]{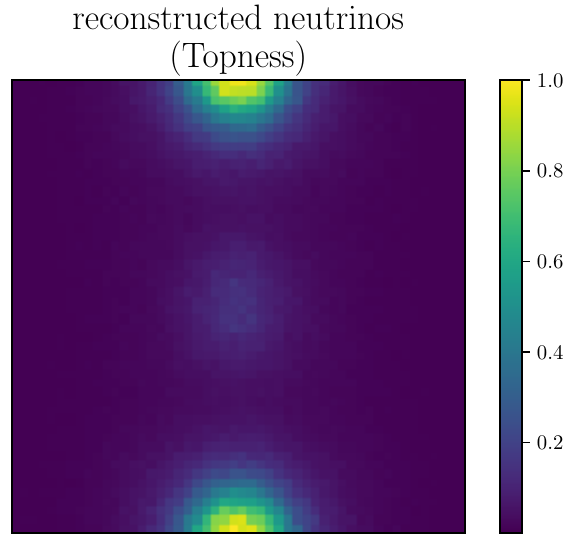} \\
\includegraphics[scale=0.292]{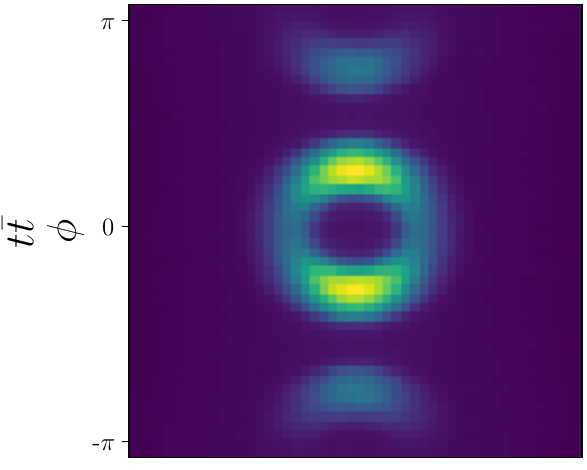}
\includegraphics[scale=0.292]{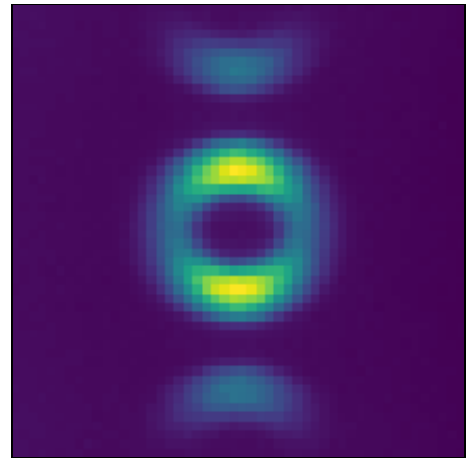}
\includegraphics[scale=0.292]{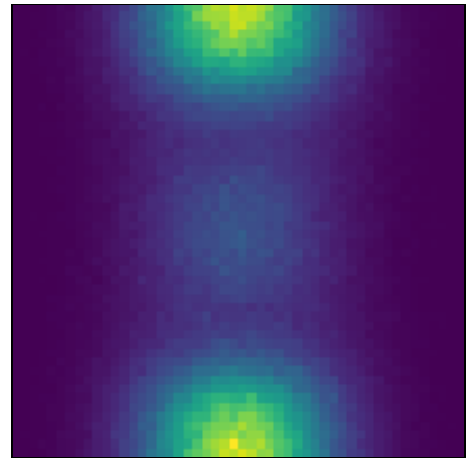}
\includegraphics[scale=0.292]{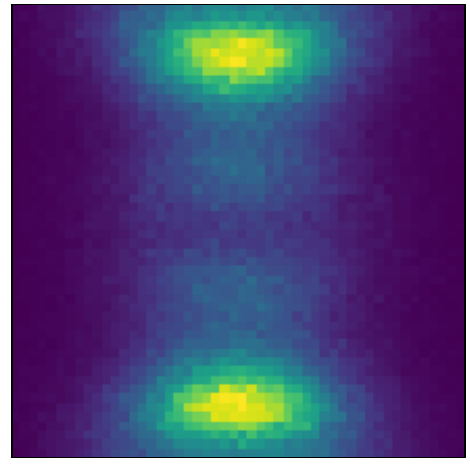}
\includegraphics[scale=0.292]{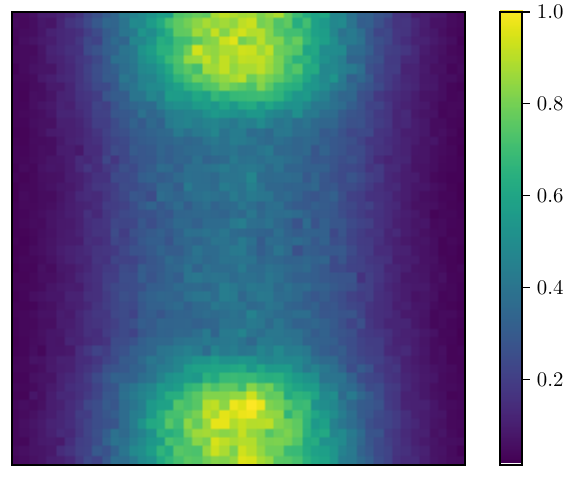}  \\
\includegraphics[scale=0.292]{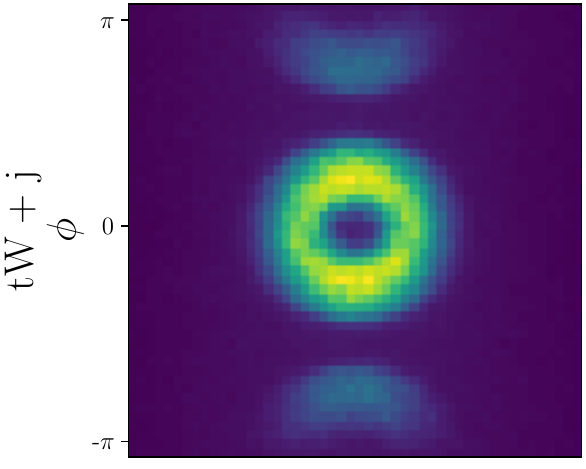}
\includegraphics[scale=0.292]{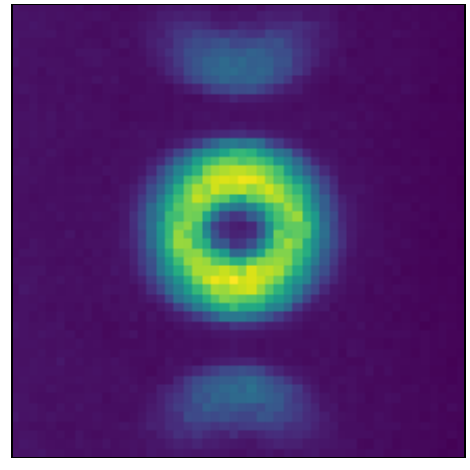}
\includegraphics[scale=0.292]{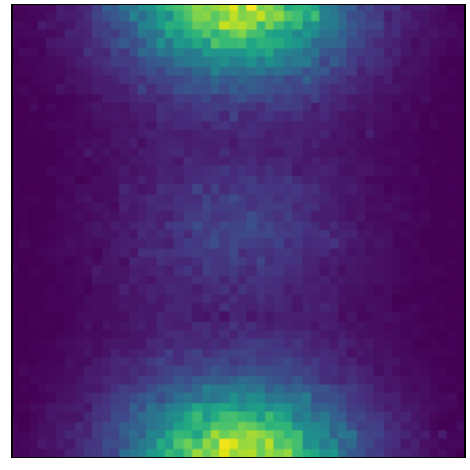}
\includegraphics[scale=0.292]{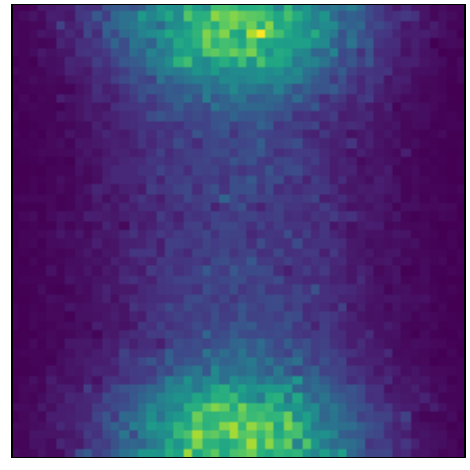}
\includegraphics[scale=0.292]{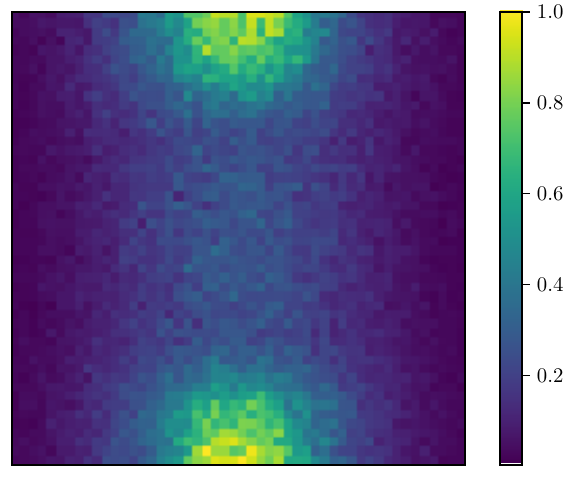}  \\
\includegraphics[scale=0.292]{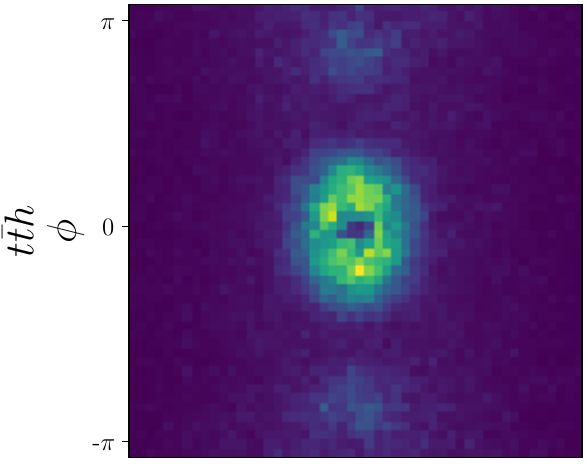}
\includegraphics[scale=0.292]{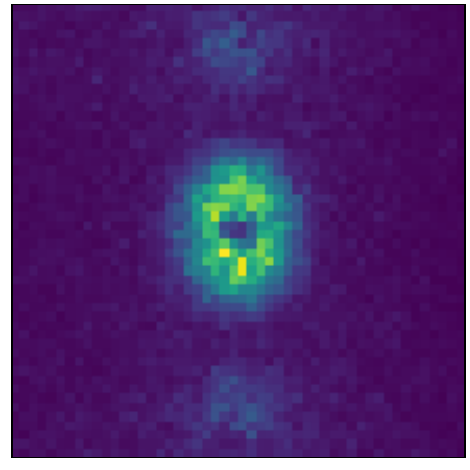}
\includegraphics[scale=0.292]{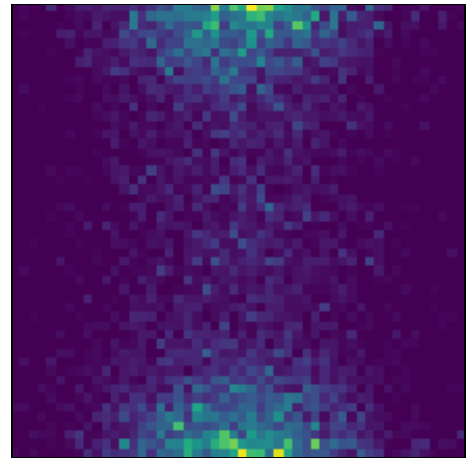}
\includegraphics[scale=0.292]{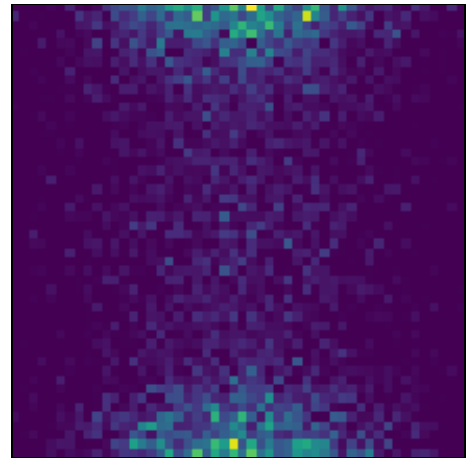}
\includegraphics[scale=0.292]{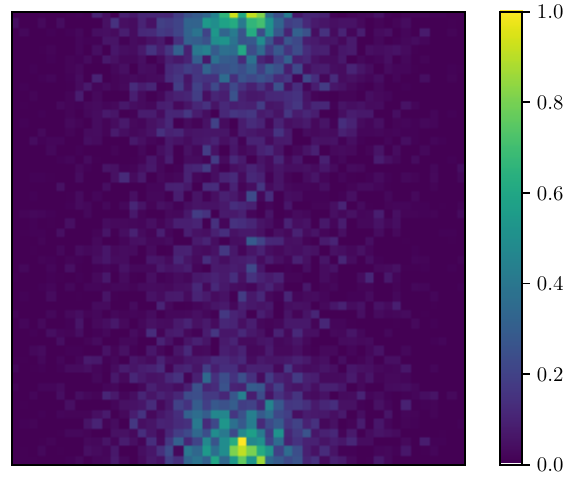} \\
\includegraphics[scale=0.292]{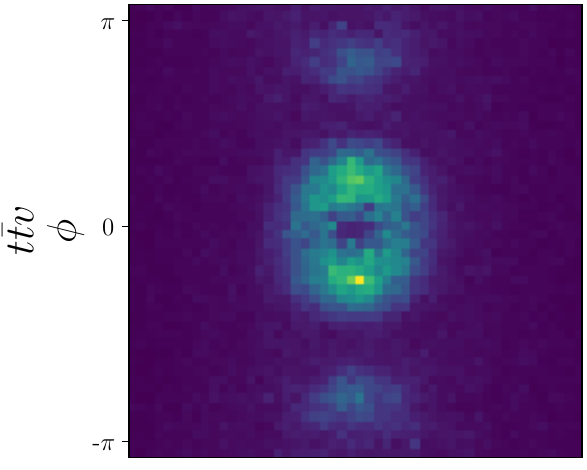}
\includegraphics[scale=0.292]{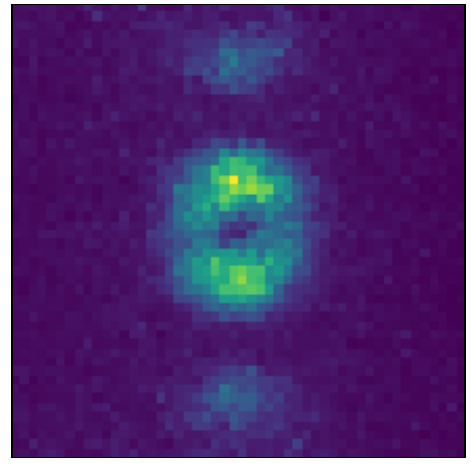}
\includegraphics[scale=0.292]{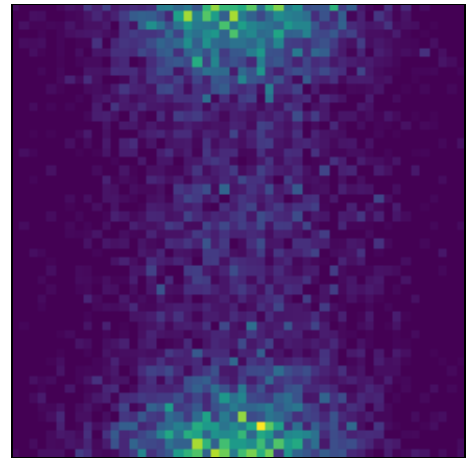}
\includegraphics[scale=0.292]{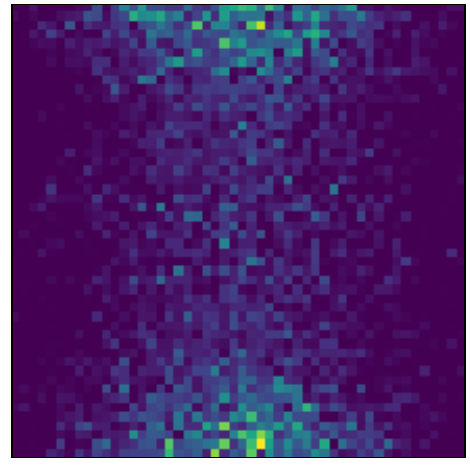}
\includegraphics[scale=0.292]{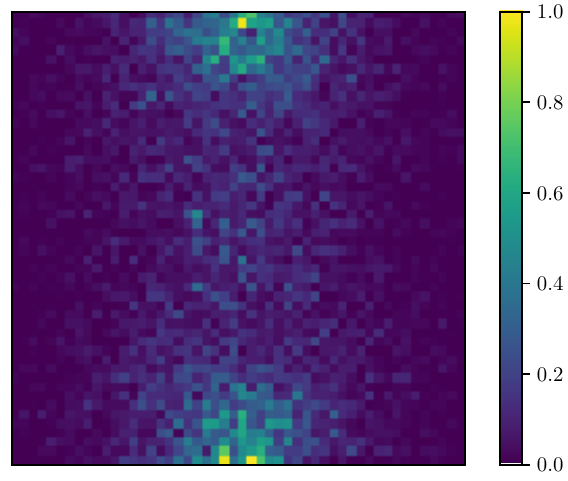} \\
\includegraphics[scale=0.292]{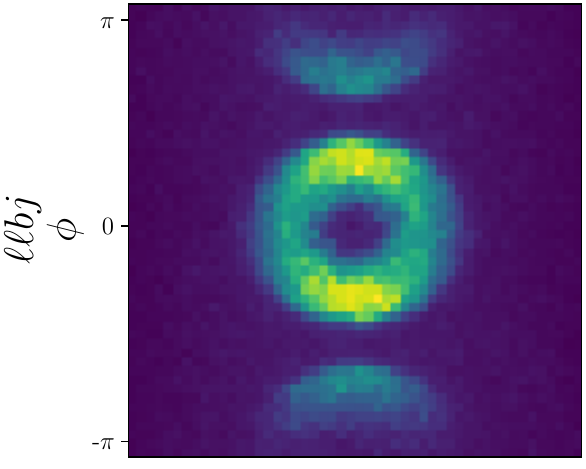}
\includegraphics[scale=0.292]{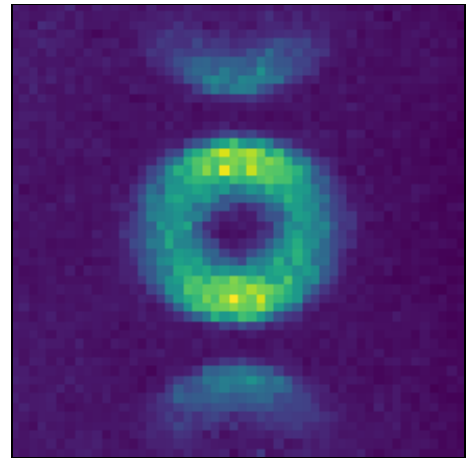}
\includegraphics[scale=0.292]{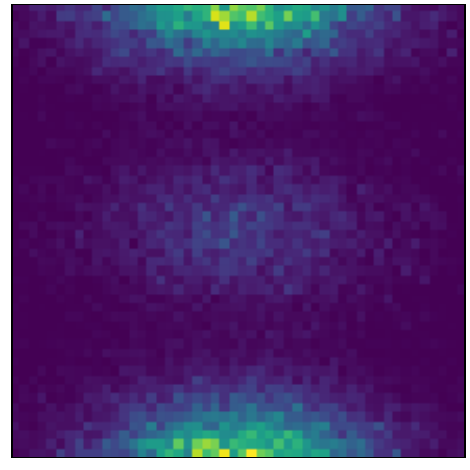}
\includegraphics[scale=0.292]{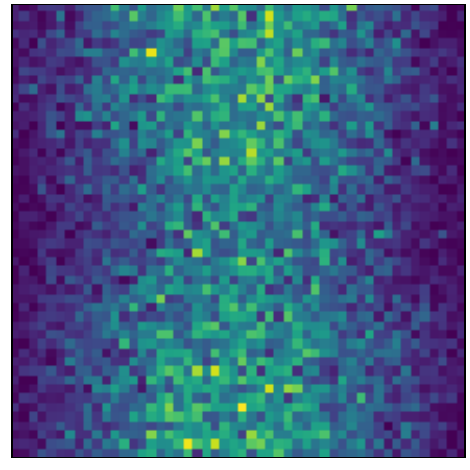}
\includegraphics[scale=0.292]{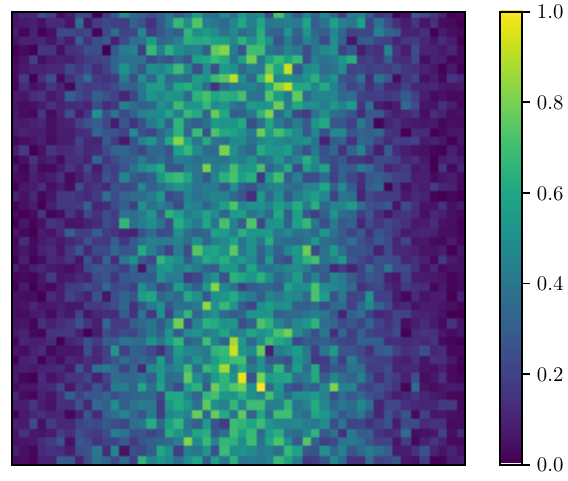} \\
\hspace*{-0.025cm}
\includegraphics[scale=0.292]{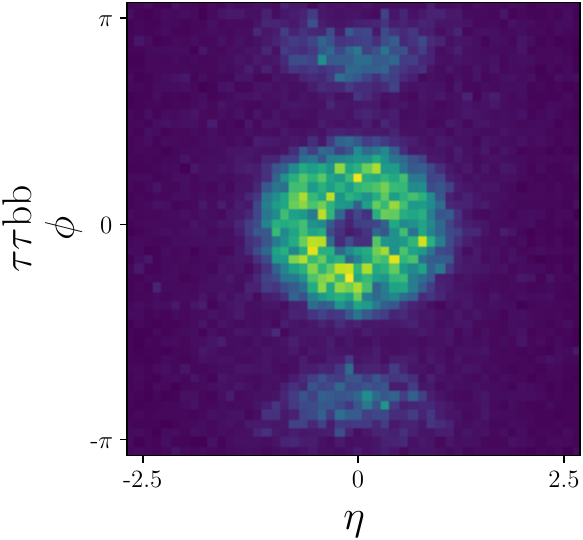} \hspace*{-0.1cm}
\includegraphics[scale=0.292]{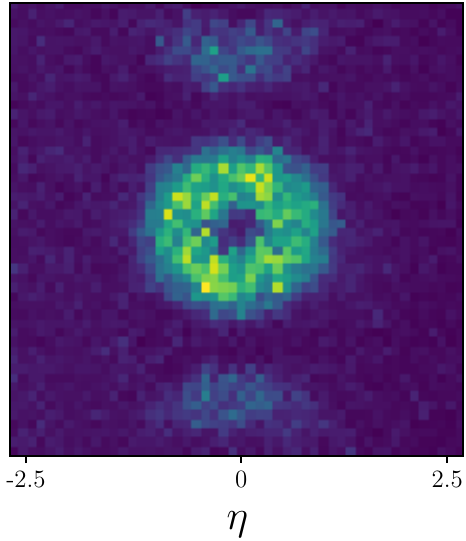}\hspace*{0.01cm}
\includegraphics[scale=0.292]{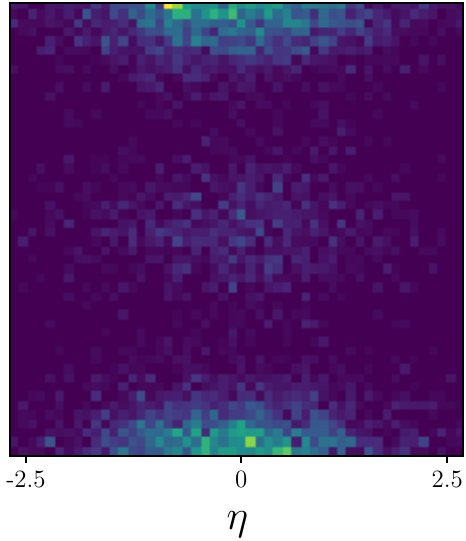}\hspace*{0.02cm}
\includegraphics[scale=0.292]{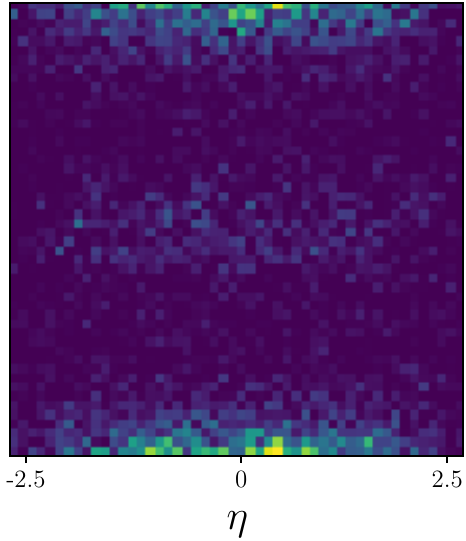}\hspace*{0.015cm}
\includegraphics[scale=0.292]{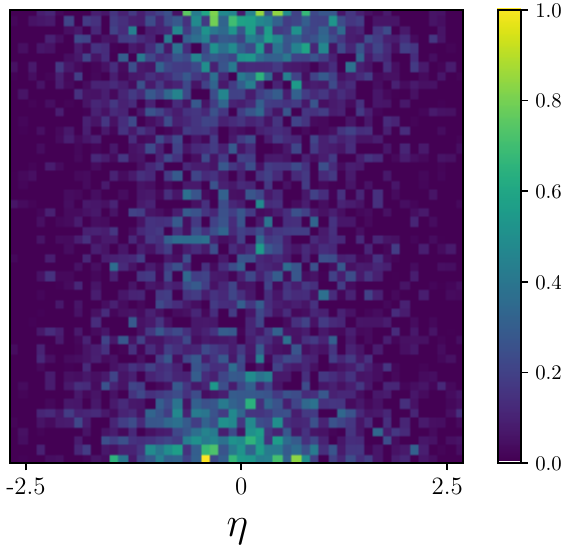} 
\vspace*{-0.2cm}
\caption{The cumulative average of various particle images for the signal and the different background processes after the baseline selection. The particles images are shown in the order from left to right: charged hadrons (1st column), neutral hadrons (2nd), isolated leptons (3rd), reconstructed neutrinos using Higgsness (4th), and reconstructed neutrinos using Topness (5th) for the signal ($hh$ in the first row), $t \bar{t}$ (2nd), $tW + j$ (3rd), $t \bar{t} h$ (4th), $t \bar{t} V$ (5th), $\ell \ell b j$ (6th), and $\tau \tau b b$ (7th).
The origin of the $(\eta, \phi)$ is taken to be the center of the reconstructed two $b$-tagged jets. }
\label{fig:All_images}
\end{figure}

A great breakthrough for deep neural networks in the image recognition opens up a possibility for a better background rejection when the energy and momentum of the final state particles are converted into image pixels. Jet images \cite{Cogan:2014oua} are based on the particle flow information \cite{CMS:2009nxa} in each event. 
We divide the particle flow into charged and neutral particles. The charged particles include charged hadrons, while the neutral particles consist of neutral hadrons as well as non-isolated photons. Leptons are removed from the both samples. Since these particles are spread over the ($\eta$, $\phi$) plane, it is challenging to identify key features such as color-flows and associated hadronization patterns of the signal and backgrounds. It is therefore instructive to rearrange them to make these features more accessible and allow for a robust identification. Here we define the origin of the ($\eta$, $\phi$) coordinate to be the center of the reconstructed $b$-tagged jets. All particles are translated accordingly in the ($\eta$, $\phi$) plane. Jet images are discretized into $50 \times 50$ calorimeter grids within a region of $-2.5\le \eta \le 2.5$ and $-\pi \le \phi \le \pi$. The intensity of each pixel is given by the total transverse momentum of particles passing through the pixel. The final jet images have a dimension of $(2 \times 50 \times 50)$ where 2 denotes a number of channels, charged and neutral particle images, which are shown in the first and second columns in Fig. \ref{fig:All_images} for the signal and backgrounds. In the case of the signal ($hh$ in the first row), the two $b$-tagged jets decayed from the color-singlet Higgs. Therefore their hadronization products are in the direction of the two $b$-tagged jet (toward the center). The empty region around the origin is due to $\Delta R_{bb} > 0.4$ requirement. On the other hand, the dominant background ($t\bar t$ in the second row), the jet images tend to be wider than the signal, as two top quarks are color-connected to the initial states. The cumulative distributions clearly demonstrate their differences.

In order for neural networks to fully take into account the spatial correlation between images of final state particles, we project two isolated leptons into the discretized $50 \times 50$ calorimeter grids as well. Combined with the jet images, we have a set of images for visible particles whose data structure is represented by
\begin{eqnarray}  
V^{(C, N, \ell)}_{\text{image}} &=& \big( 3 \times 50 \times 50  \big)  \; ,
\label{eq:3images}
\end{eqnarray}
where 3 denotes charged ($C$), neutral ($N$) particle images, and lepton ($\ell$) images, which are shown in the third column in Fig. \ref{fig:All_images} for the signal and backgrounds. In the case of the signal (first row), leptons are scattered around $\phi\approx\pm\pi$, which is opposite to the direction of the two $b$-tagged jets (origin), while for the dominant background ($t\bar t$ in the second row), leptons are more spread. This is consistent with the observation made in Refs. \cite{Kim:2018cxf,Kim:2019wns} using the $\hat s_{\text{min}}$ variable or invariant mass. The double Higgs production resembles the pencil-like (two leptons and two $b$-quarks are back-to-back approximately), while $t\bar t$ production is more or less isotropic. The lepton image also explains a shadow in the $(0, \pm\pi)$ region of two hadron images (first and second column) in the signal and backgrounds. See the Appendix for more information on the event shapes. 

Similarly, one can create images of the two reconstructed neutrinos using Topness and Higgsness, which are shown in the fourth and fifth columns in Fig. \ref{fig:All_images}. As expected from the kinematics, the neutrino images resemble lepton images, which would help the signal-background separation, in principle. To assess importance of these neutrino images in the signal sensitivity, we consider a complete set of images for all final state particles whose data structure is represented by
\begin{eqnarray}  
V^{(C, N, \ell, \nu_{\text{H}}, \nu_{\text{T}})}_{\text{image}} &=& \big( 5 \times 50 \times 50  \big)  \; ,
\label{eq:5images}
\end{eqnarray}
where 5 denotes a number of channels including all jet, lepton, and neutrino images. 
As clearly shown in Fig. \ref{fig:All_images}, the kinematic correlation among the decay products are mapped onto these images, including the missing transverse momentum. To catch the non-trivial correlations, more complex and deeper NNs will be considered. 
Each neural network takes a different set of input features for a classification problem between the signal and backgrounds. 
More details of NN architectures will be described in the next section.

%% file: 4_analysis.tex
With the increasing collision rate at the LHC, a task in collider analysis requires the significant dimensional reduction of the complex raw data to a handful of observables, which will be used to determine parameters in Lagrangian. Deep learning-based approaches offer very efficient strategies for such dimensional reduction, and have become an essential part of analysis in high energy physics.

However, important questions still remain regarding how to best utilize such tools. Specifically we ask {\it i) how to prepare input data, ii) how to design suitable neural networks for a given task, iii) how to account for systematic uncertainties, and iv) how to interpret results.} 
In this section, we scrutinize some of these questions by exploring various neural networks in the context of double Higgs production. Here, we discuss the essence of each network and summarize results briefly, leaving more detailed comparison in Section \ref{sec:results}.
Implementations of neural networks used in this paper are based on {\tt Pytorch} Framework \cite{NEURIPS2019_9015} and can be found from {\tt https://github.com/junseungpi/diHiggs/}.
The events, which pass the baseline selection described in Section~\ref{sec:eventgen}, are divided into 400k training and 250k test data sets.

To make a fair comparison of different NN structures, we consider the discovery reach of the signal at the LHC by computing the signal significance ($\sigma_{dis}$) using the likelihood-ratio~\cite{Cowan:2010js}
\begin{equation}
  \sigma_{dis} \equiv
    \sqrt{-2\,\ln\bigg(\frac{L(B | S \!+\!B)}{L( S \!+\!B| S \!+\!B)}\bigg)} \, , \label{Eq:SigDis} 
\end{equation}
where $  L(x |n) =  \frac{x^{n}}{n !} e^{-x}$, and $S$ and $B$ are the expected number of signal and background events, respectively.

\subsection{Deep Neural Networks}
\label{sec:dnn}

\begin{figure}[t]
  \centering
\includegraphics[width=0.7\textwidth,clip]{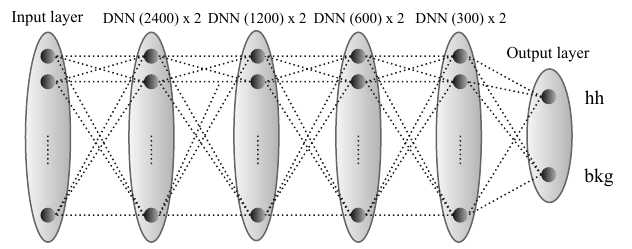}
\caption{A schematic architecture of the fully-connected NN (FC or DNN) used in this paper.}
\label{fig:FC}
\end{figure}

A fully-connected layer (FC or DNN) is the basic type of neural networks where every neuron in each layer is connected to all neurons in two consecutive layers. An input layer is composed of the combination of four-momenta of reconstructed particles in Eqs.~(\ref{eq:vis_4momenta}-\ref{eq:inv}), or kinematic variables in Eqs.~(\ref{eq:11kin}-\ref{eq:21}). It is followed by 8 hidden layers with the decreasing number of neurons from 2400 to 300 as shown in Fig. \ref{fig:FC}, and ReLU (Rectified Linear Unit) function \cite{pmlr-v15-glorot11a} is used to activate each neuron. 
The final hidden layer is connected to the output layer that contains two neurons with each representing a label of 1 for the signal and 0 for the backgrounds. A softmax activation function is used for the output layer.
We use Adam optimizer \cite{Kingma:2014vow} and a learning rate of $10^{-4}$ to minimize the binary cross entropy loss function.
To prevent overfitting, we add the $L_2$ regularization term to the loss function by setting {\tt weight\_decay}=$5 \times 10^{-4}$ in {\tt Pytorch} implementation. When training the DNN, we use a mini-batch size of 20 and the epochs of 30~\footnote{There are some common features used for all NNs in this paper. Except for CapsNets, we use the $L_2$ regularization for all NNs, which shifts the loss function, $L \rightarrow L +  \frac{1}{2} \lambda \|\textbf{W}\|^2$, where $\textbf{W}$ represents all weights, and the $\lambda$ denotes {\tt weight\_decay}. Throughout this paper, DNN hidden layers will appear repetitively in most of neural networks. For each layer, we apply the ReLU activation function.
Also the configuration of the output layer is the same for all other neural networks throughout this paper, unless otherwise mentioned.}. \begin{figure*}[tb]
\begin{center}
\includegraphics[width=0.47\textwidth,clip]{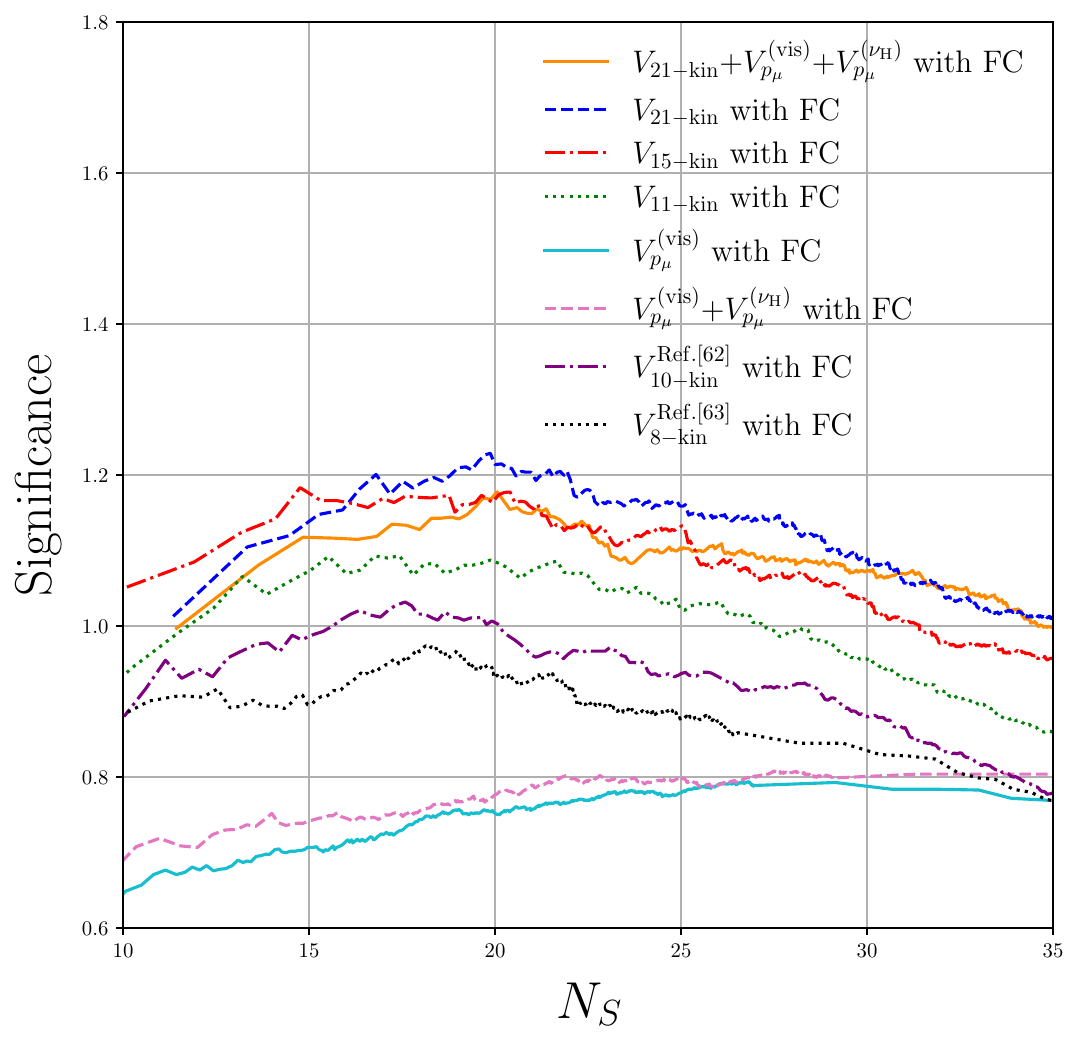}  \hspace*{0.1cm}
\includegraphics[width=0.47\textwidth,clip]{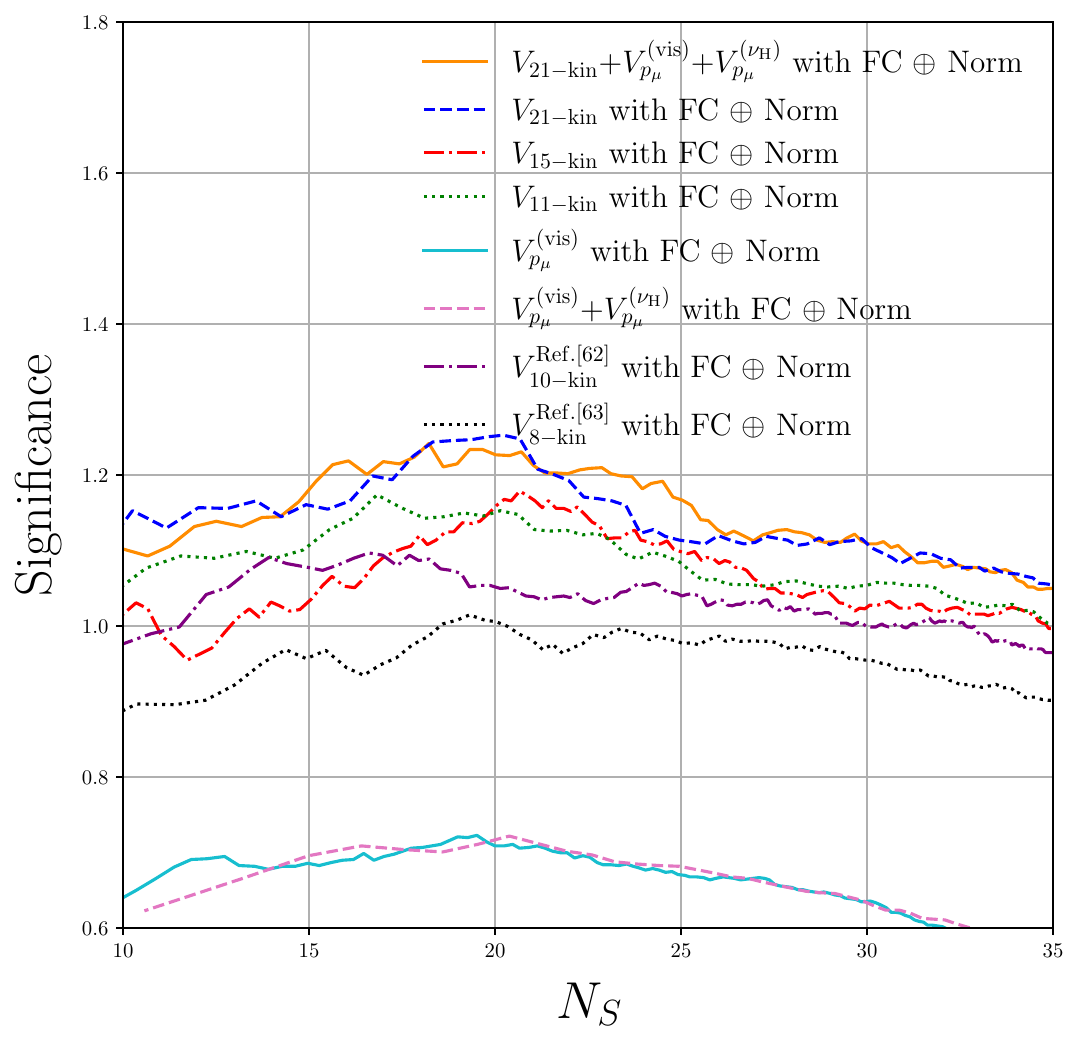}  \hspace*{0.05cm}
\caption{Significance of observing double Higgs production at the HL-LHC with ${\cal L}$ = 3 ab$^{-1}$ for DNNs with bare inputs (left) and DNNs with normalized inputs (right). }
\label{fig:FCresults}
\end{center}
\end{figure*}

Fig. \ref{fig:FCresults} shows the final significance as a function of the number of signal events ($N_s$) with bare inputs (left) and normalized\footnote{
We perform the linear transformation, $x_i \to x_i^\prime = a x_i + b$ for each input $x_i$ such that $x_i^\prime \in [0, 1]$.} inputs (right) for various combinations of input features. First observation is that DNN with the normalized input features lead to a slightly higher significance than that with the bare input for most NNs. The variation in the significances for different input is slightly narrower with the normalized features.  
Secondly, it is clear that when the DNN is trained with four-momenta of visible particles ($V^{(\text{vis})}_{ p_{\mu} }$), its performance does not stand out (see cyan-solid curves). 
Even when additional four-momenta of reconstructed neutrinos ($V^{(\nu_{\text{H}})}_{ p_{\mu} }$) are supplemented, there is no clear impact on the significance (see magenta-dashed curves). This result indicates that the simple DNN is unable to efficiently identify features of the signal and backgrounds with the primitive input data, given a finite number of training samples and the depth of DNN. 
On the other hand, addition of the human-engineered kinematic variables plays a very important role. 
$V_{\text{11-kin}}$, $V_{\text{15-kin}}$ and 
$V_{\text{21-kin}}$ are introduced in section \ref{sec:input}. 
$V_{10-kin}^{\rm Ref.[62]}$ and $V_{8-kin}^{\rm Ref.[63]}$ are sets of kinematic variables used in Ref. \cite{Adhikary:2017jtu} and Ref. \cite{Sirunyan:2017guj}, respectively. 
Note that the performance of DNN increases, when using more kinematic variables. 
We find that when the DNN is trained with 11 kinematic variables ($V_{\text{11-kin}}$)
the significance increases up to 
10\%-50\% compared to the results using the four-momenta for a wide range of signal number of events (see the green dotted curve in the left panel.). 
Interestingly, 15 kinematic variables ($V_{\text{15-kin}}$), which include the kinematic variables using the momentum of the reconstructed neutrinos, provide an additional steady $\sim10\%$ improvement on the significance (see the red and green curves in the left panel).
It is worth noting that the 6 high-level variables ($V_{\text{6-kin}}$) adds the orthogonal set of information to $V_{\text{15-kin}}$, which enables the DNN to better disentangle the backgrounds from the signal and brings additional $\sim10\%$ improvement. Finally, as mentioned previously, while the relative improvement is diminished with the normalized input features (as shown in the right panel), the importance of the kinematic variables still remain.

\subsection{Convolutional Neural Networks}
\label{sec:cnn}

\begin{figure*}[t!]
\centering
\begin{center}
\includegraphics[width=1.\textwidth,clip]{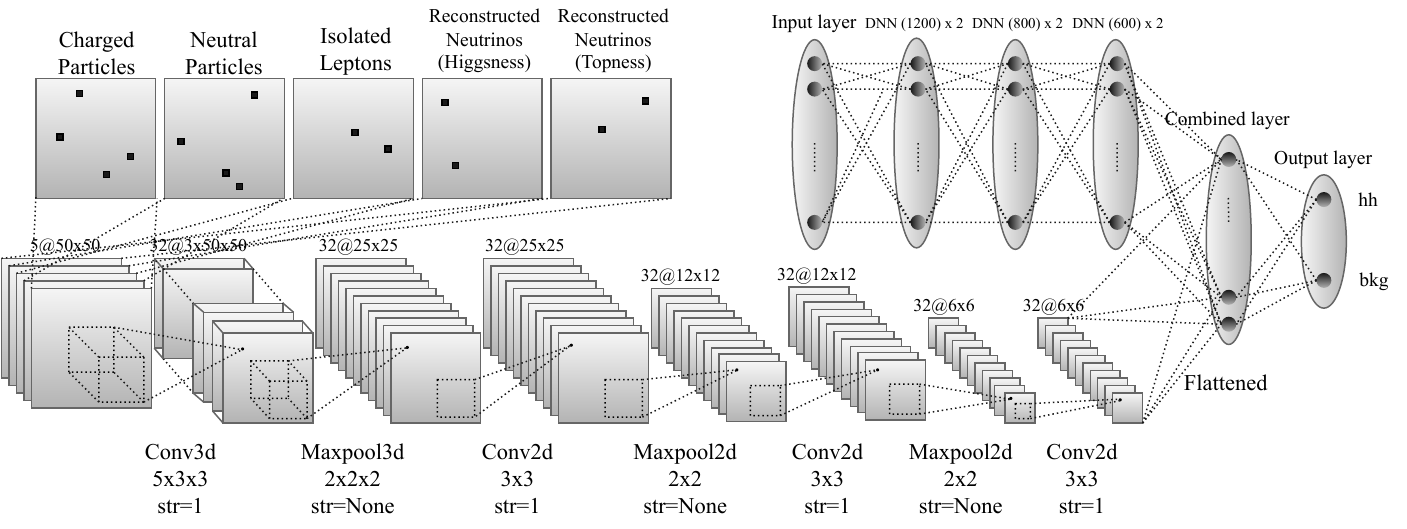}
\caption{A schematic architecture of the convolutional neural network (CNN) used in this paper. The separate DNN chain in the right-upper corner is used only when the kinematic variables are included.}
\label{fig:CNN} 
\end{center}
\end{figure*}

When the final state is fully represented by a set of images as in Eq.(\ref{eq:3images}-\ref{eq:5images}),
deep neural networks specialized for the image recognition provide useful handles.
One of the most commonly used algorithms is a convolutional neural network (CNN) as shown in Fig. \ref{fig:CNN}.
The input to the CNN is the 3D image of $V^{(C, N, \ell, \nu_{\text{H}}, \nu_{\text{T}})}_{\text{image}}$ ($V^{(C, N, \ell ) }_{\text{image}}$)
whose dimension is given by $ 5 \times 50 \times 50  $ ($ 3 \times 50 \times 50  $ ) where 5 (3) denotes a number of channels. 
In order to exploit the spatial correlation among different channels, we first apply the 3D convolution using the kernel size of $5 \times 3 \times 3$ ($3 \times 3 \times 3$), the stride 1, the padding size 1, and 32 feature maps \footnote{For all 2D and 3D convolutional layers, the padding size is fixed to 1. After each convolutional layer, we apply the batch normalization and ReLU activation function.}.
Next, we apply the max-pooling using the kernel size of $2 \times 2 \times 2$ without the stride and padding, which subsequently reduces the image dimension down to $ 32 \times 25 \times 25 $.

Since this is effectively the same dimension as the 2D image with 32 feature maps, in what follows,
we apply the 2D convolution using the kernel size of $3 \times 3$, the stride of 1, and 32 feature maps, followed by the max-pooling with the kernel size of $2 \times 2$. We repeat this procedure until the image dimension is reduced to $ 32 \times 6 \times 6 $.
Each of these neurons are connected to 3 hidden DNN layers with 600 neurons, and the final DNN layer is connected to the output layer.
To study the effect of the kinematic variables along with the image inputs, we slightly modify the NN structure, in which case these 3 hidden DNN layers are not used. Instead we construct the separate DNN consisting of 6 hidden layers with the decreasing number of neurons from 1200 to 600 (as shown in the right-upper corner in Fig. \ref{fig:CNN}). The last layer with 600 neurons for the kinematic variables are combined with the output of CNN of dimension $ 32 \times 6 \times 6 $. 
When training the network, we use Adam optimizer, the learning rate of $10^{-4}$,  regularization term of {\tt weight\_decay}=$5 \times 10^{-4}$, mini-batch size of 20, and epochs of 24. 

\begin{figure*}[t!]
\begin{center}
\includegraphics[width=0.6\textwidth,clip]{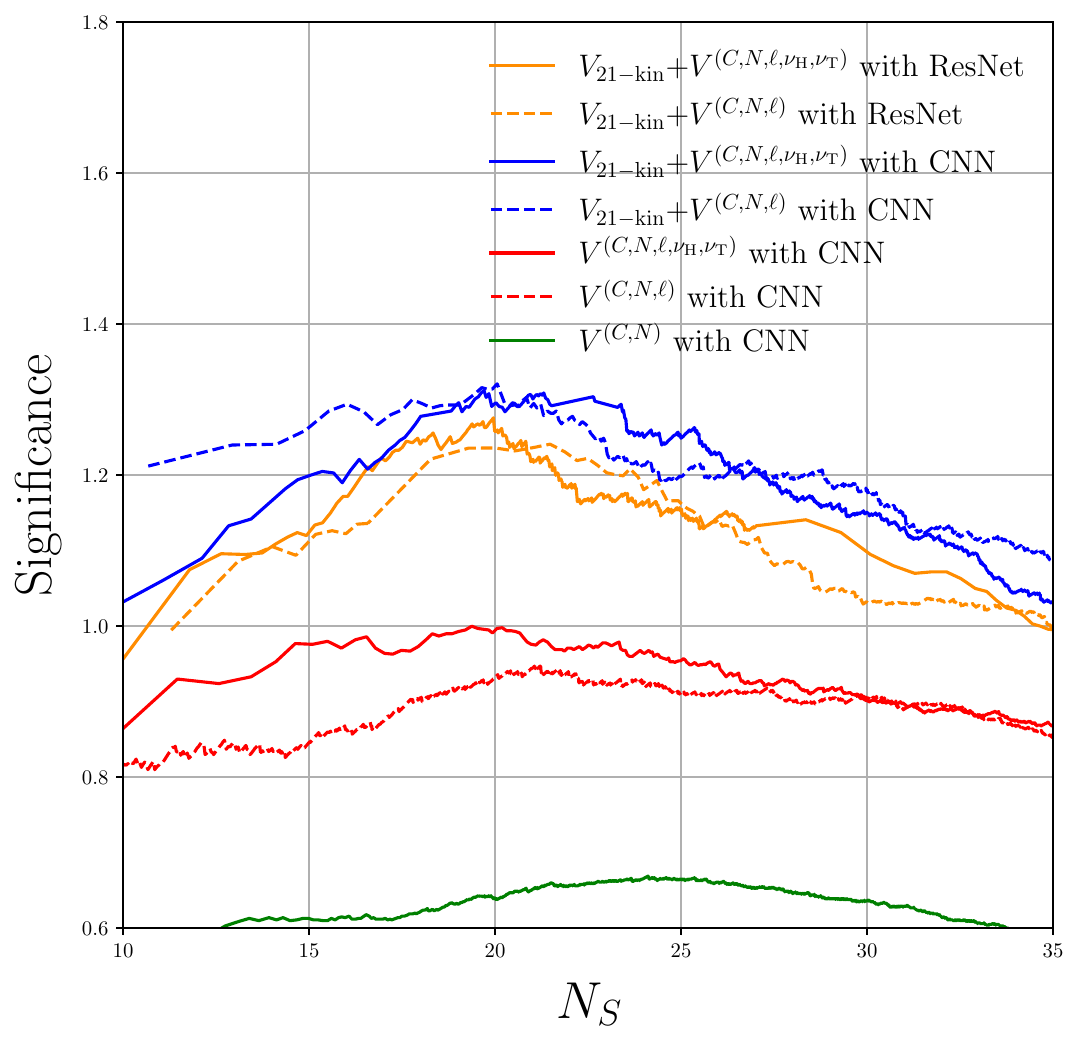} 
\caption{Significance at the HL-LHC with ${\cal L}$ = 3 ab$^{-1}$ for CNNs and ResNets. }
\label{fig:CNN_ResNet}
\end{center}
\end{figure*}

Fig. \ref{fig:CNN_ResNet} shows the performances of CNNs with various inputs.
First, we roughly reproduce results (green, solid) presented in Ref. \cite{Kim:2019wns}, taking $V^{(C,N)}_{\text{image}}$ as input. It is important to check this, as all event generations and simulations are performed completely independently. 
Adding lepton images, the overall significance of $0.9\sim 1$ can be achieved with $V^{(C,N,\ell)}_{\text{image}}$ image information without kinematic variables (red, dotted), which is substantially larger than that with $V^{(C,N)}_{\text{image}}$.
Albeit not a substantial impact, addition of neutrino images ($V^{(C,N,\ell, \nu_H, \nu_T)}_{\text{image}}$) helps improve the significance up to $\sim 5\%$ (red, solid).
Finally, addition of $V_{\text{21-kin}}$ kinematic variables increases the significance substantially, making CNN with $V_{\text{21-kin}} + V^{(C,N,\ell, \nu_H, \nu_T)}_{\text{image}}$ as the best network in terms of the signal significance (blue, solid). 

\subsection{Residual Neural Networks}
\label{sec:resnet}
\begin{figure*}[t!]
\begin{center}
\includegraphics[width=0.85\textwidth,clip]{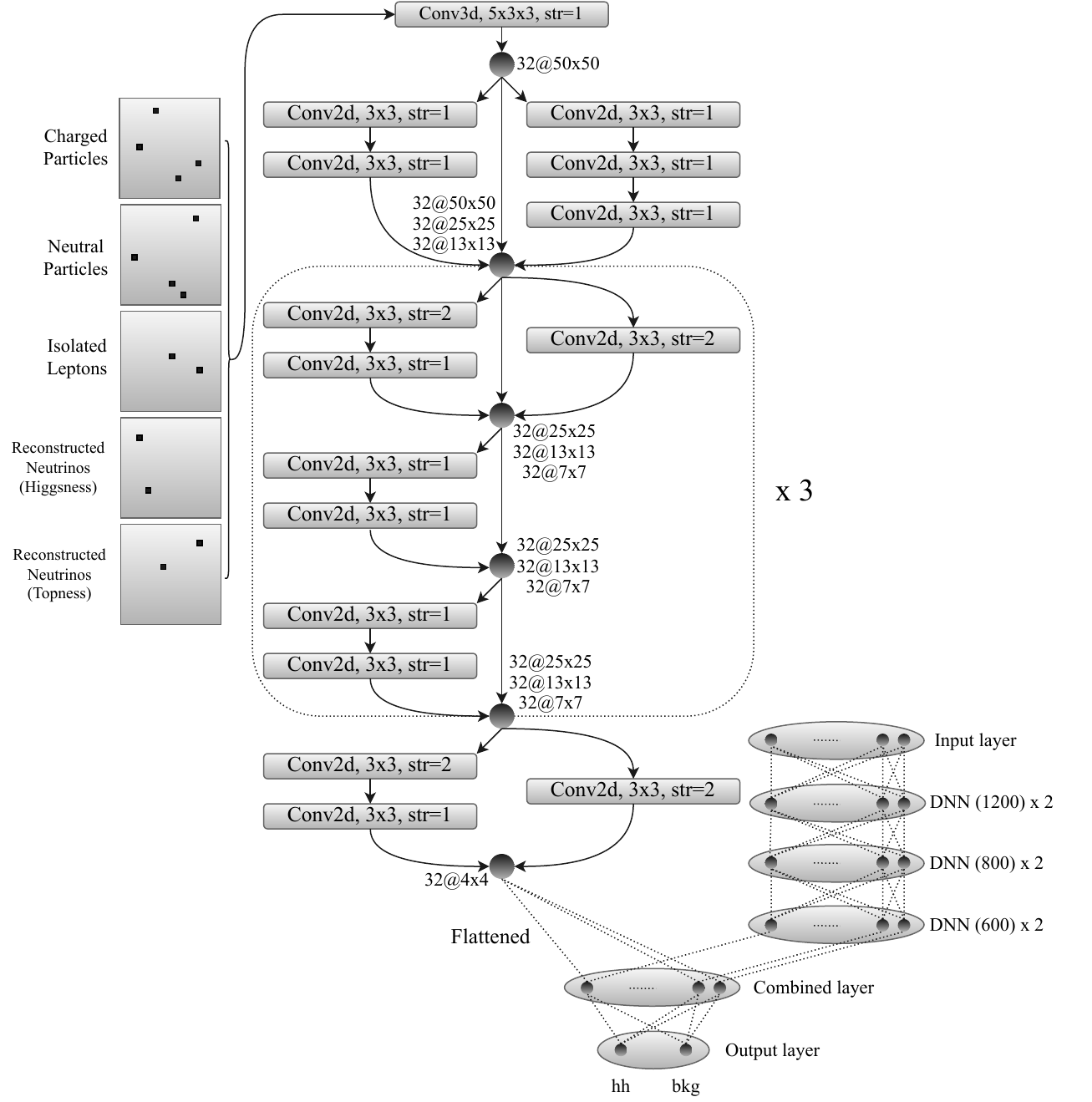}
\caption{A schematic architecture of the residual neural network (ResNet) used in this paper. The separate DNN chain in the right-bottom corner is used only when the kinematic variables are included.}
\label{fig:ResNet} 
\end{center}
\end{figure*}

The image sets that we feed into CNNs contain only few activated pixels in each channel.
Given the sparse images, the performance of the CNNs could greatly diminish.
One of possibilities to ameliorate this problem is to design neural networks at much deeper level.
As the CNN goes deeper, however, its performance becomes saturated or even starts degrading rapidly~\cite{simonyan2014very}.
A residual neural network (ResNet)~\cite{he2016deep,he2016identity} is one of the alternatives to the CNN
which introduces the skip-connections that bypass some of the neural network layers as shown in Fig. \ref{fig:ResNet}.

The input to the ResNet is the 3D image of $V^{(C, N, \ell, \nu_{\text{H}}, \nu_{\text{T}})}_{\text{image}}$ ($V^{(C, N, \ell ) }_{\text{image}}$).
We apply the 3D convolution using the kernel size of $5 \times 3 \times 3$ ($3 \times 3 \times 3$), the stride of 1, and 32 feature maps. Note that we apply the batch normalization and ReLU activation function after each convolutional layer, but we do not use the max-pooling.
We introduce the three-pronged structure: i) three series of 2D convolutions using the kernel size of $3 \times 3$, the stride of 1, and 32 feature maps, ii) two series of 2D convolutions using the same configurations, and iii) the skip-connection. 
All three paths are congregated into the single node, which enables the ResNet to learn various features of convoluted images, while keeping the image dimension unchanged. 
One way to reduce the dimensionality of the image is to change the striding distance, and we consider the two-pronged structure: i) the 2D convolution using the stride of 2, and ii) the 2D convolution using the stride of 2 followed by another 2D convolution using the stride of 1. Both layers are congregated again into the single node.
These are basic building blocks of our ResNet, and repeated several times in hybrid ways,
until the image dimension is brought down to $ 32 \times 4 \times 4 $. 
In parallel to the ResNet pipeline, we construct the DNN consisting of 6 hidden layers with the decreasing number of neurons from 1200 to 600. The inputs to the DNN are the  kinematic variables. This step is similar to what has been done in CNN when including the kinematic variables in addition to the image inputs.
The final neurons of two pipelines are congregated into the same layer, and subsequently connected to the output layer.
We use the learning rate of $10^{-4}$, the regularization term of {\tt weight\_decay}=$5 \times 10^{-4}$, the mini-batch size of 20, and the epochs of 11.

We obtain the overall significance of $\sim 1 -1.25$ can be achieved as shown in Fig. \ref{fig:CNN_ResNet}, which marks very high significance along with CNNs. 
The impact of neutrino images turns out to be mild, when including kinematic variables. This is partially because the reconstructed momentum of neutrinos (and the corresponding images) are byproducts of the {\it Higgsness} and {\it Topness} variables, which are already included in the variables of $V_{\text{6-kin}}$. Additional neutrino images, therefore, are redundant information in the ResNets and CNNs.

\subsection{Capsule Neural Networks}
\label{sec:capsnet}

\begin{figure*}[t!]
\begin{center}
\includegraphics[width=1\textwidth,clip]{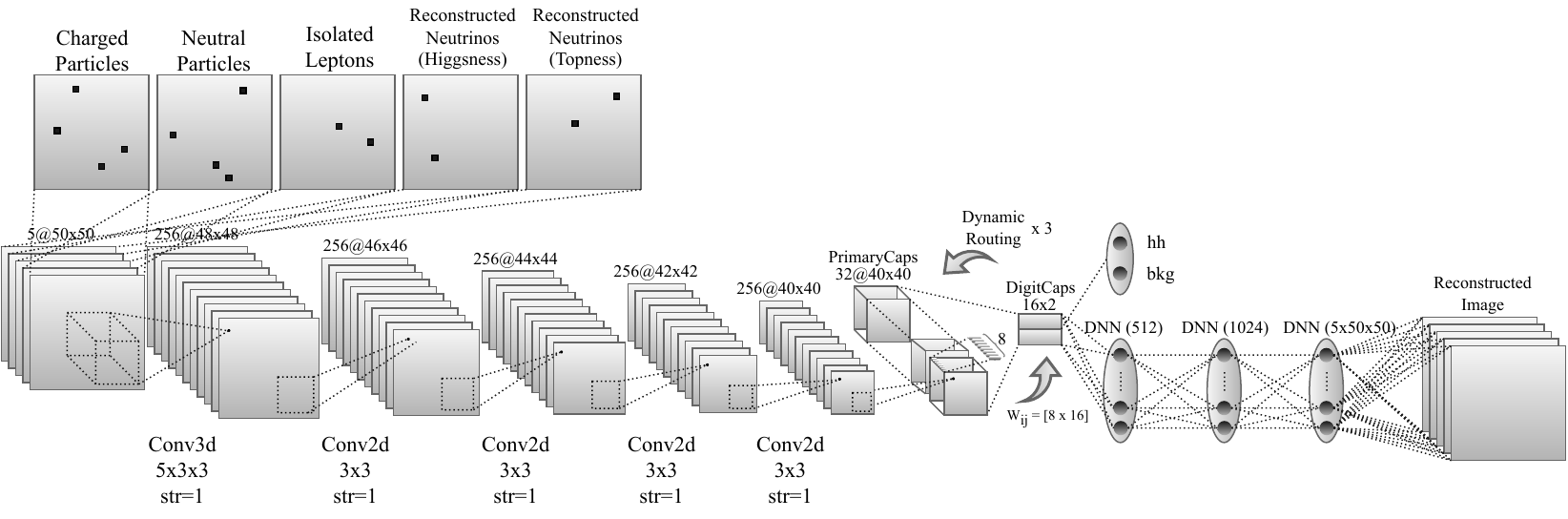}
\caption{A schematic architecture of the capsule network model (CapsNet) used in this paper. }
\label{fig:Caps} 
\end{center}
\end{figure*}

The max pooling method in the CNN selects the most active neuron in each region of the feature map, and passes it to the next layer. This accompanies the loss of spatial information about where things are,
so that the CNN is agnostic to the geometric correlations between the pixels at higher and lower levels.
A capsule neural network (CapsNet) ~\cite{NIPS2017_2cad8fa4,Diefenbacher:2019ezd} was proposed to address the potential problems of the CNN, and Fig. \ref{fig:Caps} shows the schematic architecture of the CapsNet used in our analysis.

The input to the CapsNet is the 3D image of $V^{(C, N, \ell, \nu_{\text{H}}, \nu_{\text{T}})}_{\text{image}}$ ($V^{(C, N, \ell ) }_{\text{image}}$). We apply the 3D convolution using the kernel size of $5 \times 3 \times 3$ ($3 \times 3 \times 3$), the stride of 1, and 32 feature maps. Note that we do not use the max-pooling. 
We apply the series of 2D convolution using the kernel size of $3 \times 3$, the stride of 1, and 256 feature maps, until the image dimension is reduced to $ 256 \times 40 \times 40 $. 
The output neurons of 2D convolution are reshaped to get a bunch of 8-dimensional primary capsule vectors,
which contain the lower-level information of the input image. For each feature map, there are $40 \times 40$ arrays of primary capsules, and there are 32 feature maps in total. To sum up, there are $N_{caps} = 40 \times 40 \times 32 = 51200$ primary capsules formed in this way.
We denote primary capsule vectors as $\textbf{u}_{i}$ where the index $i$ runs from 1 to $N_{caps}$.
Each primary capsule is multiplied by a $16 \times 8$ weight matrix $\textbf{W}_{ij}$ to get a 16-dimensional vector $\hat{\textbf{u}}_{j|i}$ which predicts the high-level information of the image
\begin{eqnarray}
[\hat{\textbf{u}}_{j|i}]_{16 \times 1} = [\textbf{W}_{ij}]_{16 \times 8} [\textbf{u}_{i}]_{8 \times 1} \;, \label{eq:routing1}
\end{eqnarray}
where $j$ denotes a class label, 0 or 1. 
To construct a digital capsule vector ($\textbf{v}_{j}$), we first take the linear combination of prediction vectors $\hat{\textbf{u}}_{j|i}$ from all capsules in the lower layer and define capsules $\textbf{s}_j$ in the higher-level,
\begin{eqnarray}
[\textbf{s}_{j}]_{16 \times 1} = \sum^{N_{caps}}_{i=1} c_{ij} [\hat{\textbf{u}}_{j|i}]_{16 \times 1} \;,
\end{eqnarray}
where the summation runs over the index $i$, and $c_{ij}$ denote routing weights
\begin{eqnarray}
c_{ij} = \frac{\exp b_{ij}}{\sum^1_{j=0} \exp b_{ij}} \;,
\end{eqnarray}
where all coefficients $b_{ij}$ are initialized to 0.
The digital capsule vector is defined by applying a squash activation function to $\textbf{s}_j$
\begin{eqnarray}
\textbf{v}_{j} = \frac{||\textbf{s}_{j}||^{2}}{1+||\textbf{s}_{j}||^{2}} \frac{\textbf{s}_j}{||\textbf{s}_{j}||} \;,
\label{eq:squash}
\end{eqnarray}
where $j$ again denotes a class label, 0 or 1. 
The final length of each digital capsule vector $||\textbf{v}_{j}||$ represents 
a probability of a given input image being identified as a class of $j$.
The CapsNet adjusts the routing weights $c_{ij}$ by updating the coefficients $b_{ij}$ such that
the prediction capsules $\hat{\textbf{u}}_{j|i}$ having larger inner products with the high-level capsules $\textbf{v}_{j}$ acquire larger weights
\begin{eqnarray}
b_{ij} \leftarrow b_{ij} + [\hat{\textbf{u}}_{j|i}]_{1 \times 16} \cdot [\textbf{v}_{j}]_{16 \times 1} \;.
\label{eq:bij}
\end{eqnarray}
The procedure from Eq. (\ref{eq:routing1}) to Eq. (\ref{eq:bij}) is referred to as the routing by agreement algorithm \cite{NIPS2017_2cad8fa4}, which we repeat three times in total to adjust the routing weights $c_{ij}$.
\begin{figure*}[t!]
\begin{center}
\includegraphics[width=0.46\textwidth,clip]{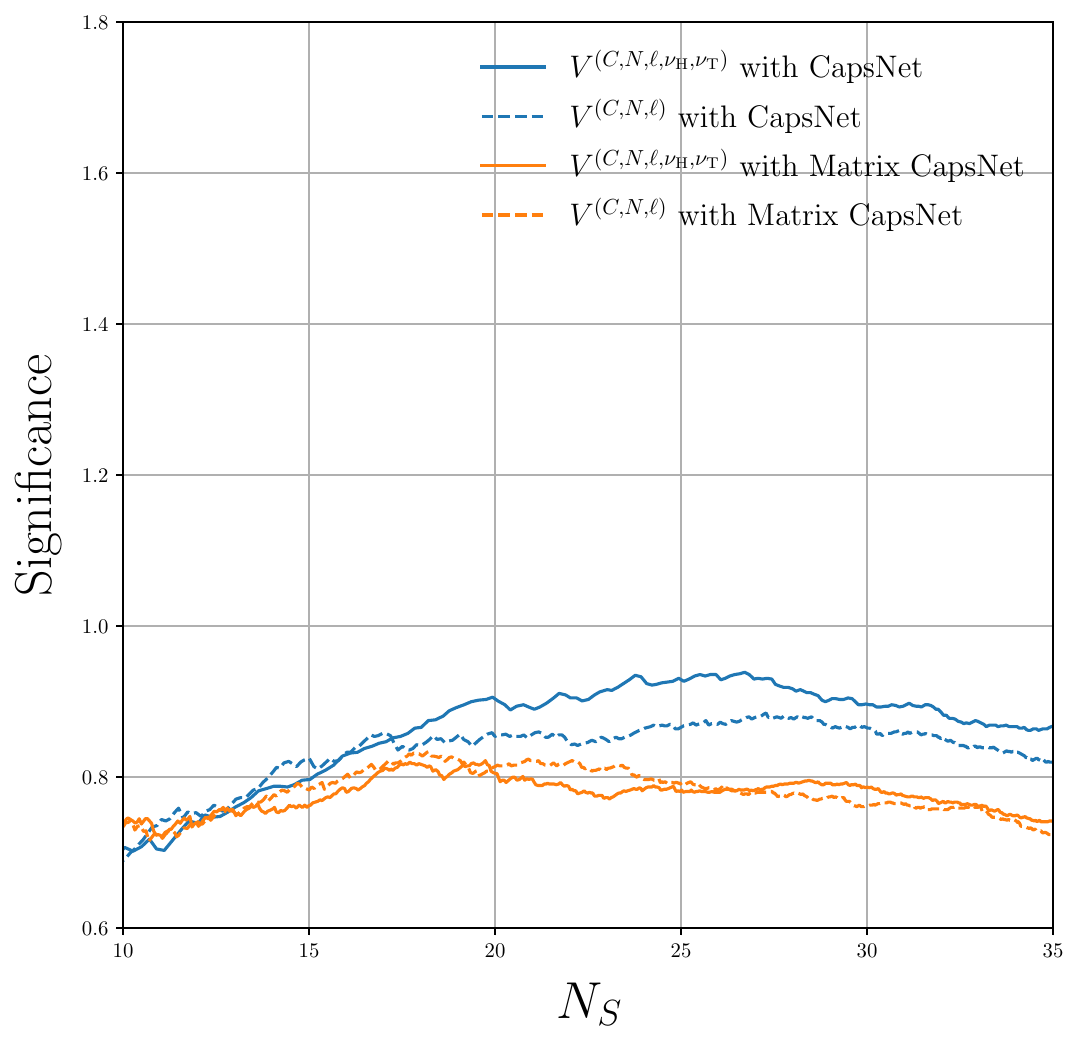} \hspace*{0.3cm}
\includegraphics[width=0.46\textwidth,clip]{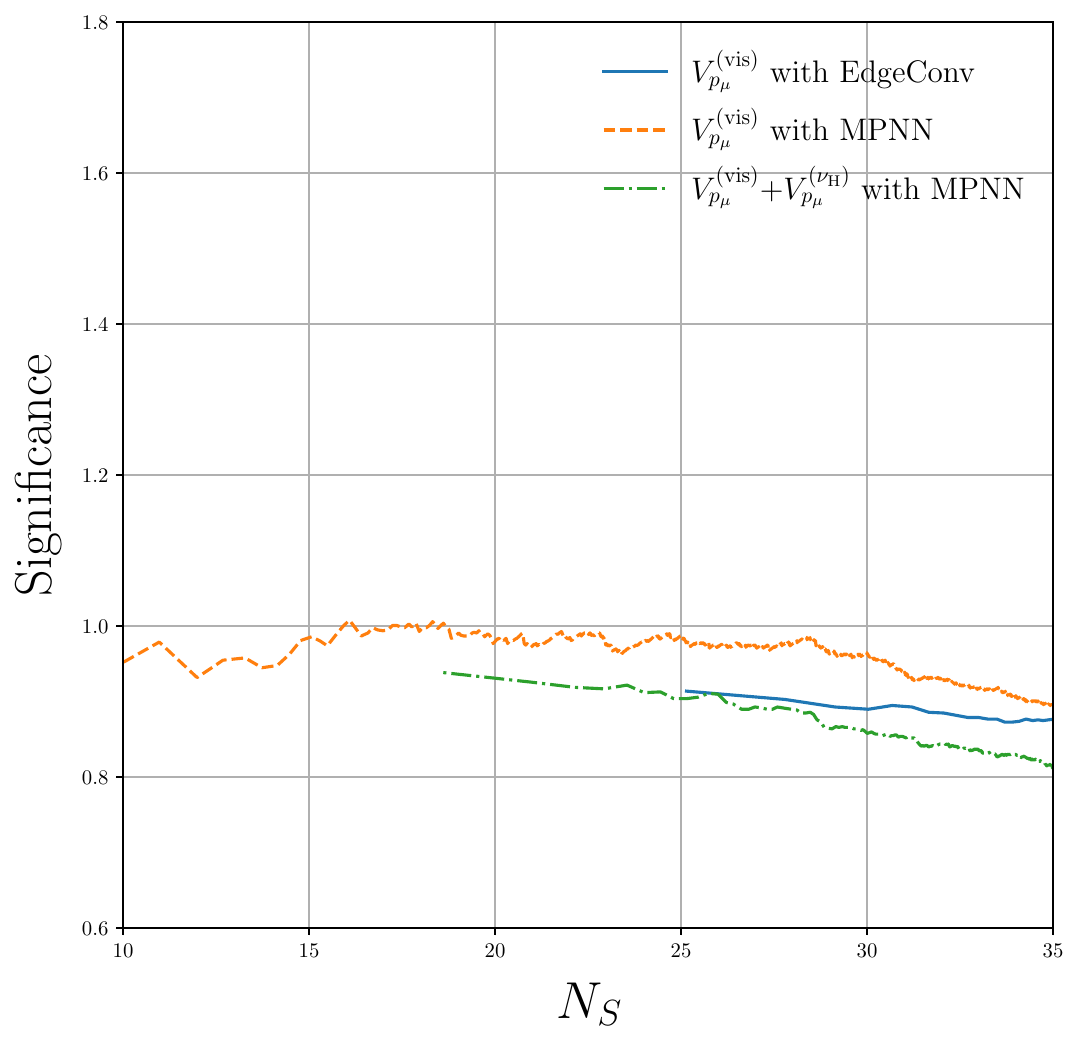} 
\caption{Significance at the HL-LHC with ${\cal L}$ = 3 ab$^{-1}$ for CapsNets/Matrix CapsNets (left) and EdgeConv/MPNN (right).}
\label{fig:CapsNet_MPNN}
\end{center}
\end{figure*}
The output digital capsule vectors $\textbf{v}_{j}$ are used to define the margin loss function
\begin{eqnarray}
  L_{j} = 
  && T_{j} \, \text{max} \Big (0,m^{+} - ||\textbf{v}_{j}|| \Big )^{2} 
+ \lambda(1-T_{j}) \, \text{max} \Big (0,||\textbf{v}_{j}||-m^{-} \Big )^{2} \;, 
\end{eqnarray} 
where $T_{1}=1$ and $T_{0}=0$ for the signal and backgrounds respectively, $m^+ = 0.9$, $m^- = 0.1$, and $\lambda = 0.5$.

Another parallel routine, called a decoder, attempts to reconstruct the input image out of
the digital capsule vectors $\textbf{v}_{j}$.
The digital capsule vectors are fed into
3 DNN hidden layers with increasing number of neurons from 512 to 12500 (7500), 
and reshaped into the input image size of $5 \times 50 \times 50$ ($3 \times 50 \times 50$).
The reconstruction loss function for the decoder is defined by the sum of squared differences in pixel intensities
between the reconstructed ($\mathcal{I}^{\text{(reco)}}_k$) and input ($\mathcal{I}^{\text{(input)}}_k$) images
\begin{eqnarray}
  L_{\text{deco}} = \frac{1}{N} \sum_{k=1} ( \mathcal{I}^{(\text{reco})}_k - \mathcal{I}^{\text{(input)}}_k)^2 \;,
\end{eqnarray} 
where the index $k$ runs from 1 to the total number of pixels in the image, and $N$ is a normalization factor
defined by the total number of pixels times the total number of training events.
In order for capsule vectors to learn features that are useful to reconstruct the original image,
we add the reconstruction loss into the total loss function
\begin{eqnarray}
  L =  L_{j}  + \alpha L_{\text{deco}} \, ,
\end{eqnarray} 
which is modulated by the overall scaling factor of $\alpha$.
Following the choice of Ref. \cite{NIPS2017_2cad8fa4}, we set $\alpha = 5.0 \times 10^{-4}$.
When training the network, we used Adam optimizer, the learning rate of $10^{-4}$,  regularization term of {\tt weight\_decay}=$0$, mini-batch size of 20, and epochs of 11.

Fig. \ref{fig:CapsNet_MPNN} shows the performances of the CapsNet where the overall significance of $0.8 \sim 0.9$ can be achieved, which is slightly lower than CNN with the same image inputs but better than CNN with $V^{(C,N)}$ (see Fig. \ref{fig:CNN_ResNet}). 
Albeit not a substantial impact, additional neutrino images help to improve the significance up to $\sim 5\%$.

\subsection{Matrix Capsule Networks}
\label{sec:matrix}
\begin{figure*}[t!]
\begin{center}
\includegraphics[width=1\textwidth,clip]{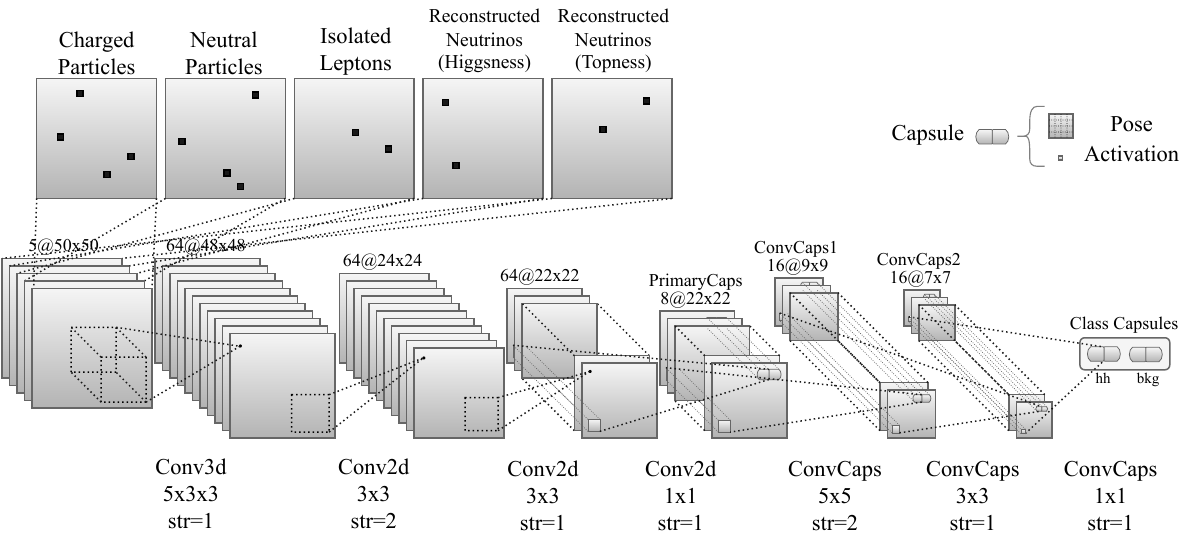}
\caption{A schematic architecture of the matrix capsule network model (Matrix CapsNet) used in this paper.}
\label{fig:Matrix_CapsNet} 
\end{center}
\end{figure*}

Regardless of the progress made in the CapsNet, there are a number of deficiencies that need to be addressed.
First, it uses vectors of length $n$ to represent features of an input image, which introduces too many $n^2$ parameters for weight matrices. 
Second, in its routing by agreement algorithm, the inner products are used to measure the accordance between two pose vectors ($cf.$ Eq.(\ref{eq:bij})).
This measure becomes insensitive when the angle between two vectors is small. 
Third, it uses the ad hoc non-linear squash function to force the length of the digital vector to be smaller than 1 ($cf.$ Eq.(\ref{eq:squash})). 

The aim of the Matrix CapsNet \cite{hinton2018matrix} is to overcome the above problems
by generalizing the concept of the capsules.
The major difference with the original CapsNet is that here each capsule is not a vector, but it is the entity which
contains a $n \times n$ pose matrix $M$ and an activation probability $a$. 
The utility of the pose matrix is to recognize objects in various angles, from which they are viewed,
and it requires a smaller number of hyper-parameters in its transformation matrix.\footnote{The same number of components in the vector of length $n$ can be contained in the $\sqrt{n} \times \sqrt{n}$ matrix. The former requires the $n \times n$ weight matrix, but the later requires the $\sqrt{n} \times \sqrt{n}$ weight matrix giving rise to a significant reduction in the amount of hyper-parameters.}

Fig. \ref{fig:Matrix_CapsNet} shows the architecture of the Matrix CapsNet used in our analysis.
The input is the 3D image of $V^{(C, N, \ell, \nu_{\text{H}}, \nu_{\text{T}})}_{\text{image}}$ ($V^{(C, N, \ell ) }_{\text{image}}$). We apply the 3D convolution using the kernel size of $5 \times 3 \times 3$ ($3 \times 3 \times 3$), the stride of 1, and 64 feature maps. 
It is followed by a series of 2D convolutions using the kernel size of $3 \times 3$, the stride of 1 or 2, and 64 feature maps, until the image dimension is reduced to $ 64 \times 22 \times 22 $. 
Next, we apply the modified 2D convolution using the kernel size of $1 \times 1$ and the stride of 1 where 
each stride of the $1 \times 1$ convolution transforms the 64 feature maps into 8 capsules.
As a result of this operation, we obtain the layer $L$ of capsules whose dimension is denoted as 
$8 \times 22 \times 22$ in unit of capsules.

Each capsule $i$ in the layer $L$ encodes low-level features of the image, 
and it is composed by the $2 \times 2$ pose matrix $M_i$ and the activation probability $a_i$.
To predict the capsule $j$ in the next layer $L+1$, which encodes high-level features,
we multiply the pose matrix $M_i$ by a $2 \times 2$ weight matrix $W_{ij}$
\begin{eqnarray}
  [V_{ij}]_{2 \times 2} = [M_i]_{2 \times 2} [W_{ij}]_{2 \times 2}  \;,
\label{eq:vote}  
\end{eqnarray} 
where $V_{ij}$ is the prediction for the pose matrix of the parent capsule $j$.
Since each capsule $i$ attempts to guess the parent capsule $j$, 
we call $V_{ij}$ the vote matrix.

On the other hand, the $2 \times 2$ pose matrix of the parent capsule $j$ is modeled by 4 parameters of $\mu^h_j$ (with $h = 1, 2, 3, 4$),
which serve as the mean values of the 4 Gaussian distribution functions with standard deviations of $\sigma^h_j$.
In this model, the probability of $V_{ij}$ belonging to the capsule $j$ is computed by
\begin{eqnarray}
P^{h}_{i|j}=\frac{1}{\sigma^{h}_{j} \sqrt{2 \pi }} \exp \bigg(-\frac{(V^{h}_{ij}-\mu^{h}_{j})^{2}}{2(\sigma^{h}_{j})^{2}} \bigg) \;,
\label{eq:probability}  
\end{eqnarray}
where $V^{h}_{ij}$ denotes the $h^{th}$ component of the vectorized vote matrix.
Using this measure, we define the amount of cost ${\mathcal{C}}$ to activate the parent capsule $j$
\begin{eqnarray}
\mathcal{C}^{h}_{ij} = -\ln (P^{h}_{i|j}) \, ,
\label{eq:cost}  
\end{eqnarray}
so that the lower the cost, the more probable that the parent capsule $j$ in the layer $L +1$ will be activated by the capsule $i$ in the layer $L$.
We take the linear combination of the costs from all the capsules $i$ in the layer $L$
\begin{eqnarray}
\mathcal{C}^{h}_{j} = \sum_{i} r_{ij} \,  \mathcal{C}^{h}_{ij} \, ,
\label{eq:costsum}  
\end{eqnarray}
where each cost is weighted by an assignment probability $r_{ij}$.
To determine the activation probability $a_j$ in the layer $L+1$, we use the following equation
\begin{eqnarray}
a_j = \text{sigmoid} \Big (\lambda (b_j -  \sum^4_{h=1} \mathcal{C}^h_j ) \Big ) \;,
\label{eq:activation}  
\end{eqnarray}
where $b_j$ and $\lambda$ are a benefit and an inverse temperature parameters, respectively.

Hyper-parameters such as $W_{ij}$ and $b_j$ are learned through a back-propagation algorithm,
while $\mu^{h}_{j}$, $\sigma^{h}_{j}$, $r_{ij}$, and $a_j$ are determined iteratively by the Expectation-Maximization (EM) routing algorithm.\footnote{It is the algorithm based on the series of steps described in Eq.(\ref{eq:vote}-\ref{eq:activation}). The algorithm is repeated several times to determine the hyper-parameters iteratively.
More details can be found in Ref. \cite{hinton2018matrix}.}
The inverse temperature parameter $\lambda$, on the other hand, is fixed to $10^{-3}$.
After repeating the EM iteration 3 times, the last $a_j$ is the activation probability, and
the final parameters of $\mu^h_j$ (with $h = 1, 2, 3, 4$) are reshaped to form the $2 \times 2$ pose matrix of the layer $L+1$.

The above prescription of computing the capsules in the layer $L+1$ from the layer $L$ can be systematically 
combined with the convolutional operation.
Recall that we ended up with the capsule layer with the dimension of $8 \times 22 \times 22$.
Here, we apply the convolutional capsule operation using the kernel size of $5 \times 5$, the stride of 2, and 16 feature maps.
This operation is similar to the regular CNN, except that it uses the EM routing algorithm to compute the pose matrices and the activation probability of the next layer.
It is followed by another convolutional capsule operation until the image dimension is brought to $16 \times 7 \times 7$ in unit of capsules.
These capsules in the last layer are connected to the class capsules, and it outputs one capsule per class.
The final activation value of the capsule of a class $j$ is interpreted as the likelihood 
of a given input image being identified as a class $j$.

The final loss function is defined by
\begin{eqnarray}
L = \frac{\sum\limits_{j \neq t}  \Big (\text{max}(0,m-(a_{t}-a_{j})) \Big )^{2}}{N_{train}} - m^{2} \, ,
\end{eqnarray}
where $a_{t}$ denotes a true class label, and $a_{j}$ is the activation for class $j$, $N_{train}$ denotes a number of training events.
If the difference between the true label $a_{t}$ and the activation for the wrong class $j ( \neq t)$ is smaller than the threshold of $m$,
the loss receives a penalty term of $(m-(a_{t}-a_{j}))^{2}$ \cite{hinton2018matrix}.
The threshold value $m$ is initially set to 0.1, and it is linearly increased by $1.6 \times 10^{-2}$ per each epoch of training.
It stops growing when it reaches to 0.9 at the final epoch.
When training the network, we used Adam optimizer, the learning rate of $10^{-4}$,  regularization term of {\tt weight\_decay}=$5 \times 10^{-5}$, mini-batch size of 20, and epochs of 20.

Fig. \ref{fig:CapsNet_MPNN} shows the performances of the Matrix CapsNet 
where the overall significance of 0.7-0.8 can be achieved. It is slightly lower than that for CapsNet. 
We find that performance of CapsNet and Matrix CapsNet is comparable or slightly worse than those using CNN with the same image inputs (see Fig. \ref{fig:CNN_ResNet}). 

\subsection{Graph Neural Networks}
\label{sec:mpnn}

Instead of using the image-based neural networks that could suffer from the sparsity of the image datasets, one could encode the particle information into the graphical structure which consists of nodes and edges. 
Graph neural networks (GNNs) \cite{1555942,4700287} take into account topological relationships among the nodes and edges in order to learn graph structured information from the data. Each reconstructed object (in Eq.(\ref{eq:vis_4momenta})) including neutrinos obtained from the Higgsness (in Eq.(\ref{eq:Hneu_4momenta})) is represented as a single node.
Each node $i$ has an associated feature vector $\textbf{x}_i$ which collects the properties of a particle.
The angular correlation between two nodes $i$ and $j$ is encoded in an edge vector $\textbf{e}_{ij}$.

\begin{figure*}[t]
\centering
\includegraphics[scale=0.57]{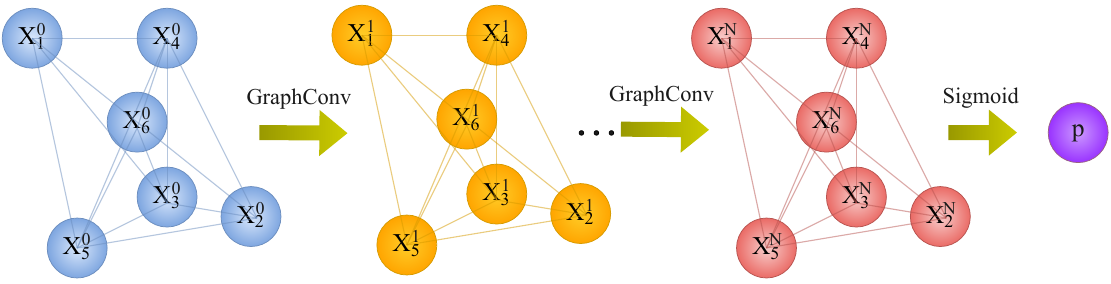}
\caption{
A schematic architecture of the edge convolutional neural network (EdgeConv) used in this paper.
}
\label{fig:EdgeConv}
\end{figure*}

The first type of GNN architectures that we consider is a modified edge convolutional neural network (EdgeConv) \cite{DBLP:journals/corr/abs-1801-07829}, which efficiently exploits Lorentz symmetries, such as an invariant mass, from the data~\cite{Abdughani:2020xfo,Abdughani:2018wrw}. Its schematic architecture is shown in Fig. \ref{fig:EdgeConv}.
All nodes are connected with each other, and the node $i$ in the input layer is represented by a 4-momentum feature vector of $\textbf{x}^0_i = (p_x,\ p_y,\ p_z,\ E)_i$, while leaving the edge vectors $\textbf{e}_{ij}$ unspecified.
The EdgeConv predicts the features of the node in the next layer as a function of neighboring features.
Specifically, each feature vector of a node $i$ in the layer $n$ is defined by
\begin{eqnarray} 
\textbf{x}_i^{n} = \sum_{j \in \mathcal{E}} \textbf{W}^n &\Big(& \textbf{x}_i^{n-1} \oplus  (\textbf{x}_i^{n-1} \circ  \textbf{x}_i^{n-1})  
 \oplus   (\textbf{x}_i^{n-1} \circ  \textbf{x}_j^{n-1})   \oplus  ( \textbf{x}_j^{n-1} - \textbf{x}_i^{n-1} )  \Big) \;,  
 \end{eqnarray}
where the symbol $\oplus$ denotes a direct sum and $\mathcal{E}$ stands for the a set of feature vectors $j$ in the layer $n-1$ that are connected to the node $i$. $\circ$ and $\textbf{W}^n$ denote the element-wise multiplication and a weight matrix, respectively. 
We do not include the bias vector and the ReLu activation function.
This completes one round of the graph convolution, and we repeat it $N=3$ times.

In the output layer, all feature vectors are concatenated and multiplied by a weight matrix, leading to a vector of dimension 2
\begin{eqnarray} 
\hat{\textbf{p}} = \textbf{W}^N \bigoplus_{i}    \textbf{x}_i^{N} \;.
\end{eqnarray}
We apply the sigmoid activation function on each component of $\hat{\textbf{p}}$
\begin{eqnarray} 
p_{k} = \frac{1}{1+e^{-\hat{p}_{k}}} \;,
\end{eqnarray}
where $k$ stands for the class label, 0 or 1, and
$p_{k}$ represents a probability which is used to classify the signal and backgrounds.
We use the cross entropy as the loss function
\begin{eqnarray}
L_k = -y_k \log p_k - (1-y_k) \log (1-p_k) \;,
\end{eqnarray} 
where $y_1 = 1$ and $y_0 = 0$ for the signal and backgrounds respectively.
When training the network, we adopted Adam optimizer with the learning rate of $9.20 \times 10^{-7}$ and the momentum parameters $\beta_1= 9.29 \times 10^{-1}$ and $\beta_2 = 9.91 \times 10^{-1}$. We used the regularization term with {\tt weight\_decay}=$3.21 \times 10^{-2}$, a multiplicative factor of the learning rate decay $\gamma = 1$, mini-batch size of 130, and epochs of 70.

\begin{figure*}[t]
\centering
\includegraphics[scale=0.57]{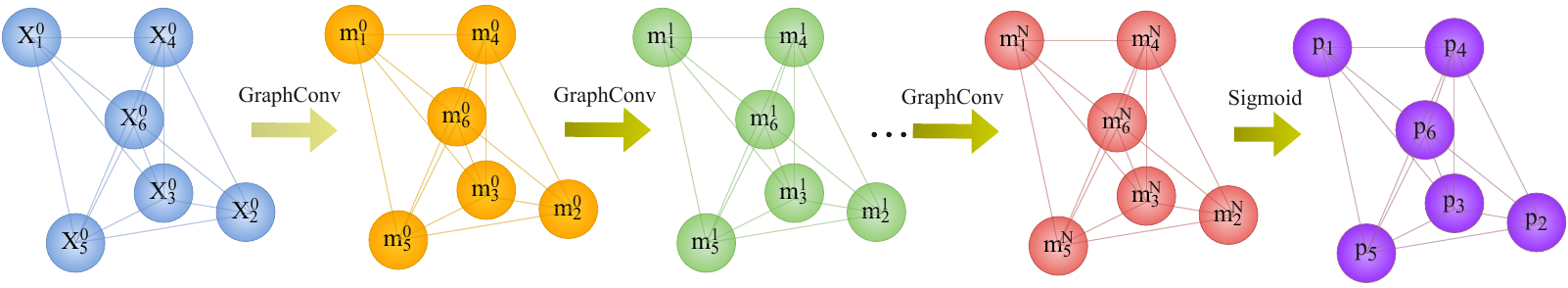}
\caption{A schematic architecture of the message passing neural network (MPNN) used in this paper.
}
\label{fig:mpnn}
\end{figure*}

The second type of GNN architectures is a message passing neural network (MPNN)  \cite{DBLP:journals/corr/GilmerSRVD17} as shown in Fig. \ref{fig:mpnn}.
The node $i$ in the input layer is represented by a feature vector of $\textbf{x}^0_i = (I_{\ell}, I_{b}, I_{\nu},\ m,\ p_T,\ E)_i$,
where $m$, $p_T$, and $E$ denote the invariant mass, transverse momentum, and energy of a particle, respectively.
The default values for $I_i$ are set to zero.
$I_{\ell} = 1$ for the hardest lepton and $I_{\ell} = -1$ for the second hardest lepton, $I_{b} = 1$ for the hardest $b$-jet and $I_{b} = -1$ for the second hardest $b$-jet, and $I_{\nu} = 1$ for the hardest neutrino and $I_{\nu} = -1$ for the second hardest neutrino (note that we are using reconstructed neutrino momenta from Higgsness). 
All nodes are connected to each other, and a single component edge vector $\textbf{e}_{ij}$ is represented by the angular separation ($\Delta R_{\textbf{x}_i, \textbf{x}_j}$) between two particles in the node $i$ and $j$.
The MPNN preprocesses each input node $i$, multiplying the $\textbf{x}^0_i$ by a weight matrix $\textbf{W}^0$
\begin{eqnarray}
\textbf{m}^0_i =  \textbf{W}^0  \textbf{x}^0_i \;.
\end{eqnarray}
Each feature vector of a node $i$ in the layer $n$ is defined by
\begin{eqnarray} 
\textbf{m}_i^{n} = \sum_{j \in \mathcal{E}} \textbf{W}^n  \Big( \textbf{m}_i^{n-1}\oplus \big(\textbf{W}^{'n}  ( \textbf{m}_j^{n-1}\oplus \textbf{e}_{ij}) \big) \Big)  ,
\end{eqnarray}  
where $\textbf{W}^{n}$ and $\textbf{W}^{'n}$ denote weight matrices. This completes one round of the graph convolution, and we repeat it $N=3$ times.

In the output layer, each feature vector of node $i$ is multiplied by a weight matrix to become the vector of length 2
\begin{eqnarray} 
\hat{\textbf{p}}_i = \textbf{W}^N \textbf{m}_i^{N} \;.
\end{eqnarray}
We apply the sigmoid activation function on each component of $\hat{\textbf{p}}_i$
\begin{eqnarray} 
p_{ik} = \frac{1}{1+e^{-\hat{p}_{ik}}} \;,
\end{eqnarray}
where $k$ stands for the class label, and
$p_{i1}$ represents a probability of a node $i$ being identified as the signal.
To identify the signal, we require the graph to pass the cut
\begin{eqnarray} 
\sum_i p_{i1} >  p^{cut}_{1}   \;,
\end{eqnarray}
where $p^{cut}_{1}$ is optimized to yield a best significance.
We use the cross entropy as the loss function
\begin{eqnarray}
L_k = \sum_i  \Big( -y_k \log p_{ik} - (1-y_k) \log (1-p_{ik})  \Big),
\end{eqnarray} 
When training the network, we used Adam optimizer, the learning rate of $1.44 \times 10^{-5}$,  regularization term of {\tt weight\_decay}=$9.67 \times 10^{-2}$, $\beta_1= 9.16 \times 10^{-1}$, $\beta_2 = 9.92 \times 10^{-1}$, $\gamma = 9.81 \times 10^{-1}$, mini-batch size of 164, and epochs of 21.

We find that the EdgeConv and MPNN show a very good performance with the basic momentum input, and show their signal significance in the right panel of Fig. \ref{fig:CapsNet_MPNN}, which clearly surpasses the performance of FC with the same input features (shown in Fig. \ref{fig:FCresults}). 
Moreover, before combining kinematic variables and image-based input features, the performance of the EdgeConv and MPNN is comparable to or slightly better than those based on NN with image only such as CapsNet, Matrix CapsNet, or CNN (see Fig. \ref{fig:CNN_ResNet} and the left panel in Fig. \ref{fig:CapsNet_MPNN}.). 
This comparison illustrates a couple of important points. First, one can further improve vanilla DNN with basic four momenta input introducing evolution of nodes and edges in MPNN/EdgeConv. Second, to bring additional improvement in the signal significance, it is crucial to consider different types of inputs such as image-based features and high-level kinematic variables in addition to basic four momenta, and develop NN architecture suitable for the corresponding input features.

%% file: 5_results.tex
\begin{figure*}[t!]
\begin{center}
\includegraphics[width=0.47\textwidth,clip]{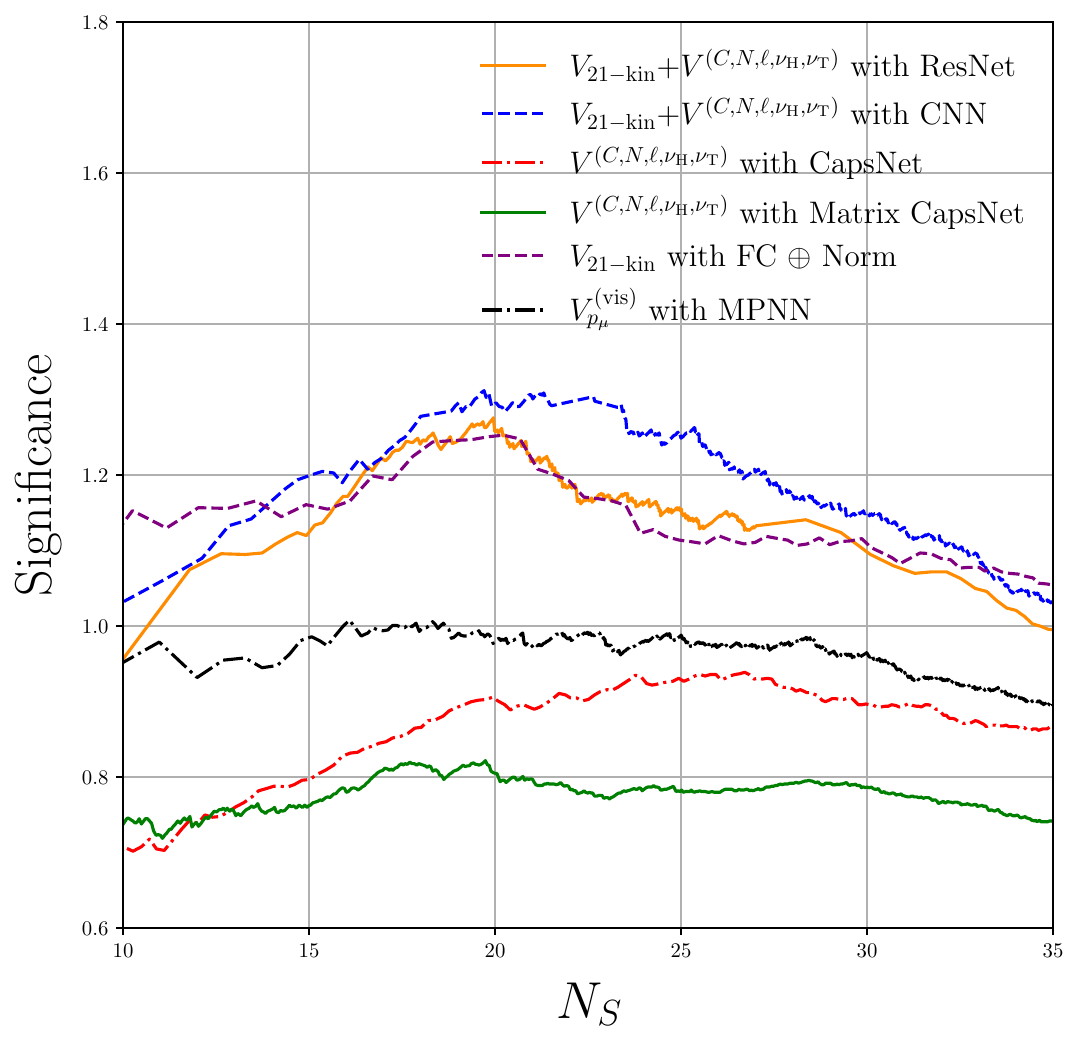}  \hspace*{0.1cm}
\includegraphics[width=0.47\textwidth,clip]{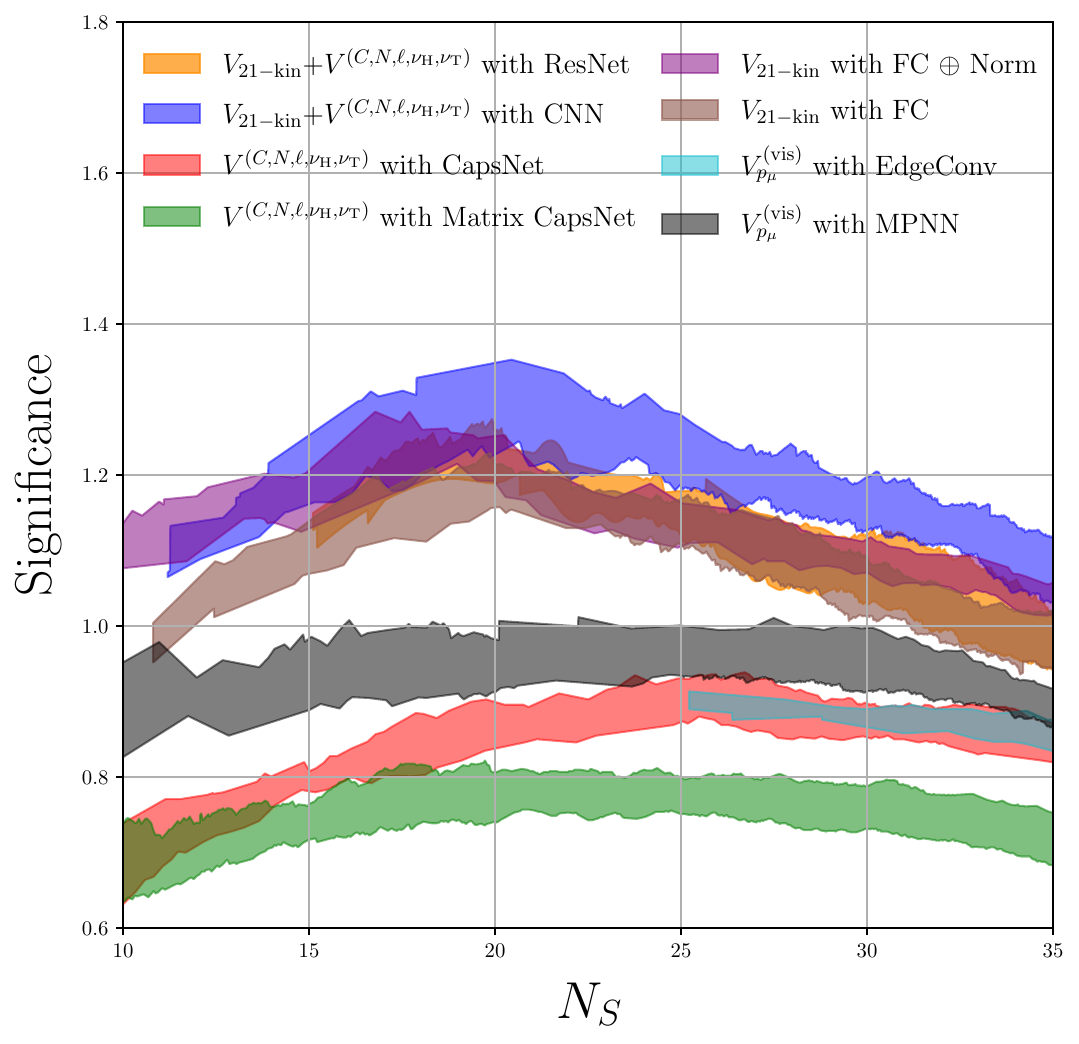} 
\caption{(left) Significance of observing double Higgs production at the HL-LHC with ${\cal L}$ = 3 ab$^{-1}$ for various NNs, taking the best model for each type. (right) Variation of the final results with 10 independent runs for the same NNs but different initial values of weights. 
\label{fig:sig_comparison}}
\end{center}
\end{figure*}
\begin{figure*}[th!]
\begin{center}
\includegraphics[width=0.6\textwidth,clip]{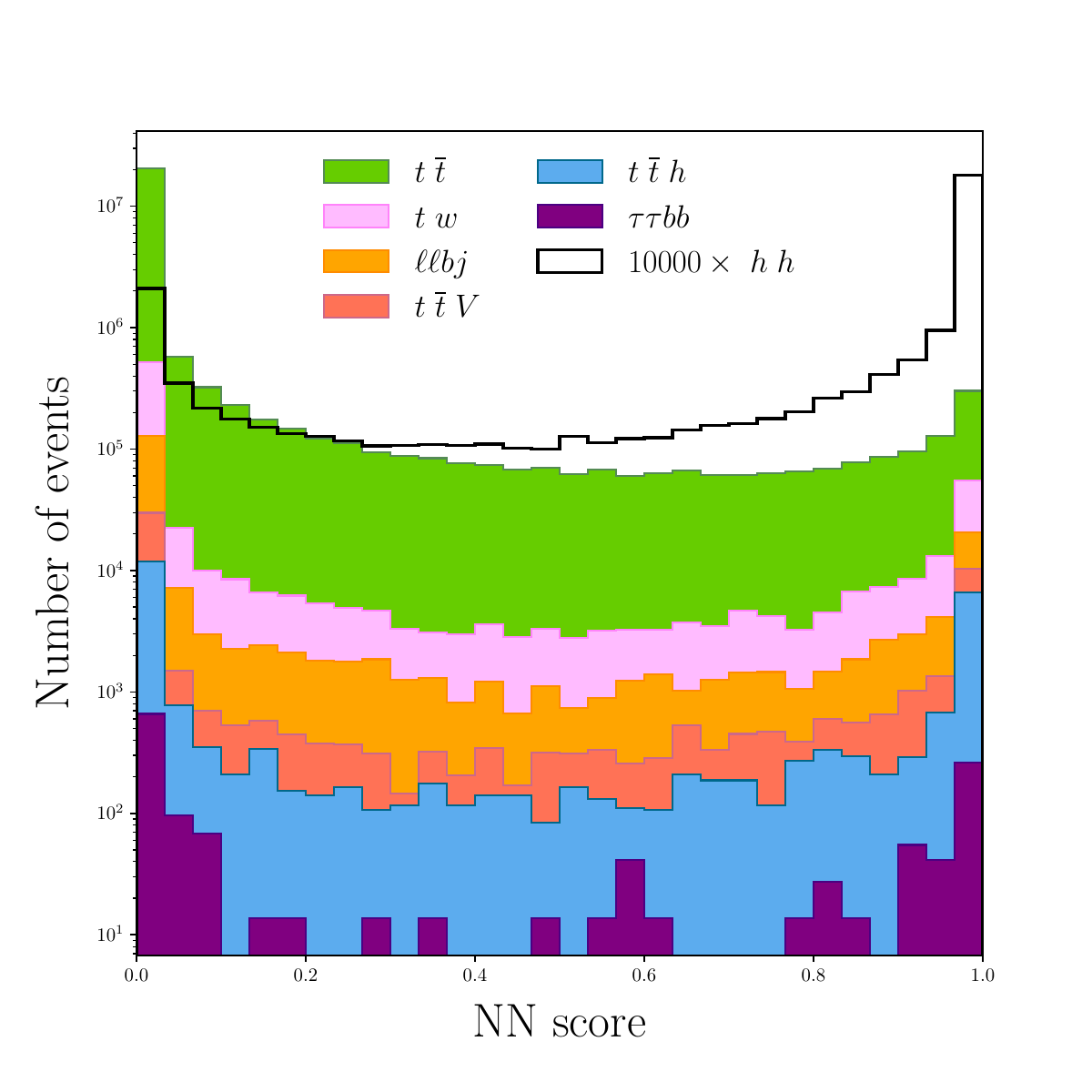} 
\caption{NN score for CNN with $V_{\text{21-kin}} $+$V^{(C,N,\ell,\nu_\text{H},\nu_\text{T})}$.
\label{fig:CNNscore}}
\end{center}
\end{figure*}
\begin{table*}[t]
\centering
\renewcommand\arraystretch{1.65}
\resizebox{\textwidth}{!}{%
\begin{tabular}{|c|c||c|c|c|c|c|c|c|c|}\hline
\textbf{Network} & \textbf{Input} & ~~~~\textbf{$hh$}~~~~ &~~~~ \textbf{$t\overline{t}$} ~~~~& ~~~\textbf{$tW$} ~~~& ~~~\textbf{$t\overline{t}h$}~~~ & ~~~\textbf{$t\overline{t}V$} ~~~& ~~\textbf{$\ell\ell b j$} ~~&~ \textbf{$\tau\tau bb$} ~& ~~~\textbf{$\sigma$}~~ ~\\ \hline \hline
 \multirow{2}{*}{ResNet} & $V_{\text{21-kin}} +V^{(C,N,\ell,\nu_{\text{H}},\nu_{\text{T}})}$  & ~0.006651~ & ~0.039796~  & ~0.016461~  & ~0.014426~  & ~0.002860~ & ~0.005352 ~& ~0.000446~  & 1.28 \\ \cline{2-10}
    					& ~~$V_{\text{21-kin}} +V^{(C,N,\ell)}$~~ & 0.006525  & 0.039796   & 0.018608 & 0.011389 & 0.003177 & 0.006690 & 0.000595 & 1.25 \\ \hline \hline
 \multirow{5}{*}{CNN}& $V_{\text{21-kin}} +V^{(C,N,\ell,\nu_{\text{H}},\nu_{\text{T}})}$ & 0.006568  &  0.031837 & 0.017176  & 0.012908  & 0.003495 & 0.007359  &  0.000298 & 1.31  \\ \cline{2-10}
  				& $V_{\text{21-kin}} +V^{(C,N,\ell)}$ & 0.006686 & 0.037143 & 0.014314 & 0.012908 & 0.003177 & 0.006690 & 0.000446 & 1.32  \\ \cline{2-10}
 				& $V^{(C,N,\ell,\nu_{\text{H}},\nu_{\text{T}})}$ & 0.006731 & 0.062347 & 0.036500 & 0.019488 & 0.006037 & 0.009367 & 0.000446 & 1.00 \\ \cline{2-10}
 				 & $V^{(C,N,\ell)}$ & 0.006806 & 0.076938 & 0.039362 & 0.020501 & 0.005402 & 0.012043 & 0.000595 & 0.94 \\ \cline{2-10}
                 & $V^{(C,N)}$ & 0.006829 & 0.169795 & 0.115224 & 0.029359 & 0.011121 & 0.005352 & 0.000744 & 0.65 \\ \hline\hline
 \multirow{2}{*}{CapsNet} & $V^{(C,N,\ell,\nu_{\text{H}},\nu_{\text{T}})}$ & 0.006645 & 0.091530 & 0.034353 & 0.021766 & 0.006355 & 0.004683  & 0.000446 & 0.91 \\   \cline{2-10}
    & $V^{(C,N,\ell)}$ & 0.006640  & 0.110101  & 0.031490 & 0.017717  & 0.006355 & 0.011374  & 0.000149   & 0.86\\ \hline\hline
 \multirow{2}{*}{~Matrix CapsNet~} & $V^{(C,N,\ell,\nu_{\text{H}},\nu_{\text{T}})}$ & 0.006580  &0.103469  & 0.048666  & 0.020248 & 0.006673 & 0.012712  & 0.000595  & 0.82 \\   \cline{2-10}
 & $V^{(C,N,\ell)}$ & 0.006792 & 0.112754 & 0.053676  & 0.019741 & 0.008579 & 0.009367  & 0.000595 & 0.82  \\ \hline\hline
 \multirow{9}{*}{FC}  & ~$V_{\text{21-kin}} +V^{(\text{vis})}_{p_\mu}+V^{(\nu_{\text{H}})}_{ p_{\mu}} $~& 0.006685  				& 0.042449  & 0.025049  & 0.015439  & 0.003813  & 0.007359  & 0.000298  & 1.18 \\  \cline{2-10}
  				& $V_{\text{21-kin}} $ & 0.006624 & 0.042449  & 0.018608 & 0.014680  & 0.003495  & 0.005352  & 0.000446  & 1.23 \\   \cline{2-10}
 				 & $V_{\text{15-kin}} $ & 0.006802  & 0.042449  & 0.024333& 0.016198 & 0.003177 & 0.011374  & 0.000298 & 1.18 \\  \cline{2-10}
  				& $V_{\text{11-kin}}$ & 0.006626 & 0.054387 & 0.022186 & 0.017210  & 0.003813 & 0.011374  & 0.000149 & 1.09 \\  \cline{2-10}
  				& $V^{(\text{vis})}_{p_\mu}$  & 0.006832 & 0.120714  & 0.071568 & 0.025563 & 0.006990  & 0.016057  & 0.000446 & 0.76 \\  \cline{2-10}
  				& $V^{(\text{vis})}_{p_\mu}+V^{(\nu_{\text{H}})}_{ p_{\mu}} $ & 0.006766 & 0.114081  & 0.057970 & 0.024803 & 0.006673 & 0.018064  & 0.000446 & 0.78 \\ \cline{2-10}
				& $V_{\text{10-kin}}^{\rm Ref.[62]}$ & 0.006550  & 0.053061  & 0.033637 & 0.022810  & 0.004448 &  0.014719 & 0.000149 & 1.01   \\  \cline{2-10}
				& $V_{\text{8-kin}}^{\rm Ref.[63]}$ &  0.006804 & 0.083571  & 0.030059 & 0.021766  & 0.005402 &   0.015388 & 0.000149 & 0.94   \\  \cline{2-10}
				\hline\hline
 \multirow{9}{*}{FC $\oplus$ Norm}  & ~$V_{\text{21-kin}} +V^{(\text{vis})}_{p_\mu}+V^{(\nu_{\text{H}})}_{ p_{\mu}} $~ & 0.006556  				& 0.045102  & 0.012167  & 0.016529  & 0.003813  & 0.008698  & 0.000149  & 1.23 \\  \cline{2-10}
  				& $V_{\text{21-kin}} $ & 0.006742 & 0.047755  & 0.015745 & 0.016859  & 0.003177  & 0.004683  & 0.000298  & 1.25 \\   \cline{2-10}
 				 & $V_{\text{15-kin}} $ & 0.006749  & 0.047755  & 0.022186& 0.023471 & 0.003495 & 0.006021  & 0.000446 & 1.17 \\  \cline{2-10}
  				& $V_{\text{11-kin}}$ & 0.006714 & 0.051734 & 0.012882 & 0.018223  & 0.003813 & 0.012712  & 0.000149 & 1.15 \\  \cline{2-10}
  				& $V^{(\text{vis})}_{p_\mu}$  & 0.006503 & 0.108775  & 0.078009 & 0.035041 & 0.007308  & 0.019402  & 0.000446 & 0.72 \\  \cline{2-10}
  				& $V^{(\text{vis})}_{p_\mu}+V^{(\nu_{\text{H}})}_{ p_{\mu}} $ & 0.006832 & 0.118061  & 0.090891 & 0.027334 & 0.009214 & 0.018064  & 0.000893 & 0.72 \\ \cline{2-10}
				& $V_{\text{10-kin}}^{\rm Ref.[62]}$ &  0.006622 & 0.062347  & 0.021470 & 0.025454  & 0.003177 &  0.009367 & 0.000298 & 1.05   \\  \cline{2-10}
				& $V_{\text{8-kin}}^{\rm Ref.[63]}$ &  0.006574 & 0.067653  & 0.027912 & 0.024132  & 0.004131 &0.006690   & 0.000446 &1.01    \\  \cline{2-10}
				\hline\hline
EdgeConv & $V^{(\text{vis})}_{p_\mu}$& 0.008383 & 0.122040 & 0.072999 & 0.029106 & 0.006673 & 0.018064 & 0.000446 & 0.91 \\ \hline\hline
 \multirow{2}{*}{MPNN} & $V^{(\text{vis})}_{p_\mu}$& 0.006522 & 0.075612 & 0.017892 & 0.019235 & 0.004766 & 0.009367 & 0.000149 & 0.99 \\  \cline{2-10}
  & $V^{(\text{vis})}_{p_\mu}+V^{(\nu_{\text{H}})}_{ p_{\mu}} $ & 0.006986 & 0.076938 & 0.051529 & 0.023032 & 0.005084 &      0.011374        &0.000446 & 0.93        \\ \hline
  \end{tabular}%
}
\caption{Signal and background cross section (in fb) cutting on NN score for various combinations of NN architectures and inputs. The NN score is chosen such that the signal number of events is approximately 20 ($N_S\approx20$.).
The significance $\sigma$ is calculated using the log-likelihood ratio for a luminosity of 3 ab$^{-1}$ at the 14 TeV LHC. Note that the efficiency of the NN score cut can be calculated by taking the ratio of cross sections in this table and those in Table \ref{tab:BGD}.}
\label{tab:significance}
\end{table*}

In this section, we summarize our results of exploration of different NN structures. 
We have tried (i) fully connected NN with four momenta and kinematic variables, 
(ii) CNN, ResNet, CapsNet and Matrix CapsNet with kinematic variables and image data, and 
(iii) EdgeConv and MPNN with four momentum. 

In the left panel of Fig. \ref{fig:sig_comparison}, we summarize the signal significance of double Higgs production at the HL-LHC with ${\cal L}$ = 3 ab$^{-1}$ for various NNs (left, top), taking the best model for each type.
We find that CNN performs the best, followed by ResNet with 21 kinematic variables and image inputs. ResNets result is very comparable to DNN with 21 kinematic variables (no image data used for fully connected NNs). These three different NNs algorithms lead to the similar performance with the signal significance 1.2-1.3 for the signal number of events around 20.
As shown in Fig. \ref{fig:CNN_ResNet} and in Table \ref{tab:significance}, addition of lepton-image ($V^{(C, N, \ell)}$) to charged and neutral hadrons ($V^{(C, N)}$) helps improve the significance (see green-solid and red-dotted). Moreover, the full set of images ($V^{(C, N, \ell, \nu_\text{H},\nu_\text{T})}$) including neutrino momentum information brings the further increase (see red-solid). This substantial improvement is attributed to the inclusion of 21 kinematic variables along with all image inputs, which lead to about 1.3 significance for the signal events $N_S=20$. This remarkable gain in the signal significance over the existing results \cite{Kim:2019wns,Abdughani:2020xfo} are due to the interplay between novel kinematics and machine learning algorithms. It is noteworthy that one can form image data out of leptons and reconstructed neutrinos, and obtain the improved result. 
We have checked that the CNN with 21 kinematic variables and image input outperforms network structures used in literature, which are labeled as $V_{10-kin}^{\rm Ref.[62]}$ and $V_{8-kin}^{\rm Ref.[63]}$ in Table \ref{tab:significance}.
FC with more kinematic variables lead to a higher significance, even without four momentum input, as illustrated in Table \ref{tab:significance}. When using $V_{\rm 11-kin}$, the significance drops, which indicates that it is crucial to choose the right kinematic variables to reach the maximum significance, and $V_{\rm 21-kin}$ are the right choice. 
The second class of algorithms include CapsNet, Matrix CapsNet and MPNN, which lead to the signal significance around 0.8-1. With four momentum input only (without any kinematic variables or images), we find that MPNN performs the best, reaching the significance of $\sim 1$. 

In the right panel of Fig. \ref{fig:sig_comparison}, we show the variation of the final results for 10 independent runs of the same NNs with different initial values of weights for various NNs (shown in the left panel). This exercise serves as an estimation of uncertainties associated with NN. As illustrated in Fig. \ref{fig:sig_comparison}, our results are stable under multiple runs, leading to similar results. 

Taking 20 for the benchmark signal number of events, in Table \ref{tab:significance}, we summarize the signal and background cross sections (in fb) as well as the significance for various combinations of NN architectures and inputs by cutting on the NN score. The significance $\sigma$ is calculated using the log-likelihood ratio (Eq. (\ref{Eq:SigDis})) for a luminosity of 3 ab$^{-1}$ at the 14 TeV LHC. Note that the signal and background efficiencies with respect to the cut on NN scores can be estimated by taking the ratio of each cross section in this table and the one in Table \ref{tab:BGD}.

As an illustration, in Fig. \ref{fig:CNNscore}, we show the NN score distribution for CNN with $V_{\text{21-kin}} +V^{(C,N,\ell,\nu_\text{H},\nu_\text{T})}$ which gives the best significance. We scan over the lower bound on the NN score and count survived number of signal and background events. For the signal number of events $N_S=20$, we obtain 220 total background events, which comprises (96, 52, 39, 10, 22, 1) events for individual backgrounds ($t\bar t$, $tW$, $t\bar t h$, $t\bar t V$, $\ell\ell b j$, $\tau\tau b b$), respectively. $t\bar t$ process makes up about 44\% of the total background after applying the cut on the NN score, while $tW$ and $t\bar t h + t\bar t V$ account for about 24\% and 22\%, respectively. $\ell\ell b j + \tau\tau b b$ background makes up roughly 10\%. 
After the baseline selection (before cutting on NN score), the $t\bar t$ contribution was 97\%, while $tW$ was $\sim$ 2\%, as shown in Table \ref{tab:BGD}.
The signal efficiency for CNN is about $\epsilon_S=23\%$, while the background rejection is $3500 = 1/\epsilon_B$, where $\epsilon_B = 2.8 \times 10^{-4}$.
Therefore the background to signal ratio is reduced to ${\sigma_{bknd}}/{\sigma_{hh}}\approx 11$ from 
${\sigma_{bknd}}/{\sigma_{hh}}\approx 9250$ (see Table \ref{tab:BGD}). 

\begin{figure}[t!]
\begin{center}
\includegraphics[width=0.7\textwidth,clip]{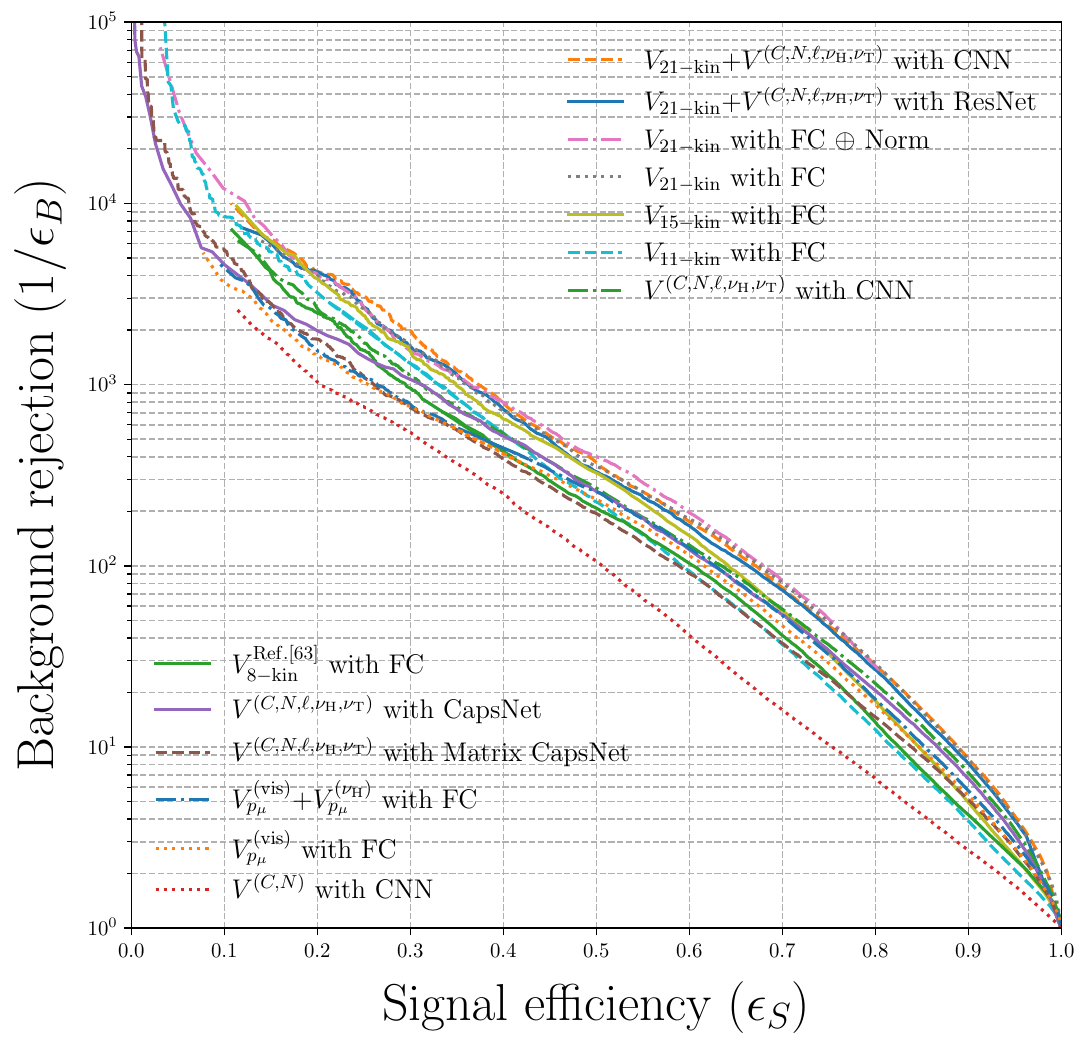} 
\caption{ROC curves for selected NN architectures evaluated on the same test sample.
\label{fig:roc}}
\end{center}
\end{figure}

We show in Fig. \ref{fig:roc} the receiver operating characteristic (ROC) curve for selected NN architectures evaluated on the same test sample, taking the signal efficiency $\epsilon_S$ for the $x$-axis and the background rejection $1/\epsilon_B$ for the $y$-axis. 
The area under the ROC curve (AUC) is another commonly used quantity to test the performance of a classification model. Here we quote a few sample AUC values in the ($\epsilon_S$, $1-\epsilon_B$) plane.
As shown in Fig. \ref{fig:roc}, the performance of CNN with $V_{\text{21-kin}} +V^{(C,N,\ell,\nu_\text{H},\nu_\text{T})}$ is the best with the AUC value 0.959, followed by
ResNet with $V_{\text{21-kin}} +V^{(C,N,\ell,\nu_\text{H},\nu_\text{T})}$ 0.956,  FC with $V_{\text{21-kin}}$ 0.958,
CNN with $V^{(C,N,\ell,\nu_\text{H},\nu_\text{T})}$ 0.952, 
CapsNet with $V^{(C,N,\ell,\nu_\text{H},\nu_\text{T})}$ 0.948, 
Matrix CapsNet with $V^{(C,N,\ell,\nu_\text{H},\nu_\text{T})}$ 0.938, 
FC with $V^{(\text{vis})}_{p_\mu}+V^{(\nu_{\text{H}})}_{ p_{\mu}} $ 0.945, 
FC with $V^{(\text{vis})}_{p_\mu}$ 0.941, 
FC with $V_{\text{15-kin}}$ 0.940,
FC with $V_{\text{11-kin}}$ 0.926, 
CNN with $V^{(C,N)}$ 0.897, etc.

%% file: 6_conclusion.tex
\begin{table*}[t]
\renewcommand\arraystretch{1.05}
\centering
\begin{tabular}{|c||c|c||c|c|}
\hline \hline
\multirow{2}{*}{Channel}			&  \multicolumn{2}{c||}{~Statistical only~}	 &  \multicolumn{2}{c|}{~Statistical + Systematic~} \\
\cline{2-5}
							& ~~\,ATLAS\,~~& ~~~~CMS~~~~&~~\,ATLAS\,~~& CMS \\
\hline \hline
$hh \to b\bar b b\bar b $  			& 1.4			& 1.2		& 0.61 		& 0.95		\\
$hh \to b\bar b \tau^+\tau^-$		& 2.5			& 1.6		& 2.1			& 1.4  		\\
$hh \to b\bar b \gamma\gamma$	& 2.1			& 1.8		& 2.0			& 1.8  		\\
$hh \to b\bar b VV (\ell\ell\nu\nu)$	& -			& 0.59	& - 			& 0.56		\\
$hh \to b\bar b ZZ (4\ell)$			&-			& 0.37	& -			& 0.37 	\\
\hline
combined						& 3.5			& 2.8  	& 3.0			& 2.6 	\\
\hline
							& \multicolumn{2}{c||}{combined} & \multicolumn{2}{c|}{combined}	\\
							& \multicolumn{2}{c||}{4.5}		&\multicolumn{2}{c|}{4.0}	 \\
\hline \hline
~combined with the new results on~  	& \multirow{2}{*}{\textbf{3.8}}	& \multirow{2}{*}{\bf{3.0}}	& \multirow{2}{*}{\bf{3.2}}	& \multirow{2}{*}{\bf{2.8}}  \\
$hh \to b\bar b VV (\ell\ell\nu\nu)$ in this study~	&		&	& & \\
\hline
							& \multicolumn{2}{c||}{combined}	&  \multicolumn{2}{c|}{combined}\\
							& \multicolumn{2}{c||}{\bf{4.8}}			& \multicolumn{2}{c|}{\bf{4.2}}		 \\
\hline \hline
\end{tabular}
\caption{\label{table:hh} Summary of significance of the individual channels and their combination at the HL-LHC. Results for the first 5 channels and their combinations are taken from Ref. \cite{Cepeda:2019klc}, while last two rows are combined results, replacing the existing result with new results presented in this paper. We use 1.3 for the significance for $hh \to b\bar b VV (\ell\ell\nu\nu)$ from this study, and assume that the systematics brings about 10\% reduction in the significance. 
As in Ref. \cite{Cepeda:2019klc}, we assume that ATLAS and CMS would have the same significance, as the results in the $b\bar b VV (\ell\ell\nu\nu)$ and $b\bar b ZZ (4\ell)$ are only performed by the CMS collaboration.} 
\end{table*}
In this paper, we investigated the discovery potential of the HL-LHC for the double Higgs production in the final state with two $b$-tagged jets, two leptons and the missing transverse momentum. We have utilized the novel kinematic variables and particle images (including reconstructed neutrinos) along with various machine learning algorithms to increase the signal sensitivity over the large backgrounds. 

Our study provides a plenty of meaningful and interesting results.
First, for the collider kinematics point of view, we have shown the importance of high-level kinematic variables in NNs.
Without suitable kinematic variables, NNs with images only can not achieve optimal results.
We also showed the minor improvement in the final significance, when using kinematic variables constructed with neutrino momenta, which are obtained via Topness and Higgsness minimization. 

Second, for various machine learning methods, we have made a dedicated comparison with different types of input features: low-level four momenta, high-level kinematic variables and particle images. 
We have observed that CNN outperforms most other NNs, and are comparable to or slightly better than DNN.
We also illustrated the importance of high-level kinematic variables in DNN as well as in CNN/ResNet. 
One of important results in our study is that the signal significance is roughly stable around $\sim 1$ for various machine learning algorithms with different choices of input features. 
We have checked that the variation in the NN performance within each algorithm is also stable, when repeating the same training with new random seeds. 
Finally, for physics point of view, we have shown that the expected significance is $\sim 1.3$ for the number of the signal events of $\sim 20$ with the improved $b$-tagging efficiency. This is due to the interplay of clever kinematic variables and flexible NNs. 

Such a high and stable significance has a large impact on the double Higgs production.
Especially, the dilepton channel would make a sizable contribution to the combined significance. 
Our study also motivates a similar analysis in the semi-leptonic channel, whose significance is known to be much smaller than that in the dilepton channel. 
Table \ref{table:hh} summarizes the signal significance of the five individual channels and their combination at the HL-LHC with 3 ab$^{-1}$, taken from Ref. \cite{Cepeda:2019klc}. 
The significances are added in quadrature, and the channels are treated as uncorrelated, assuming that the systematic uncertainties which are expected to be correlated between the experiments, such as the theory uncertainties and the luminosity uncertainty, have little impact on the individual results.
Since the results in the $b\bar b VV (\ell\ell\nu\nu)$ and $b\bar b ZZ (4\ell)$ are only performed by the CMS collaboration, the likelihoods for those two channels are scaled to 6 ab$^{-1}$ in the combination of ATLAS and CMS, leading to 4.5 without systematics and 4.0 with systematics.
Similar results are found in Ref. \cite{DiMicco:2019ngk}. 

\begin{figure}[t]
\begin{center}
\includegraphics[width=1\textwidth,clip]{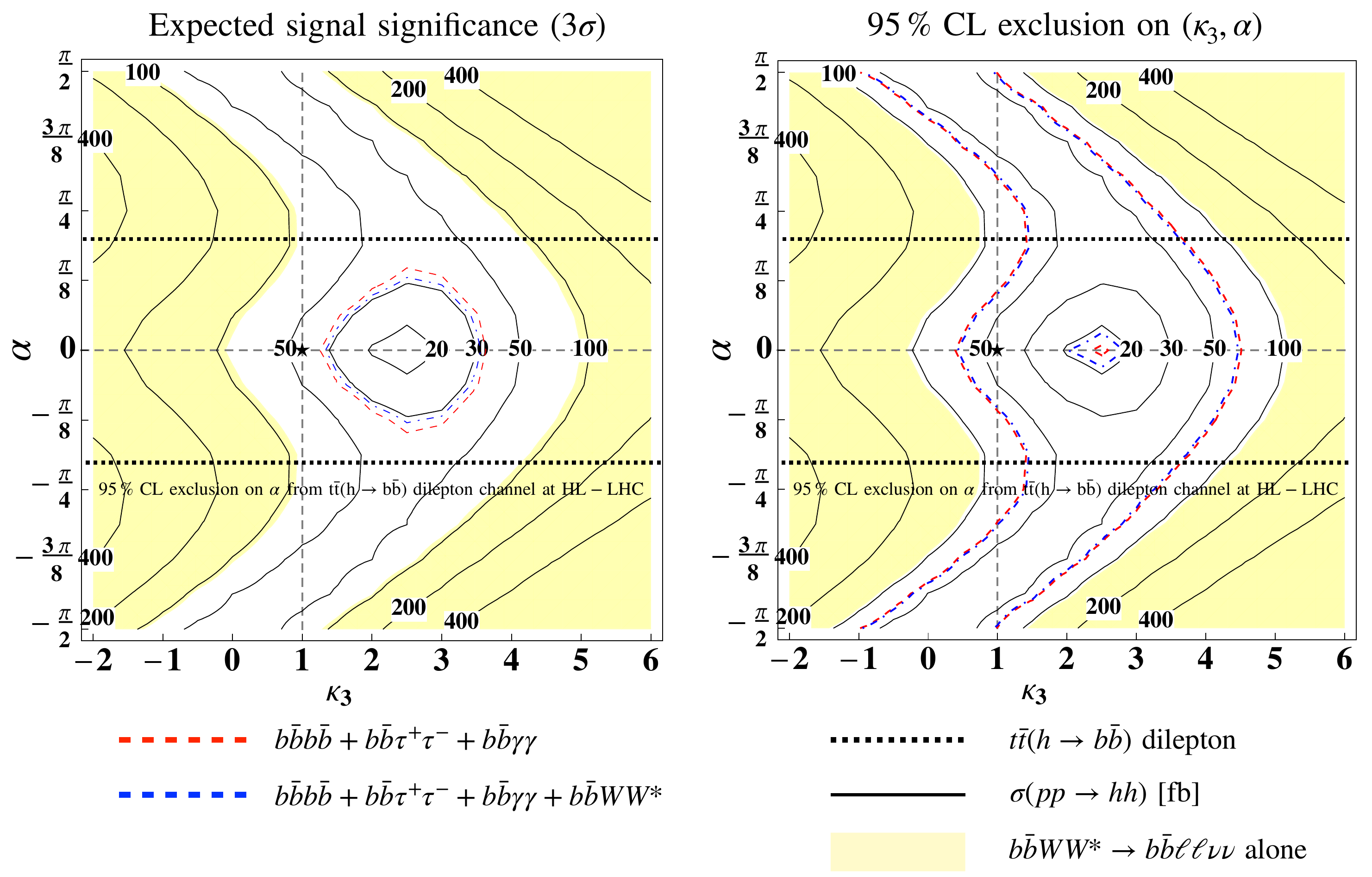}
\caption{\label{fig:exclusion}
Expected $3\sigma$ significance of observing Higgs boson pair production  (left) and 95\% C.L. exclusion (right) in the ($\kappa_3$, $\alpha$) plane at the HL-LHC with 3 ab$^{-1}$. We used the binned log-likelihood analysis with statistical uncertainties only, assuming the same efficiencies for all ($\kappa_3$, $\alpha$) values as one for $(\kappa_3, \alpha) = (1,0)$ (SM point denoted by $\star$). Contours of the double Higgs production cross section (in fb) are shown in black-solid curves. The yellow shaded region is obtained using results in this study for the dilepton channel ($hh \to b\bar b W W^* \to b\bar b \ell\ell\nu\bar\nu$). The red dashed curve is obtained combining three channels, $b\bar b b\bar b + b\bar b \tau^+\tau^- + b\bar b \gamma\gamma$ following Ref. \cite{ATLAS:2018combi}, while the blue dashed curve includes all four channels. The horizontal-black dotted line represents a sample 95\% exclusion on the CP angle from the dilepton channel of $t \bar t h$ production with $h \to b \bar b$ \cite{Goncalves:2021dcu}, $|\alpha| \lesssim 35^\circ$. 
}
\end{center}
\end{figure}
In the two bottom rows, we show the updated combined results, replacing the existing results (0.59 without systematics and 0.56 with systematics) with new results (1.3 without systematics) presented in this paper. 
The total systematic uncertainties from current analyses are about 20\% \cite{Sirunyan:2017guj,ATLAS:2019vwv}, while Ref. \cite{CMS:2015nat} considers somewhat optimistic scenarios and scans the systematic uncertainty up to 5\% for the expected relative uncertainty in the signal yield. 
The estimation of the systematic uncertainty is beyond the scope of our study. Instead, we simply consider 10\% reduction in the signal significance, to take into account the systematics in the $b\bar b VV (\ell\ell\nu\nu)$ channel, which is not an unreasonable assumption, since the significance reduction in the CMS study in the same channel is about 5\%. This naive estimate of the systematic uncertainty results in the reduction of ATLAS (CMS) significance from 3.8 to 3.2 (from 3.0 to 2.8). 
Considering both ATLAS and CMS, the expected significance including our results in this study becomes 4.8 without systematics and 4.2 with systematics. 

The double Higgs production exhibits a non-trivial interference between the box and triangle diagrams.
 In Fig. \ref{fig:exclusion}, we show the production cross section of the double Higgs (black-solid) in the 2 dimensional parameter space of ($\kappa_3$, $\alpha$), where $\alpha$ is the CP angle in the $t \bar t h $ coupling (see Ref. \cite{Li:2019uyy} for the impact of the double Higgs production on the top quark Yukawa coupling.). The \ding{72} mark represents the $hh$ production cross section corresponding to the SM point $(\kappa_3, \alpha) = (1,0)$, which happens to be near the minimum cross section at $(\sim2.5, 0)$. Although the signal cross section is too small to measure the double Higgs production with this single channel alone, one could set an exclusion limit, as shown in the right panel of Fig. \ref{fig:exclusion}. A significance ($\sigma_{excl}$) for exclusion can be calculated using a different likelihood-ratio 
\begin{equation}
  \sigma_{excl} \equiv
    \sqrt{-2\,\ln\bigg(\frac{L(S\!+\!B | B )}{L( B | B)}\bigg)} \, .
\label{Eq:SigExc} 
\end{equation}
For an exclusion at 95\% C.L., we demand $\sigma_{excl} \geq 2$, leading to $\sigma_{excl}^{bbWW}\approx 105$ fb, which is shown as the yellow shaded region in Fig. \ref{fig:exclusion}, which would be excluded at 95\% C.L. at the HL-LHC with 3 ab$^{-1}$, assuming the efficiencies are the same as that for the SM production. 
Although this is roughly a reasonable approximation, a dedicated study should be performed to obtain more precise bounds. 
After reproducing the signal significances (1.4, 2.5 and 2.1) for $b\bar b b\bar b$, $b\bar b \tau^+\tau^- $ and $ b\bar b \gamma\gamma$ channels as shown in Table \ref{table:hh} following Ref. \cite{ATLAS:2018combi}, we have confirmed the individual $2\sigma$ bounds on the double Higgs production cross section: 
$\sigma_{excl}^{bbbb}\approx 100$ fb,
$\sigma_{excl}^{bb\gamma\gamma}\approx 95$ fb, and 
$\sigma_{excl}^{bb\tau\tau}\approx 76$ fb.
Then we have calculated the combined 95\% C.L. exclusion, $\sigma_{excl}^{bbbb+bb\gamma\gamma+bb\tau\tau}\approx 66$ fb, using the binned likelihood ratio, which is shown as the red dashed curve in Fig. \ref{fig:exclusion}. 
The addition of our results in $b\bar bWW^* \to b\bar b \ell\ell \nu\bar \nu$ improves further, $\sigma_{excl}^{bbbb+bb\gamma\gamma+bb\tau\tau+bbWW}\approx 64$ fb, which is shown as the blue dashed curve. 
Independent bounds on the Top-Higgs CP phase $\alpha$ can be obtained by studying $t\bar t h$ production. As an illustration, we have added the horizontal-black dotted line, which represents 95\% exclusion on the CP angle from the dilepton channel of $t \bar t h$ production with $h \to b \bar b$ \cite{Goncalves:2021dcu}.
Similarly, the expected $3\sigma$ significance of observing Higgs boson pair production is shown in the left panel, corresponding to $\sigma_{3\sigma}^{bbWW}\approx 92$ fb,
$\sigma_{3\sigma}^{bbbb+bb\gamma\gamma+bb\tau\tau}\approx 34$ fb, and 
$\sigma_{3\sigma}^{bbbb+bb\gamma\gamma+bb\tau\tau+bbWW}\approx 31$ fb. 

The novel kinematic variables that we have introduced can be used in other channels.
For example, the semi-leptonic channel $hh \to b\bar b W W^* \to b \bar b \ell \nu_\ell j j $ is known to give a very poor signal significance \cite{Adhikary:2017jtu}, although a somewhat promising result was obtained in Ref. \cite{Goertz:2013kp} using jet substructure techniques. A recent study Ref. \cite{ATLAS:2019mwn} applied the Topness and Higgsness to $hh \to b\bar b W W^* \to b \bar b \ell \nu_\ell j j $ and showed that a suitable cut leads to 69\% of the signal surviving fraction and 1.2\% of the background surviving fraction. We expect that these new kinematic variables could help improve the signal and background separation in the semi-leptonic channel as well as other channels such as $hh \to \gamma \gamma WW^*$ and $hh \to W W^* W W^*$. 
Finally we note that machine learning methods could help to study the semi-leptonic as well as the fully hadronic channel, utilizing the effectiveness in resolving the combinatorial problems in these channels \cite{Alhazmi:2022qbf,Shmakov:2021qdz,Fenton:2020woz,Lee:2020qil,Badea:2022dzb}.

%% file: 7_appendix.tex
\section{A brief review on kinematic variables}
\label{appen:variable}

In this appendix, we provide a short review on kinematic variables used in this paper. For more details, we refer to Refs. \cite{Barr:2010zj,Barr:2011xt}. 
We consider the following 21 kinematic variables in total (Eqs. (\ref{eq:11_kin})-(\ref{eq:15_kin})), in addition to four momentum of all visible and invisible particles (Eqs. (\ref{eq:vis_4momenta})-(\ref{eq:Hneu_4momenta})),
\begin{eqnarray} 
V_{\text{21-kin}} = &\{& p_T(\ell_1), p_T(\ell_2), p_{Tbb}, p_{T\ell\ell}, \mpt, \Delta R _{bb}, \Delta R_{\ell\ell}, \Delta\phi_{bb,\ell\ell}, m_{\ell\ell}, m_{bb}, \label{eq:V21} \\ 
      & & min[\Delta R_{b\ell}], \Delta R^{\text{H}}_{\nu \nu}, m^{\text{H}}_{\nu \nu},  \Delta R^{\text{T}}_{\nu \nu}, m^{\text{T}}_{\nu \nu},  \sqrt{\hat{s}}_{\text{min}}^{(bb\ell \ell)},  \sqrt{\hat{s}}_{\text{min}}^{(\ell \ell)}, M_{T2}^{(b)}, M_{T2}^{(\ell)}, {\rm H}, {\rm T} \,  \} \, . \nonumber
\end{eqnarray}
The distributions of 16 variables (out of 21 except for $\text{min}[\Delta R_{b\ell}], \Delta R^{\text{H}}_{\nu \nu}, m^{\text{H}}_{\nu \nu},  \Delta R^{\text{T}}_{\nu \nu}, m^{\text{T}}_{\nu \nu}$) in Eq. \ref{eq:V21} are shown in Refs. \cite{Kim:2019wns,Kim:2018cxf} and we obtained similar results.

\begin{figure*}[t]
\begin{center}
\includegraphics[width=0.51\textwidth,clip]{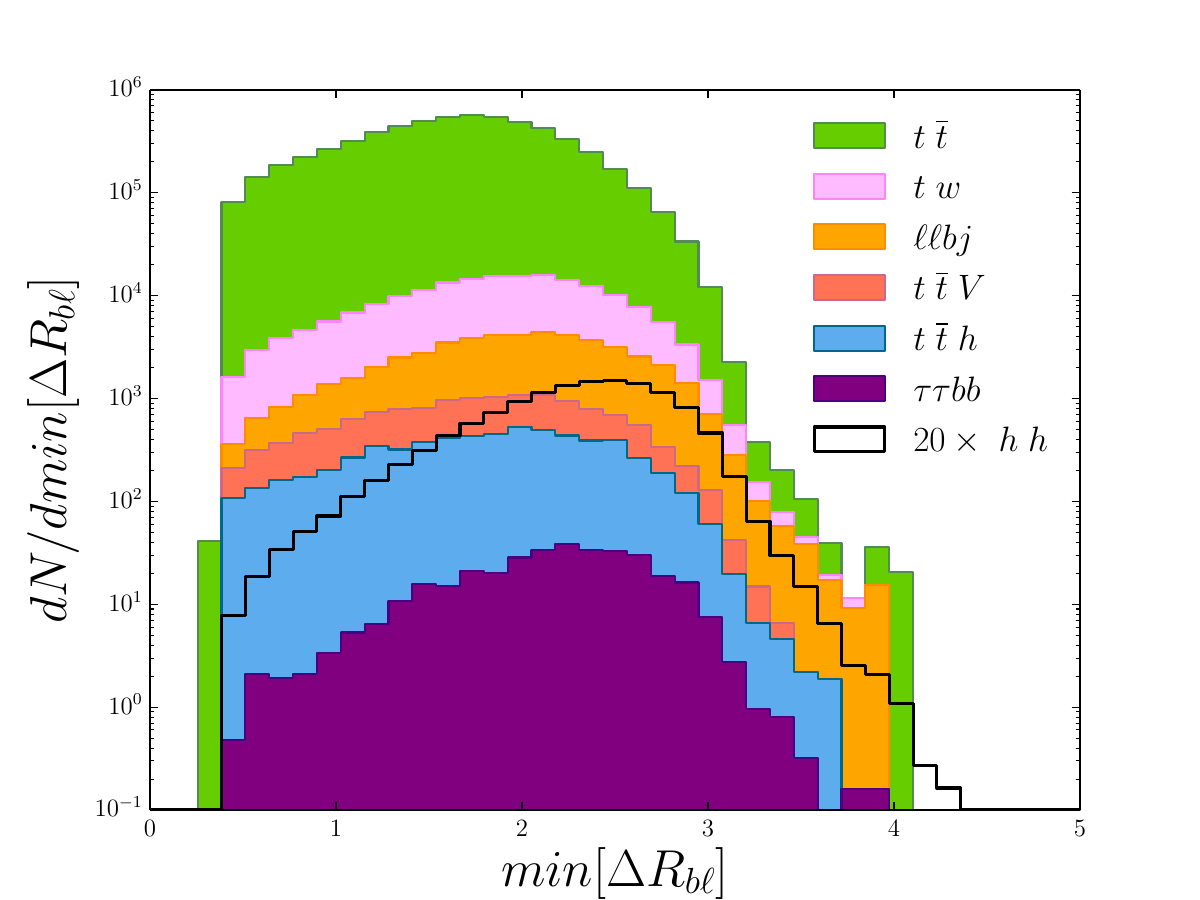} \hspace*{-0.7cm}
\includegraphics[width=0.51\textwidth,clip]{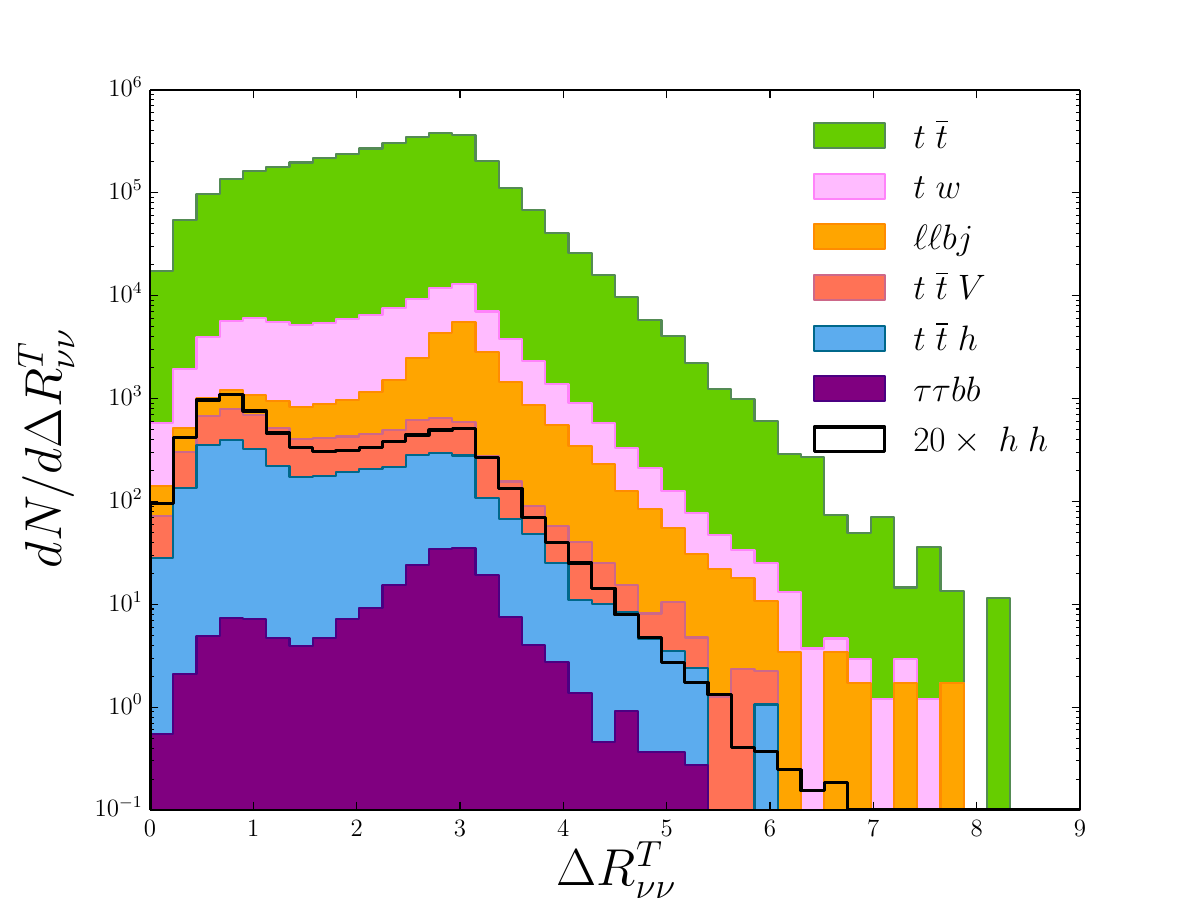} \\ 
\includegraphics[width=0.51\textwidth,clip]{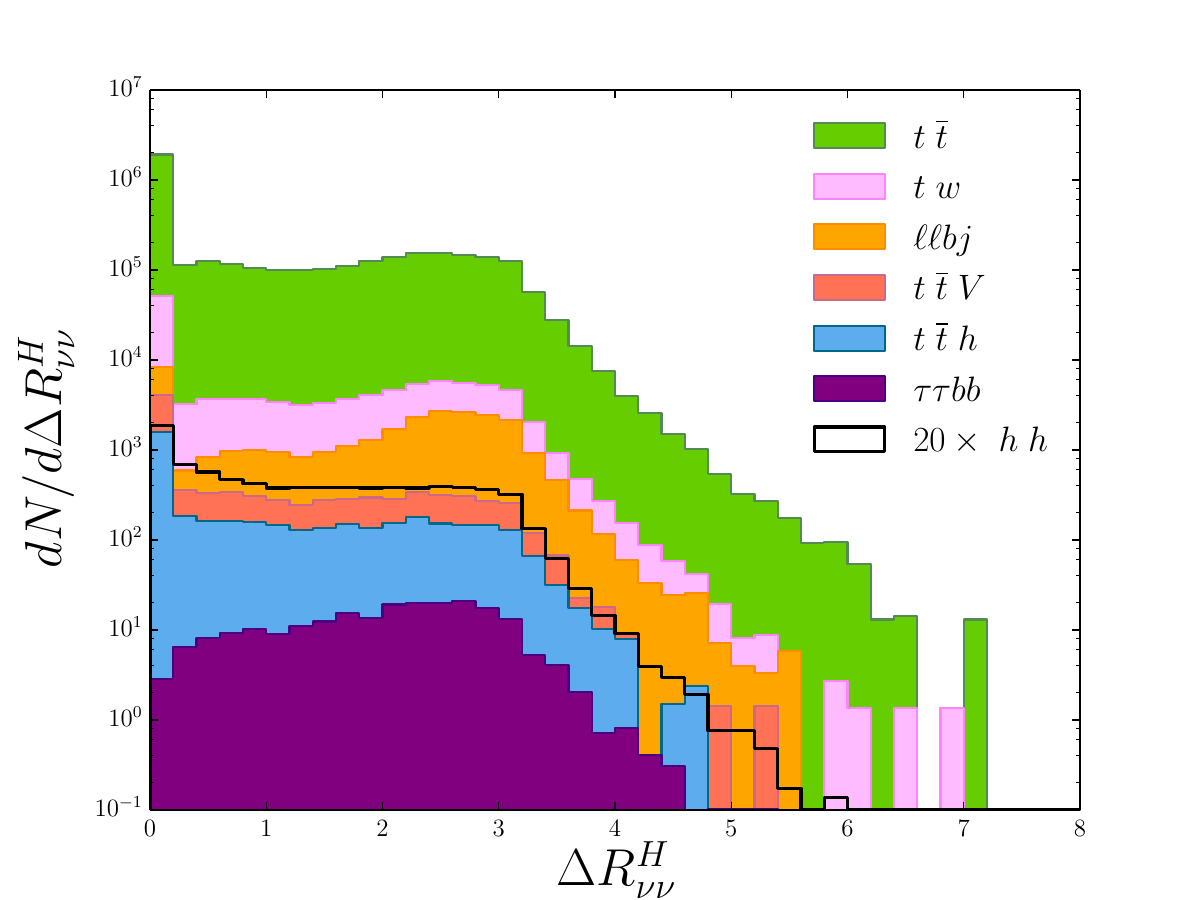}  \hspace*{-0.7cm} 
\includegraphics[width=0.51\textwidth,clip]{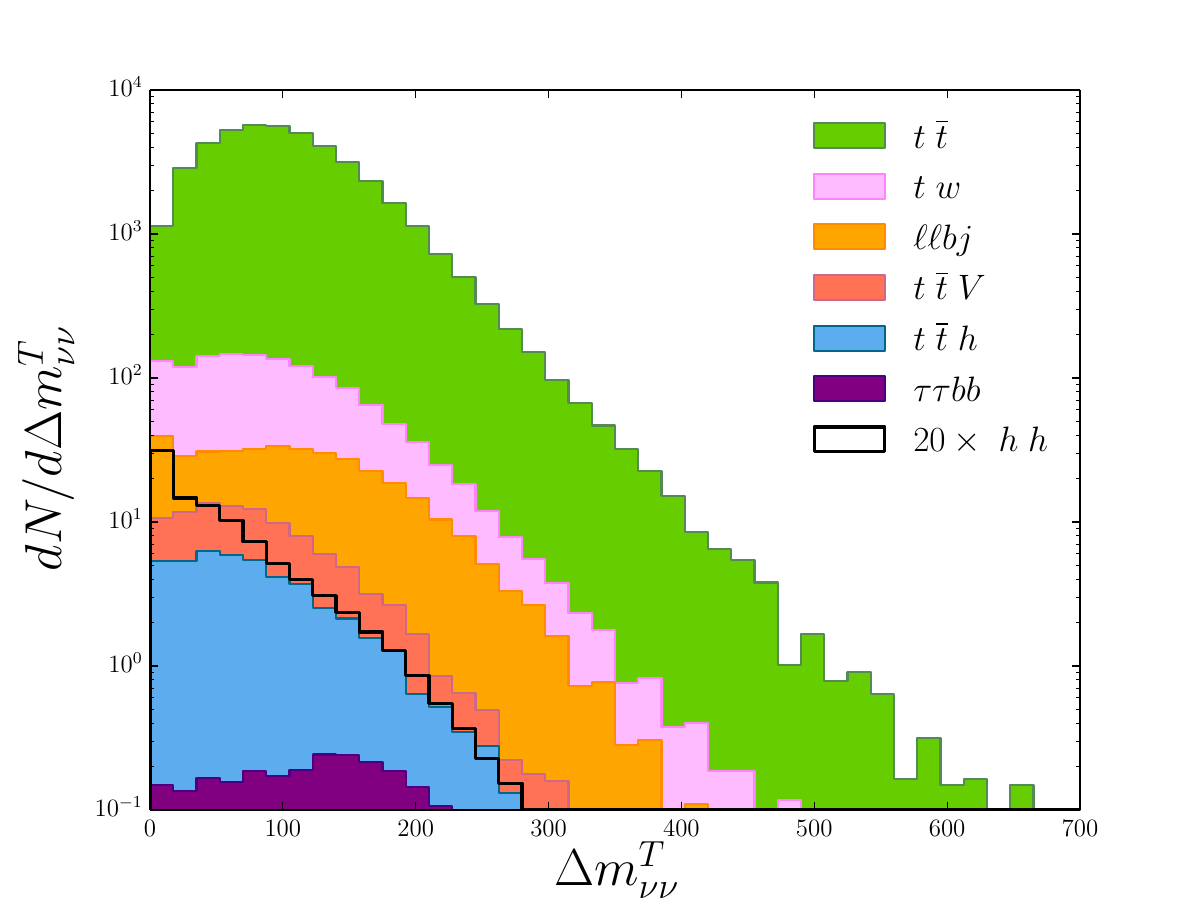} \\ 
\includegraphics[width=0.51\textwidth,clip]{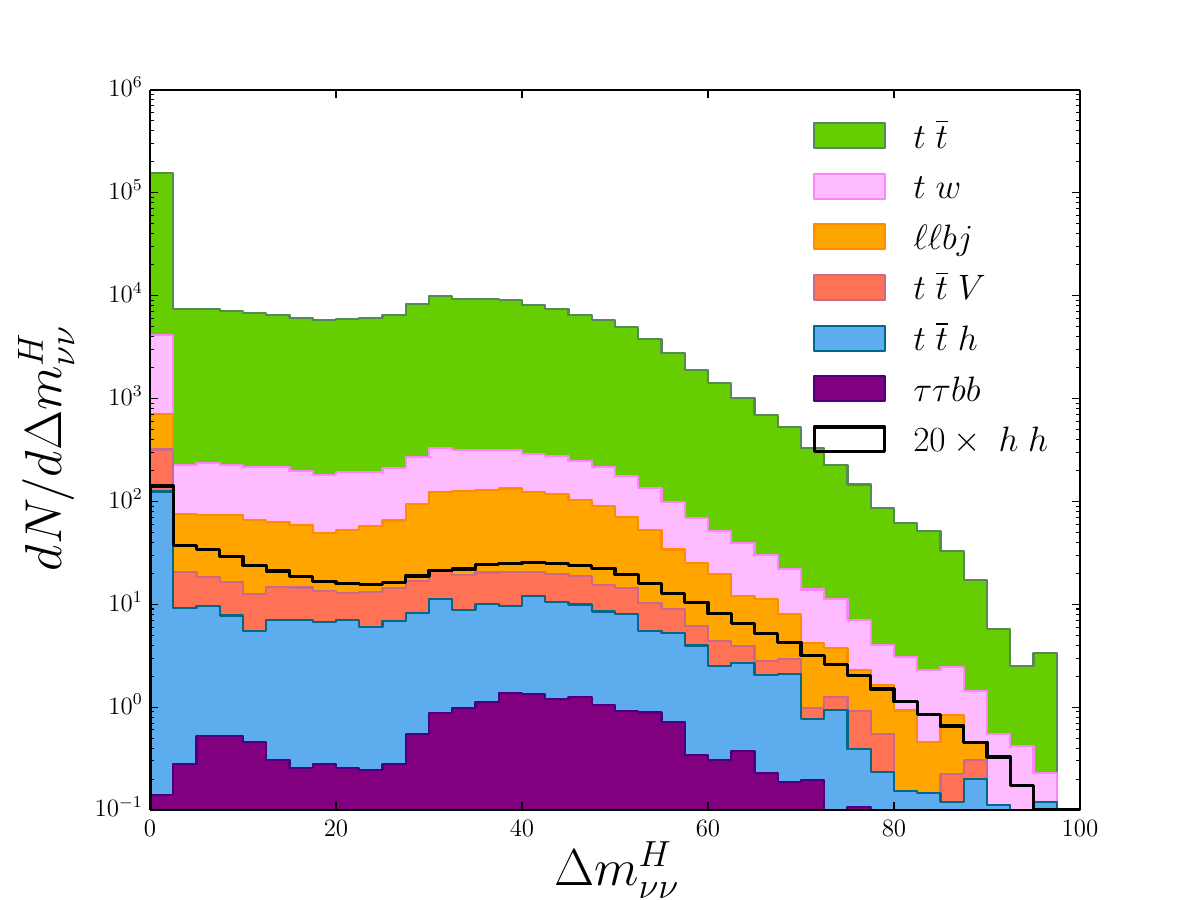} \hspace*{-0.7cm}
\includegraphics[width=0.51\textwidth,clip]{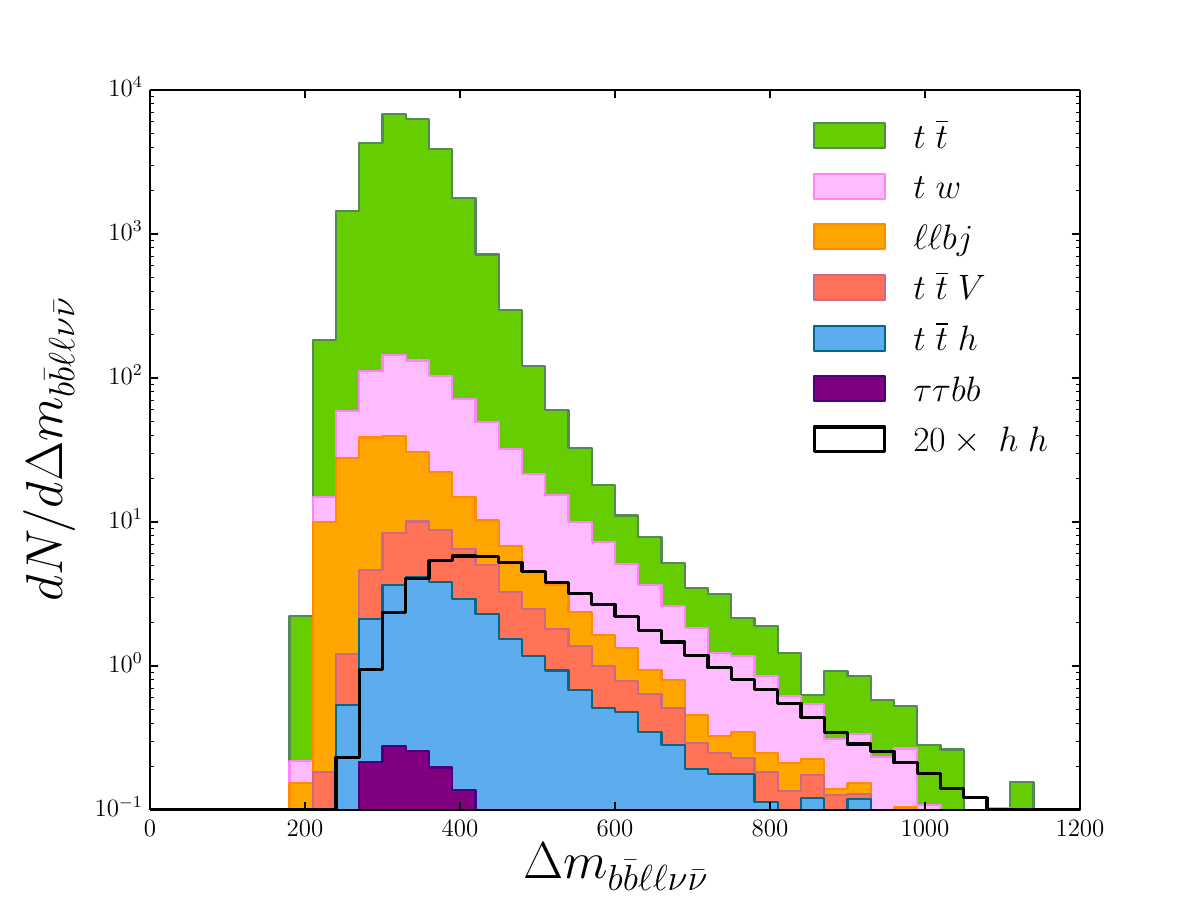} \\
\caption{$min[\Delta R_{b\ell}]$ (left, top), $\Delta R^{\text{T}}_{\nu \nu}$ (right, top), $\Delta R^{\text{H}}_{\nu \nu}$ (left, middle), $m^{\text{T}}_{\nu \nu}$ (right, middle), $m^{\text{H}}_{\nu \nu}$ (left, bottom), $m_{b\bar b \ell\ell\nu\bar\nu}$  with reconstructed neutrino momenta using Higgsness (right, bottom). 
\label{fig:5dists}}
\end{center}
\end{figure*}

\begin{itemize}
\item $p_T(\ell_1)$ and $p_T(\ell_2)$ are the transverse momentum of the hardest and the next hardest leptons, respectively. 
\item $p_{T\ell\ell}$ and $p_{Tbb}$ are the transverse momentum of the two-lepton system and the two-$b$-tagged jets, respectively. 
\item $\mpt$ is the missing transverse momentum. 
\item $\Delta R_{\ell\ell}$ and $\Delta R _{bb}$ are the angular separation between two leptons, between two $b$-tagged jets, respectively.  
The angular distance is defined by
\begin{eqnarray}
\Delta R_{ij} = \sqrt{(\Delta\phi_{ij})^2+(\Delta \eta_{ij})^2}, \label{deltaR}
\end{eqnarray}
where $\Delta\phi_{ij} = \phi_i-\phi_j$ and $\Delta\eta_{ij}=\eta_i-\eta_j$ denote the differences in the azimuthal angles and rapidities respectively between particles $i$ and $j$.
\item $\Delta\phi_{bb,\ell\ell}$ is the angular separation between the $b\bar b$ system and the two lepton system. 
\item $m_{\ell\ell}$ and $m_{bb}$ are the invariant mass of two leptons and two $b$-tagged jets, respectively. 
\item $\text{min}[\Delta R_{b\ell}]$ is the smallest angular distance between a $b$-jet and a lepton out of 4 possible combinations, which is shown in the left-upper corner in Fig. \ref{fig:5dists}. 
\item $\sqrt{\hat{s}}_{\text{min}}^{(bb\ell \ell)}$ and $\sqrt{\hat{s}}_{\text{min}}^{(\ell \ell)}$ are the minimum $\sqrt{\hat s}$ for the $(bb\ell \ell)$ subsystem and the $(\ell \ell)$ subsystem, respectively. The $\hat{s}_{\text{min}}$ variable \cite{Konar:2008ei,Konar:2010ma} is defined as 
\begin{equation} 
 \hat{s}_{\text{min}}^{({\rm v})} = m_{{\rm v}}^2 + 2 \left ( \sqrt{ |\vec P_{T}^{\rm v} |^2 + m_{\rm v}^2 }\ |\mptvec| - \vec P_{T}^{\rm v} \cdot \mptvec \right ) \, ,
 \label{smindef}
\end{equation}
where $({\rm v})$ represents a system of visible particles, and $m_{\rm v}$ ($\vec P_{T}^{\rm v}$) is corresponding invariant mass (transverse momentum). It provides a way to approximate the Mandelstam variable $\hat s$ of the system $({\rm v})$ in the presence of the missing momenta.
Here we consider two systems, ${\rm v}=\{bb\ell\ell\}$ and ${\rm v}=\{\ell\ell\}$. First, $\sqrt{\hat s}^{(bb\ell\ell)}_{\text{min}}$ represents a minimum energy required to produce two Higgs bosons (two top quarks) in the signal ($t\bar{t}$) events. Second, $\sqrt{\hat s}^{(\ell\ell)}_{\text{min}}$ represents a minimum energy required to produce two $W$ bosons. In case of the $t\bar{t}$ background, the peak of the distribution appears near $2m_W$. In case of the signal, the peak appears around the Higgs mass. The distributions of the variables $\hat{s}_{\text{min}}^{(bb\ell\ell)}$ and $\hat{s}_{\text{min}}^{(\ell\ell)}$ are shown in Refs. \cite{Kim:2019wns,Kim:2018cxf}.

\item $M_{T2}^{(b)}$ and $M_{T2}^{(\ell)}$ are the stransverse of ($b\bar b$) and $(\ell^+\ell^-)$ subsystem, respectively. 
The $M_{T2}$ \cite{Lester:1999tx} defined as
\begin{equation} 
M_{T2} (\tilde m) \equiv \min\left\{\max\left[M_{TP_1}(\vec{p}_{T \nu },\tilde m),\;M_{TP_2} (\vec{p}_{T \bar\nu },\tilde m)\right] \right\} \, , 
\label{MT2def}
\end{equation}
where $\tilde m$ is the test mass for the daughter particle, and the minimization over the transverse masses of the parent particles $M_{TP_i}$ ($i=1, 2$) is performed by scanning the unknown transverse neutrino momenta $\vec{p}_{\nu T}$ and $\vec{p}_{\bar\nu T}$, under the $\mptvec $ constraint. See Refs.~\cite{Barr:2011xt,Kim:2017awi,Cho:2014naa,Konar:2009wn,Konar:2009qr,Baringer:2011nh,Kim:2015uea,Goncalves:2018agy,Debnath:2017ktz} for more information and other variants of $M_{T2}$.

When $M_{T2}$ is applied to the $b \bar b$ visible system we abbreviate it as $M_{T2}^{(b)}$. In this case, the daughter particles are the $W$ bosons, and hence we set $\tilde m = m_W = 80$ GeV. By construction, $M_{T2}^{(b)}$ is bounded by the mass of the corresponding parent particle, so that its distribution exhibits a sharp drop at around $M_{T2}^{(b)} = m_t$, for $t\bar{t}$ events. Double Higgs production, on the other hand, does not obey this bound, which in turn can be used to disentangle two different samples.

When $M_{T2}$ is applied to the $\ell^+\ell^-$ visible system, we abbreviate it as $M_{T2}^{(\ell)}$. In this case, we set $\tilde m = m_\nu = 0$. The $M_{T2}^{(\ell)}$ distribution for the di-leptonic $\tau \tau b b$ events drops at around $\sim m_{\tau}$. This suggests that $M_{T2}^{(\ell)}$ can be used to discriminate the $\tau \tau b b$ background.

\item Topness measures a degree of consistency of a given event with $t\bar{t}$ production.
Topness itself is a minimized chi-square value constructed by using four on-shell constraints, $m_t$, $m_{\bar t}$, $m_{W^+}$ and $m_{W^-}$, and 6 unknowns (the three-momenta of the two neutrinos, $\vec p_{\nu}$ and $\vec p_{\bar\nu}$) 
\begin{eqnarray}
\chi^2_{ij} &\equiv& \min_{\tiny \mptvec = \vec p_{\nu T} + \vec p_{ \bar\nu T}}  \left [ 
\frac{\left ( m^2_{b_i \ell^+ \nu} - m^2_t \right )^2}{\sigma_t^4} \,   +
\frac{\left ( m^2_{\ell^+ \nu} - m^2_W \right )^2}{\sigma_W^4}  \, \right.  \label{eq:tt} \nonumber \\
&& \hspace*{1.9cm}\left . + \frac{\left ( m^2_{b_j \ell^- \bar \nu} - m^2_t \right )^2}{\sigma_t^4} \, +
\frac{\left ( m^2_{\ell^- \bar\nu} - m^2_W \right )^2}{\sigma_W^4}   \right ]  \; ,
\label{eq:Tness1}
\end{eqnarray}
subject to the constraint, $ \mptvec = \vec p_{\nu T} + \vec p_{ \bar\nu T}$. 
Due to the twofold ambiguity in paring a $b$-jet and a lepton, we define Topness as the smallest chi-square value
\begin{eqnarray}
T &\equiv&  { \min} \left ( \chi^2_{12} \, , \, \chi^2_{21} \right ) \, .
\label{eq:Tness2}
\end{eqnarray}
 
\item Similarly, Higgsness measures a degree of consistency of a given event with double Higgs production defined by
\begin{eqnarray}\nonumber
H &=& 
\min_{\tiny \mptvec = \vec p_{\nu T} + \vec p_{ \bar\nu T}}  \left [
 \frac{\left ( m^2_{\ell^+\ell^-\nu \bar\nu} - m^2_h \right )^2}{\sigma_{h_\ell}^4}    \right. 
 + \frac{ \left ( m_{\nu  \bar\nu}^2 -  m_{\nu\bar\nu, peak}^2 \right )^2}{ \sigma^4_{\nu}}
   \nonumber \\
 && \hspace*{1.1cm}   + {\min} \left ( 
\frac{\left ( m^2_{\ell^+ \nu } - m^2_W \right )^2}{\sigma_W^4} + 
\frac{\left ( m^2_{\ell^- \bar \nu} - m^2_{W^*, peak} \right )^2}{\sigma_{W_*}^4}  \, , \right. \label{eq:hww}  \\ \nonumber
 && \hspace*{2.3cm}    \left.  \left .
\frac{\left ( m^2_{\ell^- \bar \nu} - m^2_W \right )^2}{\sigma_W^4} + 
\frac{\left ( m^2_{\ell^+ \nu} - m^2_{W^*, peak} \right )^2}{\sigma_{W_*}^4}  
\right )  \right ]   \, ,
\label{eq:Hness}
\end{eqnarray}
where $m_{W^*}$ is bounded above $m_{W^*} \leq m_h- m_W$ and its location of the peak can be estimated by
\begin{eqnarray} 
m_{W^*}^{peak} &=& \frac{1}{\sqrt{3}} \sqrt{ 2 \left ( m_h^2 + m_W^2 \right ) - \sqrt{m_h^4 + 14 m_h^2 m_W^2 + m_W^4}} \approx 40 \text{ GeV} \, .
\end{eqnarray}
The location of the peak in the invariant mass $m_{\nu\bar\nu}$  distribution of two neutrinos appears at $m_{\nu\bar\nu}^{peak} \approx 30$ GeV. 
The definitions of Topness and Higgsness involve $\sigma$ hyperparameters which stand for experimental uncertainties and particle widths. In our numerical study, we use $\sigma_t=5$ GeV, $\sigma_W=5$ GeV, $\sigma_{W^*}=5$ GeV, $\sigma_{h_\ell}=2$ GeV, and $\sigma_\nu = 10 $ GeV. Our results are not sensitive to these numerical values. The Topness and Higgsness are shown in Fig. \ref{fig:TH}. See Ref. \cite{Alves:2022gnw} for the heavy Higgs decaying to two $W$ bosons. 
\begin{figure*}[t]
\begin{center}
\includegraphics[width=0.54\textwidth,clip]{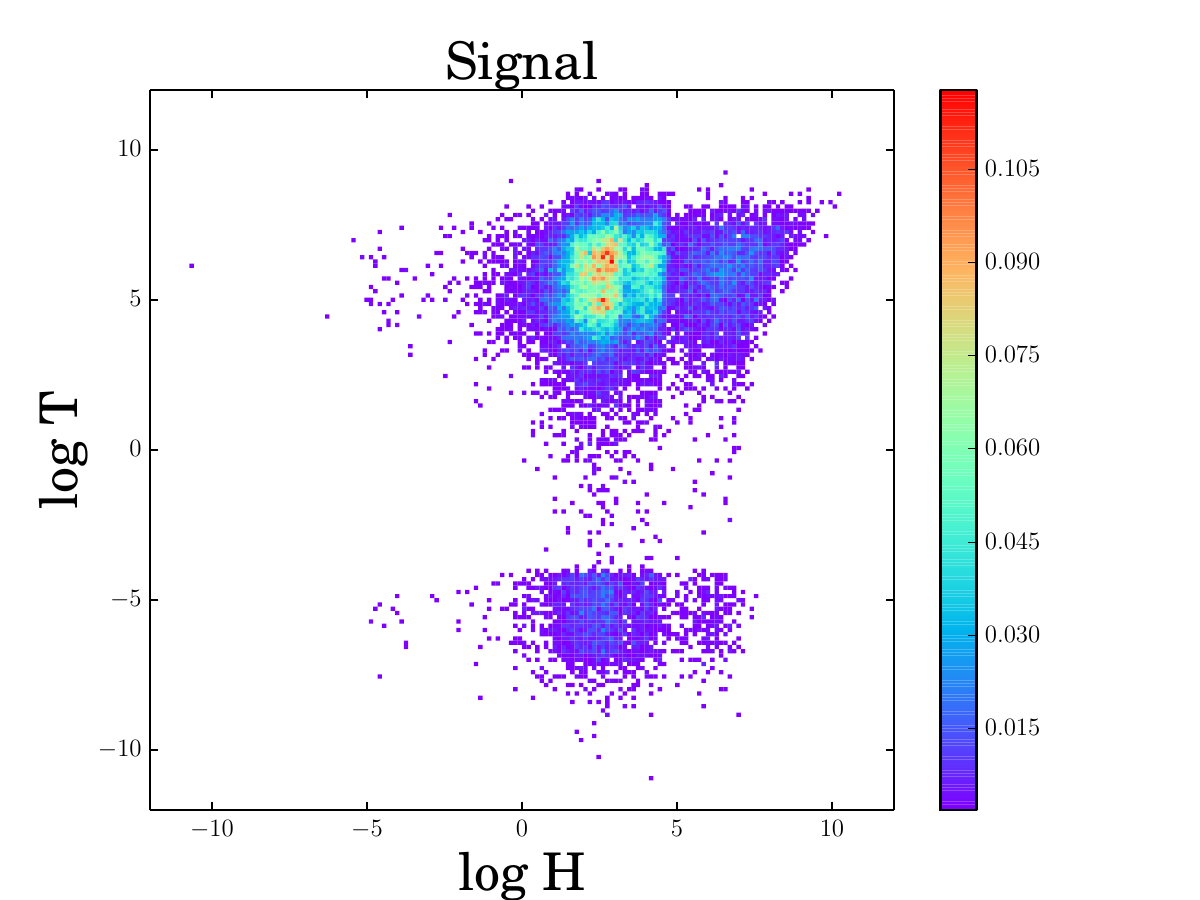}  \hspace*{-1.5cm}
\includegraphics[width=0.54\textwidth,clip]{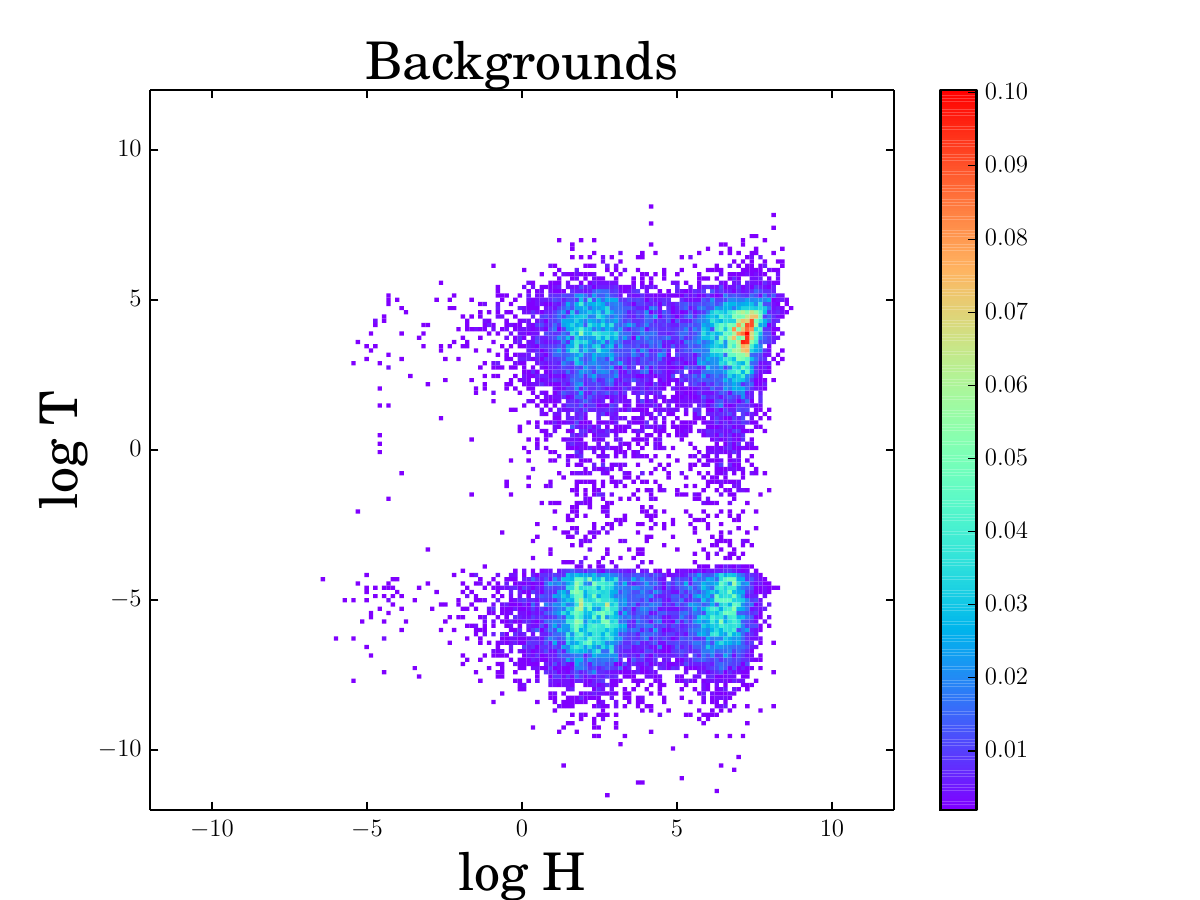}  
\caption{$\log T$ vs $\log H$ for signal (left) and all backgrounds (right). 
\label{fig:TH}}
\end{center}
\end{figure*}

\item $\Delta R^{\text{H}}_{\nu \nu}$ (left-middle panel in Fig. \ref{fig:5dists}) and $ \Delta R^{\text{T}}_{\nu \nu}$ (right-top panel in Fig. \ref{fig:5dists}) are the angular separations of the two neutrinos which are reconstructed using Higgsness and Topness, respectively. It is expected that they resemble $\Delta R_{\ell\ell}$. 
\item $m^{\text{H}}_{\nu \nu}$ (left-bottom panel in Fig. \ref{fig:5dists}) and $m^{\text{T}}_{\nu \nu}$ (right-middle panel in Fig. \ref{fig:5dists}) are the invariant mass of the two neutrinos which are reconstructed using Higgsness and Topness, respectively. It is expected that they resemble $m_{\ell\ell}$. 
\end{itemize}

Finally we would like to comment on the use of the reconstructed neutrinos. 
Some of the 21 kinematic variables which are defined in the laboratory frame show the global properties of the signal and background processes. For example, distributions of invariant masses ($m_{\ell\ell}$ and $m_{bb}$) and angular variables ($\Delta R_{\ell\ell}$, $\Delta\phi_{bb,\ell\ell}$, $\text{min}[\Delta R_{b\ell}]$, etc) show that the pencil-like production of $hh$ and isotropic production of $t \bar t$.
These are better measured via shape variables in the CM frame. 
Now that we have obtained approximate momenta of the missing neutrinos, we are able to Lorentz-boost to the CM frame of the production event by event. Among many shape variables, we consider the sphericity and thrust variables. 
\begin{figure*}[t!]
\begin{center}
\includegraphics[width=0.325\textwidth,clip]{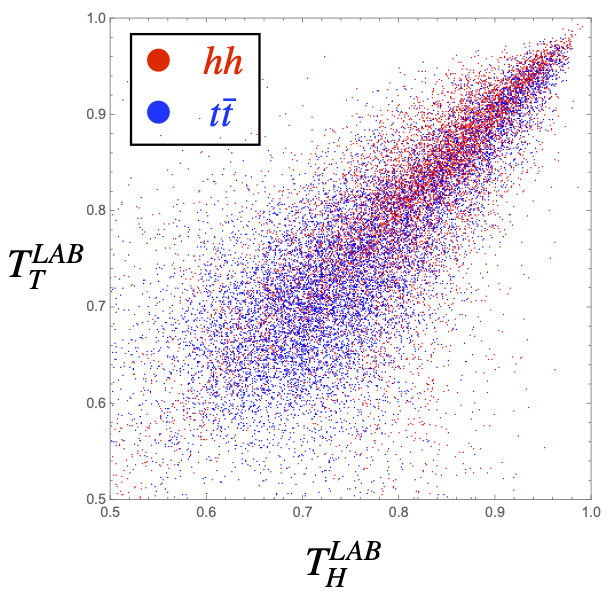}\hspace*{-0.1cm} 
\includegraphics[width=0.325\textwidth,clip]{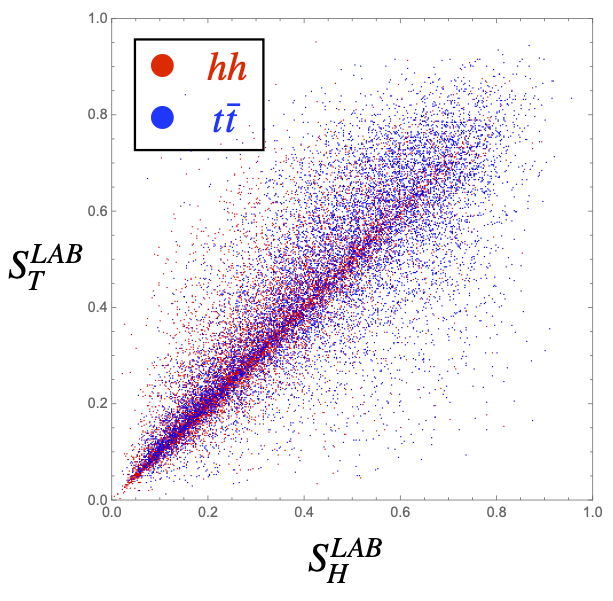}\hspace*{-0.1cm} 
\includegraphics[width=0.325\textwidth,clip]{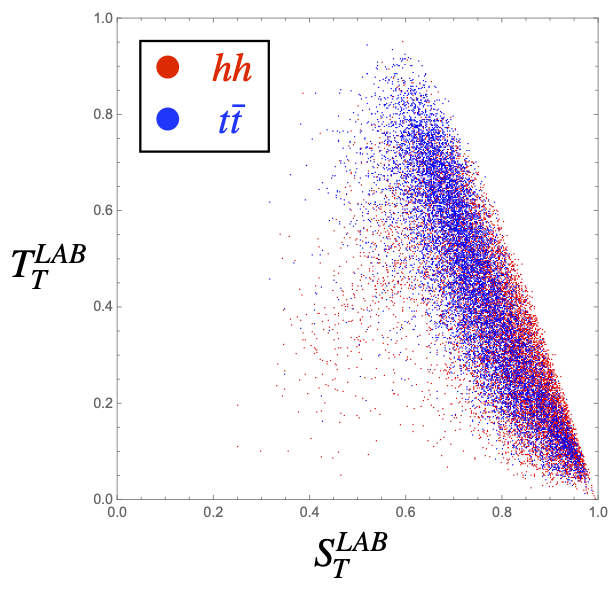} \\
\includegraphics[width=0.325\textwidth,clip]{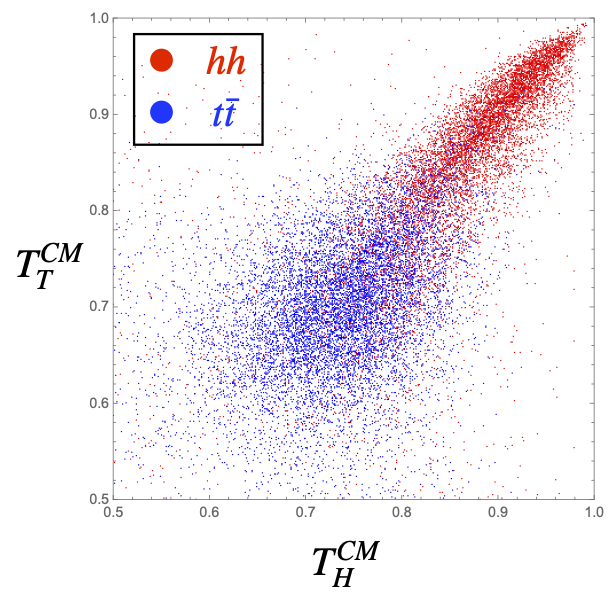}\hspace*{-0.1cm} 
\includegraphics[width=0.325\textwidth,clip]{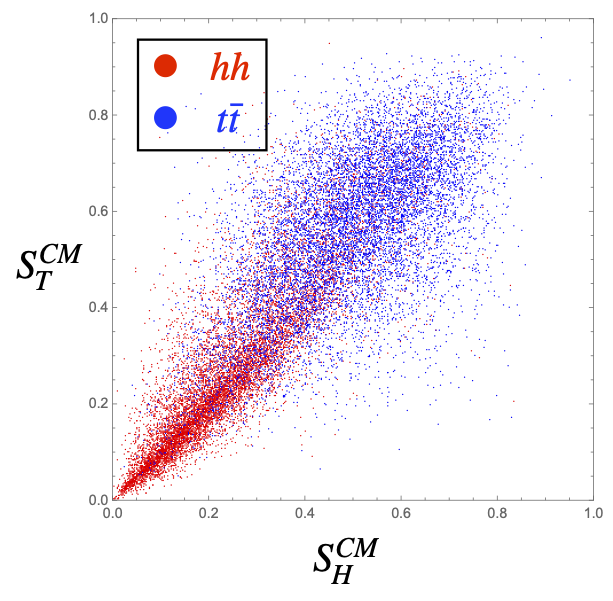}\hspace*{-0.1cm} 
\includegraphics[width=0.325\textwidth,clip]{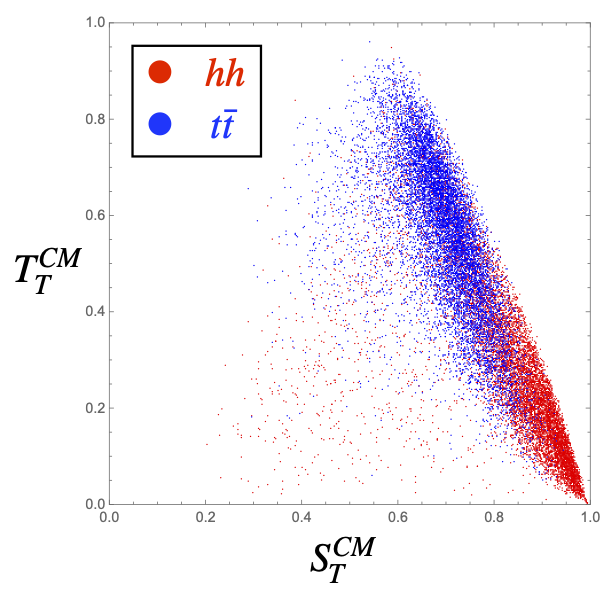}
\caption{Thrust (left), sphericity (middle) and thrust vs. sphericity (right) in the lab frame (top) and CM frame (bottom) for the $hh$ (red) and $t\bar t$ (blue) productions. 
The quantities with subscript `H' (`T') are obtained using Higgsness (Topness).
\label{fig:shapes} }
\end{center}
\end{figure*}
%

The thrust ($T$) is defined as
\begin{eqnarray}
T &=& \max_{ \vec n} \left ( \frac{\sum_i \Big | \vec p_i \cdot \vec n \Big | }{ \sum_i \Big | \vec p_i \Big |} \right ) \, ,
\end{eqnarray}
with respect to the direction of the unit vector $ \vec{n} $ which maximizes $ T $, identified as the thrust axis $\vec n_T$.
This definition implies that for $T = 1$ the event is perfectly back-to-back while for $T = 1/2$ the event is spherically symmetric. 
The sphericity ($S$) provides global information about the full momentum tensor ($M$) of the event via its eigenvalues:
\begin{eqnarray}
    M_{i j} &=& \frac{\sum_{a=1}^{n_j} p_{i a} p_{j a}}{\sum_{a=1}^{n_j} |\vec p_a|^2} \, , 
\end{eqnarray}
where $i,j$ are the spatial indices and the sum runs over all jets. The ordered eigenvalues $\lambda_i$ ($\lambda_1 > \lambda_2 > \lambda_3$) with the normalization condition $\sum_i \lambda_i = 1$ defines the sphericity
\begin{eqnarray}
S   &=& \frac{3}{2} \Big ( \lambda_2 + \lambda_3 \Big )\, .
\end{eqnarray}
The sphericity axis $\vec n_S$ is defined along the direction of the eigenvector of $\lambda_1$. 
The sphericity measures the total transverse momentum with respect to the sphericity axis defined by the four momenta in the event. In other words, the sphericity of an event is a measure of how close the spread of energy in the event is to a sphere in shape. The allowed range for $S$ is $0 \leq  S < 1$. For a pencil-like production, $S\approx0$, while $S\approx 1$ for isotropic production.

We obtain neutrino momenta using Higgsness and Topness, which in fact show similar distributions statistically, as shown in the neutrino images. Fig. \ref{fig:shapes} shows the thrust (left), sphericity (middle) and thrust vs. sphericity (right) in the lab frame (top) and CM frame (bottom) for the $hh$ (red) and $t\bar t$ (blue) productions. 
The numerical values of ($T_H^{LAB}$, $S_H^{LAB}$, $T_H^{CM}$, $S_H^{CM}$) are obtained using Higgsness, 
while ($T_T^{LAB}$, $S_T^{LAB}$, $T_T^{CM}$, $S_T^{CM}$) are obtained using Topness. 
Two plots in the right panel show the correlation of the thrust and sphericity, taking Topness minimization for the neutrino momenta. We do not show results with Higgsness, as they are very similar.